\documentclass[11pt]{article}
\usepackage{times,amsmath,amsthm,amsopn,amsfonts,cite,mathrsfs,latexsym,color,graphicx}
\usepackage{amssymb}
\usepackage{amsmath}
\usepackage{amsfonts}
\usepackage{amstext}
\usepackage{amsbsy}
\usepackage{graphics}
\usepackage{wrapfig}
\usepackage{float}
\usepackage{stmaryrd} %for \boxbox symbol... substitute \boxdot when this package is not included
\usepackage[active]{srcltx} % for inverse search
\usepackage{verbatim} %for the \begin{comment}, \end{comment}
\newcommand{\grad}{\vec{\nabla}}

\newcommand{\Og}{\Omega}

\DeclareMathOperator{\erfc}{erfc}
\DeclareMathOperator{\erf}{erf}
\makeatletter						
\def\Eqlfill@{\arrowfill@\Relbar\Relbar\Relbar}
\newcommand{\extendEql}[1][]{\ext@arrow 0359\Eqlfill@{#1}}
\makeatother

\title{Detection of small low emission sources - case studies}
\author{A.~Olson, A.~Ciabatti, Y.~Hristova, P.~Kuchment, J.~Ragusa, W.~Charlton, M.~Allmaras
\footnote{aolson@math.tamu.edu, yuliagh@umd.umich.edu, kuchment@math.tamu.edu, jean.ragusa@tamu.edu, wcharlton@tamu.edu, moritz.allmaras@siemens.com}}
\date{}
\begin{document}
\maketitle
\begin{abstract}
%\textcolor[rgb]{1.00,0.00,0.00}{\textbf{Enter abstract}}
The article is devoted to a further study of the Compton camera method of passive detection of small amounts of special nuclear materials, developed by the authors in their previous work. Various cargo scenarios, detector errors, and other issues are addressed.
\end{abstract}

%%%%%%%%%%%%%%%%
\section{Introduction}
%%%%%%%%%%%%%%%%

%\marginpar{\textcolor[rgb]{1.00,0.00,0.00}{\textbf{Extend Introduction. Description of results}}}
%the results for a 30 second exposure time for just the background (top row) and with the HEU source (bottom row).
%Detected particle counts: 145,391 for background only and 210,178 for background plus HEU source (with 134 ballistic)
%In the nuclear contraband scenarios we consider here, a small amount of Highly Enriched Uranium (HEU) is hidden in
%a significantly larger volume. This is representative, for instance, of an attempt to smuggle HEU in shipping cargo containers.
%In our application, detection of gamma
%radiation is to be carried out in a passive manner, i.e., by monitoring the radiation reaching the detectors, without any active probing of the container using external radiation sources.
%The work reported here tests the inverse problem method developed in \cite{AHKK}, utilizing more realistic scenarios.

In \cite{AHKK}, some of the authors of this work developed and conducted first numerical testing of algorithms that allow the detection of small,
very low emission sources embedded in a much larger background (see also \cite{Xun} for an alternative Bayesian technique).
The algorithms developed required using some directional information for particles hitting the detectors.
Since collimation is not a viable option at the extremely low SNR levels (on the order of $.1\%$) considered, it was suggested to use the so called Compton type cameras,
which are currently available for $\gamma$ particles detection and are being developed for neutrons \cite{shield,SpCharlton}.

The techniques of \cite{AHKK}, developed and tested under some rather restrictive assumptions, require further detailed numerical scrutiny under more complex cargo scenarios than the ones of \cite{AHKK}. Indeed:
\begin{itemize}
\item Some assumptions of \cite{AHKK} (e.g., unstructured random background without cargo influence) are not always fulfilled in real world situations.
\item Some of the arguments were not entirely rigorous, e.g., the ones that estimated theoretically the sensitivity and specificity levels of the detection method under a random variable independence assumption, which was fulfilled only approximately.
\item The forward modeling was done in a way that did not allow for complex cargo scenarios nor spatially varying absorption and scattering cross sections.
\item Uncertainty in the direction detection (which is better for $\gamma$ cameras, but can go up to standard deviation of $12$ degrees for neutron Compton type cameras) was not taken into account.
\end{itemize}
Overcoming these problems is the goal of the current text. Clearly, there is no silver bullet here, which would assure detection under all possible circumstances.
Thus, a thorough study of the efficiency of the previously developed technique is called for. Due to technical constraints of numerical implementation, we only address the $2D$ case here.
The implications of this restriction are discussed in Section \ref{S:remarks}. The $3D$ case, as well as other extensions, will be treated in a subsequent publication.

The outline of the article is as follows: Necessary physics preliminaries are provided in Section \ref{S:physics}. Section \ref{S:forward} contains the description of the technique employed to solve the radiation transport problem through a cargo container (forward model).
The next Section \ref{S:math} addresses briefly the inverse problem algorithm of \cite{AHKK}.
The main Section \ref{S:cases} provides the detection results for various cargo container scenarios and their discussions.
Final remarks and conclusions are provided in Section \ref{S:remarks}.
In particular, while in the examples of Section \ref{S:cases} the presence of a source was detected by finding its location, numerical evidence is provided of possible indications of presence of a source without detecting its actual position.

%%%%%%%%%%%%%%%%%%%%%
\section{Physics preliminaries}\label{S:physics}
%%%%%%%%%%%%%
In this Section, we discuss the physics of gamma radiation pertaining to the application of interest.

The technique devised in \cite{AHKK} is based on the availability of at least a small number of uncollided (``ballistic'') particles emanating from a small localized source, versus other particles
(background particles and source particles that have scattered).
Of particular interest as uncollided particles are the gamma emitted from characteristic lines of Uranium.
By having relatively narrow energy bins around such lines, one can enhance the signal-to-noise ratio for gamma radiation
detected in such energy ranges. However, the four main lines for Uranium-235 have somewhat low energies
(143.8, 166.3, 185.7, and 205.3 keV) and can be attenuated easily by some amount of shielding material. For instance, particles emitted
from the 185-keV line of U-235 (the line with the strongest activity) have an average mean-free-path in high-Z materials
of about 0.69 \cite{PANDA}, and thus can be significantly attenuated.
On the other hand, Uranium-238 possesses gamma lines at higher energies (i.e., a line at 1.001 MeV) for which the
mean-free-path in high-Z materials is significantly larger (13.3 mean-free-paths) \cite{PANDA}. Therefore, it will be advantageous
to consider, as our signal of interest, the uncollided gamma rays from the 1.001 MeV line of U-238.

However, other gamma particles may also be present in the energy range of interest. In our application, background radiation will
be stemming from the concrete base located some distance below the container. Significant gamma lines (above the 1.001 MeV line of U-238)
should be taken into consideration as background radiation impinging the container. These lines include: the 1.461 MeV line
from Potassium-40, the 1.12 MeV and 1.76 MeV lines from Bismuth-214, and the 2.61 MeV line from Thallium-208 (Bismuth and
Thallium are daughter products from the decay of Uranium-238 and Thorium-232, respectively, and are naturally found in concrete).

Assuming a concrete slab thickness of 40 cm and 60 cm gap between the concrete surface and the bottom of the container, the background
radiation emitted from the concrete slab and impinging the lateral and bottom sides of the container can be easily computed.
% (using the general purpose N-particle code MCNP \cite{MCNP}, for instance).

Gamma rays due to the HEU source, any legitimate commercial source (if any), and the background radiation entering the container
will undergo scattering and absorption within the volume of the container. Absorption and scattering within the container will reduce
the amount of uncollided HEU source particles reaching the detectors placed around the container; recall that the signal of
interest in our study are the gamma rays from the 1.001 MeV line of U-238. Scattering will also cause gammas from energies higher
than 1.001 MeV to down-scatter into the energy of interest, adding noise to the signal.

%Knowing the source of ballistic particles of interest (gamma rays from the 1.001 MeV line of U-238),

%%%%%%%%%%%%%%%%%%%%%%%%%%%%%
\section{Forward radiation transport modeling}\label{S:forward}
%%%%%%%%%%%%%%%%%%%%%%%%%%%

The radiation transport within the cargo container is governed by the linear Boltzmann equation, written below using
the multigroup approximation:
\begin{equation} \label{eq:photon_transport}
\vec{\Og} \cdot \grad \Psi^g(\vec{r},\vec{\Og}) + \Sigma_t^g(\vec{r}) \Psi^g(\vec{r},\vec{\Og})
= \sum_{g^\prime=1}^G \sum_{\ell=0}^L \Sigma_{s,\ell}^{g^\prime \rightarrow g}(\vec{r})
  \sum_{m=-\ell}^{m=+\ell} \Phi^{g^\prime}_{\ell,m}(\vec{r}) + Q^g(\vec{r},\vec{\Og})
%\quad \forall \vec{r} \in \mathcal{D}
\end{equation}
for all positions $\vec{r} \in \mathcal{D}$, all directions $\vec{\Og} \in S^2$, and all energy groups $g \in [1,G]$, where
$\mathcal{D}$ is the volume of the cargo container, $S^2$ the unit sphere, $G$ the total number of energy groups,
$\Psi^g$ the photon angular flux in energy group $g$, $\Sigma_t^g$ the total interaction cross section in group $g$,
$\Sigma_{s,\ell}^{g^\prime \rightarrow g}$ the $\ell^{\text{th}}$-Legendre moment of the scattering cross section from
group $g^\prime$ to group $g$, $L$ is the maximum anisotropy expansion order, and $Q^g$ is the volumetric external source
of photons in group $g$ (stemming from the HEU source and any other volumetric source, if present). The moments of the
angular flux are given by
\begin{equation}
\Phi^g_{\ell,m}(\vec{r})  = \int_{4\pi}d\Og Y_{\ell,m}(\vec{\Og})  \Psi^g(\vec{r},\vec{\Og})
\end{equation}
where $Y_{\ell,m}$ is the spherical harmonic of order $\ell$ and degree $m$.
Eq.~(\ref{eq:photon_transport}) is supplied with boundary conditions:
\begin{equation} \label{eq:bc}
 \Psi^g(\vec{r},\vec{\Og}) = h^g(\vec{r},\vec{\Og})
\quad \forall \vec{r} \in \partial\mathcal{D}^{-}
\end{equation}
where $\partial\mathcal{D}^{-}$ is the incoming boundary defined as
$\partial\mathcal{D}^{-} = \{ \vec{r} \in \partial\mathcal{D} \text{ such that } \vec{\Og}\cdot\vec{n}(\vec{r}) < 0\}$
with $\vec{n}(\vec{r})$ the outward unit normal vector at position $\vec{r}$. The boundary condition function $h^g$
describes the background radiation impinging on the container due to the presence of a large concrete slab underneath the
container.
Photon cross sections for various representative materials have been generated using NJOY-99 \cite{NJOY}. The materials
compositions are given in Table~\ref{tab:MaterialCompositions}. The multigroup structure employed is provided
in Table~\ref{tab:EnergyGroupStructure}. Energy groups numbered 1, 18, 25, 32, 37 are purposefully narrow to represent accurately
background radiation lines and the U-238 1.001 MeV line.

\begin{table}
		\centering
		\begin{tabular}{|l||l|} \hline
Plastic   & C (33.3\%), H(66.7\%) \\ \hline
Wood      & O (25\%), C (25\%), H (50\%) \\ \hline 	
Cotton    & O (25\%), C (25\%), H (50\%) \\ \hline 	
Iron      & Fe (100\%) \\ \hline 	
HEU       & Uranium (100\%) \\ \hline 	
Concrete  & H (0.5\%), O (48.7\%), Na (1.7\%), Mg (2.5\%), Al (4.5\%), Si (30.8\%), K (1.9\%), \\
          & Ca (8.1\%), Fe (1.2\%), traces of Th-232 and U-238 and their daughters \\ \hline 	
Air       & N (78.1\%), O (20.9\%), Ar (0.94\%), H (0.06\%) \\ \hline 	
Fertilizer& K (13.9\%),  N (46.8\%), O (30\%), P (9.3\%) \\ \hline 	
    \end{tabular}
	\caption{Material Compositions}
	\label{tab:MaterialCompositions}
\end{table}

\begin{table}
		\centering
  	\begin{tabular}{|c||c|c|} \hline
		Group index   &  Lower bound (MeV)  & Upper bound (MeV)  \\ \hline \hline
                         1  &                 2.61449     &              2.61451  \\ \hline
                         2  &                 2.56137     &              2.61449  \\ \hline
                         3  &                 2.50824     &              2.56137  \\ \hline
                         4  &                 2.45512     &              2.50824  \\ \hline
                         5  &                   2.402     &              2.45512  \\ \hline
                         6  &                 2.34887     &                2.402  \\ \hline
                         7  &                 2.29575     &              2.34887  \\ \hline
                         8  &                 2.24262     &              2.29575  \\ \hline
                         9  &                  2.1895     &              2.24262  \\ \hline
                        10  &                 2.13638     &               2.1895  \\ \hline
                        11  &                 2.08325     &              2.13638  \\ \hline
                        12  &                 2.03013     &              2.08325  \\ \hline
                        13  &                   1.977     &              2.03013  \\ \hline
                        14  &                 1.92388     &                1.977  \\ \hline
                        15  &                 1.87076     &              1.92388  \\ \hline
                        16  &                 1.81763     &              1.87076  \\ \hline
                        17  &                 1.76451     &              1.81763  \\ \hline
                        18  &                 1.76449     &              1.76451  \\ \hline
                        19  &                 1.71391     &              1.76449  \\ \hline
                        20  &                 1.66333     &              1.71391  \\ \hline
                        21  &                 1.61275     &              1.66333  \\ \hline
                        22  &                 1.56217     &              1.61275  \\ \hline
                        23  &                 1.51159     &              1.56217  \\ \hline
                        24  &                 1.46101     &              1.51159  \\ \hline
                        25  &                 1.46099     &              1.46101  \\ \hline
                        26  &                 1.40421     &              1.46099  \\ \hline
                        27  &                 1.34743     &              1.40421  \\ \hline
                        28  &                 1.29065     &              1.34743  \\ \hline
                        29  &                 1.23387     &              1.29065  \\ \hline
                        30  &                 1.17709     &              1.23387  \\ \hline
                        31  &                 1.12031     &              1.17709  \\ \hline
                        32  &                 1.12029     &              1.12031  \\ \hline
                        33  &                 1.09047     &              1.12029  \\ \hline
                        34  &                 1.06065     &              1.09047  \\ \hline
                        35  &                 1.03083     &              1.06065  \\ \hline
                        36  &                 1.00101     &              1.03083  \\ \hline
                        37  &                 1.00099     &              1.00101  \\ \hline
		\end{tabular}
	\caption{Energy group structure}
	\label{tab:EnergyGroupStructure}
\end{table}

For demonstration purposes, calculations will be carried out in two-dimensional space.
The photon transport equation, Eq.~(\ref{eq:photon_transport}), is discretized using standard techniques:
\begin{enumerate}
\item An $S_n$ product Gauss-Legendre-Tchebychev angular quadrature \cite{RASA} is employed (with a few polar angles but a very high
number of azimuthal angles to resolve properly the angular distribution in the 2D domain).
\item Spatial discretization based on a standard bilinear discontinuous finite element technique with upwinding at cell interfaces \cite{HIRE,MOWA}.
\item Transport sweeps and Source Iteration are employed to converge the scattering source and determine the angular photon flux, solution of
Eq.~(\ref{eq:photon_transport}) \cite{Lewis_book}.
\end{enumerate}
Once the transport equation, Eq.~(\ref{eq:photon_transport}), has been iteratively solved, the outgoing angular photon flux at any
boundary edge in 2D is recorded for subsequent utilization in the inverse problem algorithm.

%%%%%%%%%%%%%%%%%%%%%%%%%%%%%
\section{Mathematics of the detection technique}\label{S:math}
%%%%%%%%%%%%%%%%%%%%%%%%%%%%%%%

We now recap some of the analysis and algorithms of the previous paper \cite{AHKK} (a different, Bayesian approach was developed in \cite{Xun}).

%%%%%%%%%%%%%%%%%%%%%%%%%%%%%%%
\subsection{2D with Uniform Random Background}
%%%%%%%%%%%%%%%%%%%%%%%%%%%%%5
Here we consider a $2D$ version of the problem. We assume that a small, low emission source of radioactive material (e.g., HEU) is placed inside of a significantly larger cargo. The ballpark linear size ratio is assumed to be on the order of $100$ - several meters long cargo container vs. several centimeters long source. The cargo
will be depicted in our simulations as a square region surrounded by arrays of detectors. A significantly larger, in comparison with the source emission, uniform and isotropic random radiation background is present.
Our goal is to detect the presence of a source, or to claim its absence. Since we are going to consider situations of extremely low signal-to-noise ratio,
it is not expected that detection would be possible, unless the detectors are, at least to some degree, direction sensitive.

Collimating the detectors, as typically done in medical emission imaging (e.g., SPECT \cite{BGH,Natterer_old,Natterer_new}), provides the incoming directions of detected particles.
However, collimation would most surely decimate the already low source signal to imperceptible levels.
In fact, a similar situation occurs in SPECT, but the signal-to-noise ratio there is still much larger than in our intended application.

One way to retain the SNR while obtaining some directional information is to use Compton $\gamma$-cameras \cite[and references therein]{AHKK,BaskoZengGull_patent,BaskoZengGull97_1,BaskoZengGull98,CreeBones,Gunter08}.
Compton cameras do not use collimation, but correspondingly provide less specific directional information: a surface cone of possible particle directions is detected,
rather than a specific direction. Analogous detectors (albeit based on somewhat different principles) are currently being developed for neutron detection \cite{shield,SpCharlton}.

Neutron and $\gamma$ radiations are the main signals one can rely upon in determining the presence of illicit HEU.
We will use the name ``Compton type camera'' without making a distinction between the types of particles considered.
The reader can think, for instance, in terms of $\gamma$-photons being detected (the assumption adopted further in this text).

In \cite[see also references therein]{AHKK}, some ``electronic collimation'' techniques were developed and tested for converting Compton data into data that would have come from physically collimated detectors.
There is some limited SNR loss attached to this conversion (see Section \ref{S:remarks}).
We will thus operate in the main body of this text as if we could collect the true directional information of incoming particles.

The detection technique suggested in \cite{AHKK} is based upon the assumption of existence of at least a small number of ballistic particles from the source reaching the detector arrays.
The scattered source particles are considered to be a part of the uniform background.
Correspondingly, the SNR  of interest is defined as the ratio of the number of detected ballistic source particles to the number of all other detections.
We will be working with signal-to-noise ratios of the order of $s=0.001$, or $0.1\%$, which is in line with numerical simulations of a shielded HEU source placed in a cargo container.

It was shown in \cite{AHKK} that the standard analytic tomographic techniques fail in this situation.
It was argued, however, that a simple backprojection (see Section \ref{S:backproj}) might create reliably detectable bumps (measured in standard deviations from the mean) at the source location.
A simple probabilistic analysis showed that a reliable (with $95\%$ confidence) detection can be expected, if the total number $N$ of detected particles reaches
\begin{equation}\label{E:rule}
N \;\;\propto\;\; \left(\frac{8}{\mathrm{SNR}}\right)^2 p(1-p),
\end{equation}
where $p$ is the ratio of the radius of the possible source to the one of the cargo container ($p=0.01$ in our simulations).
It was also shown how to estimate specificity, and it was demonstrated that one can achieve simultaneously high levels of both.

%%%%%%%%%%%%%%%%%%%%%%%%%%%
\subsection{The Backprojection Method}\label{S:backproj}
%%%%%%%%%%%%%%%%%%%%%%%%%%%
In this section we briefly describe the backprojection algorithm used in \cite{AHKK}. In that paper, uniform random background noise and a small low emission point source were simulated.
It was assumed that a square container of interest was surrounded along all four sides by detector arrays which record the location (more precisely, the detector bin)
of collision of each particle as well as its exact incoming direction. A backprojection image of the container was created by back propagating each particle from the
detector into the numerical domain along its incoming line (which, in general, is different from the original zig zag trajectory of the particle).
A fast rasterization algorithm was used to add a value of $1$ to all pixels intersected by the line. Thus, in the final image,
the value at every pixel equaled the number of lines intersecting it. Note that if a source is present, all backprojected ballistic source particles pass through the source location.

After backprojection was completed, the image was tested for unusual peaks, that is, pixels with values several standard deviations above the mean.
Since high deviations are very unlikely to appear in a uniform random background, their presence signals a possible source.
Confidence values were assigned to detected sources following a probabilistic analysis similar to that in Section \ref{S:probab}.
However, it was assumed in \cite{AHKK} that exact incoming directions of particles could be measured,
while in the current work we use a more realistic detector model which takes into account angular errors.
In \cite{AHKK}, based on the assumption of absence of angular errors, it was shown that high confidence of detection and high sensitivity are to be expected when
the total number of detected particles satisfies relation (\ref{E:rule}).

It is worth mentioning that in the same paper a more delicate situation was also considered, where only a ``gate'' of three arrays was present.
The issue in this case was that backprojected data was non-uniform anymore, due to many particles being undetected.
In this case, the mean and standard deviation were computed locally in a neighborhood of each pixel. A similar situation will also arise in the current work,
the background being non-uniform not only due to missing detectors but also as a consequence of external radiation sources (e.g. concrete flooring) and different materials filling the cargo.

%%%%%%%%%%%%%%%%%%%%%%%%%%%
\subsection{Using Compton type cameras}\label{S:compton}
%%%%%%%%%%%%%%%%%%%

As it has already been mentioned, several methods were developed in \cite{AHKK} to convert Compton type data to one that would come from truly collimated detectors.
Such transforms have been considered in the literature before (e.g. \cite{BaskoZengGull98,BaskoZengGull97_1,CreeBones,TN_fixedaxis,Smith05,MaxFranProst09}).
The caveat here is that Compton data is significantly over-determined, since the space of cones with apex at a detector is three-dimensional in $2D$ and five-dimensional in $3D$.
The solution adopted in most works addressing Compton camera imaging was to restrict the data to a two-dimensional set in $2D$ and a three-dimensional one in $3D$.
However, this approach is unfeasible in the detection problem discussed here, since restriction would essentially eliminate all useful signal.
Thus, methods involving all Compton data had to be developed and numerical implementation needed to be done in a way that avoided the speed and memory problems arising due to multi-dimensionality.
The main conversion method considered in \cite{AHKK} was equivalent to backprojection of each particle along its cone of possible incoming directions.
Thus, the values at every pixel in the backprojected image equaled the number of cones\footnote{Note that in the two dimensional setting,
a cone simply consists of two lines intersecting at a point on the detector array.} intersecting it.

In this text we restrict ourselves to using the directional data that would have been obtained by converting from Compton camera measurements.

%%%%%%%%%%%%%%%%%%
\subsection{Incorporation of angular errors }\label{S:probab}
%%%%%%%%%%%%%%%%%

Here we introduce a modification to the treatment of \cite{AHKK} that incorporates possible errors in detection of directions of incoming particles.

Consider a circular domain of interest of radius $R$ surrounded by detector arrays. Suppose the detectors are collimated physically or electronically, so that directional information of all detected particles can be determined.
We assume that the angular information recorded by the detectors has a Gaussian error with standard deviation $\alpha$ degrees. Let us partition the source region into square pixels of dimension $r$ by $r$ and define $p = r/R$.
Then the area contains approximately $M = \pi/p^2$ pixels. We also denote the total number of detected particles by $N$.
Suppose $n$ lines (backprojected particle trajectories) intersect a given pixel $B$ and write $n = n_s + n_b$ where $n_s$ is the number of ballistic particles from the source and $n_b$ is the number of background particles.
Under the conditions of our scenarios, one can apply the Central Limit Theorem and consider the number $n$ of lines intersecting a pixel being normally distributed with the mean $\mu=pN$ and standard deviation $\sigma=\sqrt{Np(1-p)}$.

Let us now choose a threshold value $k_t$ such that the probability of a normal variable to reach more than $k_t$ standard deviations above the mean in at least one out of $M\sim 10^4$ independent realizations\footnote{$10^4$ corresponds to the number of pixels.}
is very small. We use in our experiments the number $k_t=4.3$. Note that a useful $k_t$ will change depending on the size of the pixels relative
to the size of the whole domain, or equivalently, on the value of $p$. (We will revisit this shortly.) We make the claim that a source has been detected if an abnormally high number $n$
of lines pass through a pixel $B$, that is, if $n > n_t := \mu + k_t \sigma$.

The probability of at least $n_t$ lines crossing $B$ due to random reasons is approximately $r_t := 0.5 \erfc(k_t/\sqrt{2})$. Taking into account the number of pixels $M$,
the probability of at least $n_t$ lines crossing at least one of the pixels due to random reasons (a ``\textbf{false positive}'') can be approximated by
\begin{equation}\mbox{fp rate} = 1 - (1-r_t)^{M}.\end{equation}
The \textbf{true negative} rate is then given by $\mbox{tn rate} = 1 - \mbox{fp rate}$ or
\begin{equation}
\kappa := \mbox{tn rate} = \mbox{confidence} := (1 - r_t)^{M} = \left[1 - 0.5 \erfc\left(k_t / \sqrt{2}\right)\right]^{M}.
\end{equation}
Now coming back to the value of $k_t$, suppose we want to have a minimum confidence level $\kappa \in (0,1)$. Then $k_t$ must satisfy
\begin{equation}
k_t > \sqrt{2} \; \erfc^{-1} \left[2\left(1 - \kappa^{1/{M}}\right)\right].
\end{equation}
For example, if we need a true negative rate of at least $99\%$ and we have $M= 10^4$ pixels, then we need to choose $k_t > 4.753$. If we are satisfied with a true negative rate of only $90\%$ and have just $M = 10^3$ pixels, then $k_t > 3.706$ will suffice. Recall that $M$ is dependent on $p$, so we can instead write in the inequality for $k_t$ as
$$
k_t > \sqrt{2} \; \erfc^{-1} \left[2\left(1 - \kappa^{p^2/\pi}\right)\right].
$$
Notice that this consideration is still valid when errors in direction determination are present

Now consider the probability of missing a source that is present. Here errors might play significant role. Assume that we have an \emph{a priori} ballpark estimate for $n_s$.
In $\sigma$ units, $n_s = k_s \sigma$ where $k_s$ is a positive constant and $\sigma$ is the standard deviation of the distribution of the number of background trajectories passing through a pixel.
However, since our detector has errors in its measurements, not all of the detected source particles detected are back projected through the pixel containing the source.
Let $\alpha$ be the standard deviation of the error of the angular measurement. Then if a particle is emitted from a source inside a pixel $B$ and is detected on the circular array,
the probability that it will be back projected through $B_r$ is related to the size of the pixel $r$, the distance from the boundary $d$, and the standard deviation of the angular error $\alpha$.
Since the location of the source is not known \emph{a priori}, we do not know what $d$ is. Consider the worst case example where $d = R$ (if the source is close to the boundary, detection is easier).
Then the detected particle is back projected through $B$ if and only if $\theta < \arctan (r/R) = \arctan p$.
%\begin{center}
%\includegraphics[height=2.5in]{Backproject.png}\qquad
%\includegraphics[height=2.5in]{C_p-alpha.png}
%\end{center}
Let $C(p,\alpha)$ be defined as
$$
C(p,\alpha):= \frac{1}{2}\erf\left(\frac{\arctan p}{\sqrt{2}\alpha}\right).
$$
This number is the probability that given the normal distribution $N(0,\alpha)$ of angular errors, the back projection of a particle passes through the pixel $B$ (that is, if $\theta < \arctan p$).
Then, due to the error, only about
$
n_s' = C(p,\alpha) n_s = k_s' \sigma
$
background particles may be projected through pixel $B_r$. Our test detects a source at $B$ if $n = n_s' + n_b > \mu + k_t \sigma$, so it will miss it if $n_b < \mu - (k_s' - k_t)\sigma$.
The probability of this happening (the false negative rate) is
\begin{equation}
\mbox{fn rate} = 0.5 \erfc\left(\frac{k_s' - k_t}{\sqrt{2}}\right).
\end{equation}
For example, if $k_s' - k_t \ge 3$, then the probability of missing the source is at most $0.135\%$.

The true positive rate (given by $\mbox{tp rate} = 1 - \mbox{fn rate}$) is what we call the {\bf sensitivity}. If we require a particular sensitivity $\zeta$, then in order to detect a source with this sensitivity we need
\begin{equation}
k_s' - k_t \ge \sqrt{2}\, \erfc^{-1} [2(1-\zeta)].
\end{equation}
Recalling the requirement for $k_t$ given a confidence $\kappa$, this means
\begin{equation}
k_s' > K(\kappa,\zeta,M) := \sqrt{2} \; \erfc^{-1} \left[2\left(1 - \kappa^{1/M}\right)\right] + \sqrt{2} \, \erfc^{-1} [2(1-\zeta)].
\end{equation}
To detect a source with sensitivity $\zeta$, we need $n_s' > K(\kappa,\zeta,M)\, \sigma$. We claim that for any SNR $s > 0$ and $k_t > 0$ it is possible to detect the source, provided that the total number of detected particles $N$ is sufficiently large.

Given a signal to noise ratio $s$ and if $p \ll 1$, the number of detected source particles back projected near the source is $n_s' \approx C(p,\alpha) s\, N$ and $\sigma \approx \sqrt{N p(1-p)}$. So we require
\begin{equation}
C(p,\alpha) s\, N > K(\kappa,\zeta,M) \sqrt{N p(1-p)}.
\end{equation}
Since the left hand side grows linearly with $N$ and the right hand side grows only as $\sqrt{N}$, the inequality will hold for sufficiently large $N$.

We can expect to reliably detect the source with given sensitivity and specificity, if the total particle count $N$ at the detectors is on the order of
\begin{equation}
N \propto \left(\frac{K(\kappa,\zeta,M)}{C(p,\alpha)s}\right)^2 p(1-p)
\end{equation}
or higher. Or, writing this without using coefficients $K(\kappa,\zeta,M)$ and $C(p,\alpha)$,
\begin{equation}
N \;\;\propto\;\; 8 \left[\frac{\erfc^{-1} \Bigl(2\left(1 - \kappa^{1/M}\right)\Bigr) + \erfc^{-1} \bigl(2(1-\zeta)\bigr)}
{\erf\left(\frac{\arctan p}{\sqrt{2}\alpha}\right) s}\right]^2 p(1-p).\label{E:rule_error}
\end{equation}

%%%%%%%%%%%%%%%%%%%
\subsection{3D case}\label{3D}
%%%%%%%%%%%%%%%%%%

Similar approach and algorithms were also developed for the fully $3D$ Compton data \cite{Allm}. This part still requires additional work, both on forward and inverse computations, and is thus postponed until the next publication.

The relation of the $2D$ model to the true $3D$ situation is discussed in Section \ref{S:remarks}.

%%%%%%%%%%%%%%%%%%%%%%%
\section{Case studies}\label{S:cases}
%%%%%%%%%%%%%%%%%%%%%%%%%%%
In this section, we study the performance of the algorithm for various cargo/background/detector configurations and various SNR levels.

In all subsections below, it is assumed that a small (with $p=0.01$) source is placed into a larger cargo. The cargo configurations vary from case to case. We start with the simplest situation of spatially and angularly isotropic random background\footnote{This often a non-realistic assumption, except probably for containers filled with fertilizer.} and then proceed to more difficult scenarios.

As it was explained in \cite{AHKK}, even for very small values of SNR, increasing observation time (and thus the number of detected particles), one could eventually reach the detection level. The question is, though, whether this observation time is practical (observing a car at a border crossing for a week is clearly unreasonable). Thus, in all cargo scenarios (except the next section with uniformly random background), we show the observation time instead of the number of detected particles. Doing so, we use the real parameters of cargo materials and HEU source (see section \ref{S:physics}).

%%%%%%%%%%%%%%%%%%%%%%%%%%%%%%%%%%%%%
\subsection{Uniform random background}
%%%%%%%%%%%%%%%%%%%%%%%%%%%%%%%%%%%

We start with the simplest situation considered in \cite{AHKK}. Namely, we assume that the background radiation is uniform and isotropic in the space of all possible rays intersecting the cargo container. We thus ran $20$ (sometimes $100$) random realizations of the radiation for each box, labeled by the pair of values of $N$ and SNR.
Four arrays of $48$ detectors, each with $180$ angular bins, were used in each case. The background corresponded to the concrete base, as described in Section \ref{S:physics}.

\begin{figure}[H]
\begin{center}
 \includegraphics[width=6in]{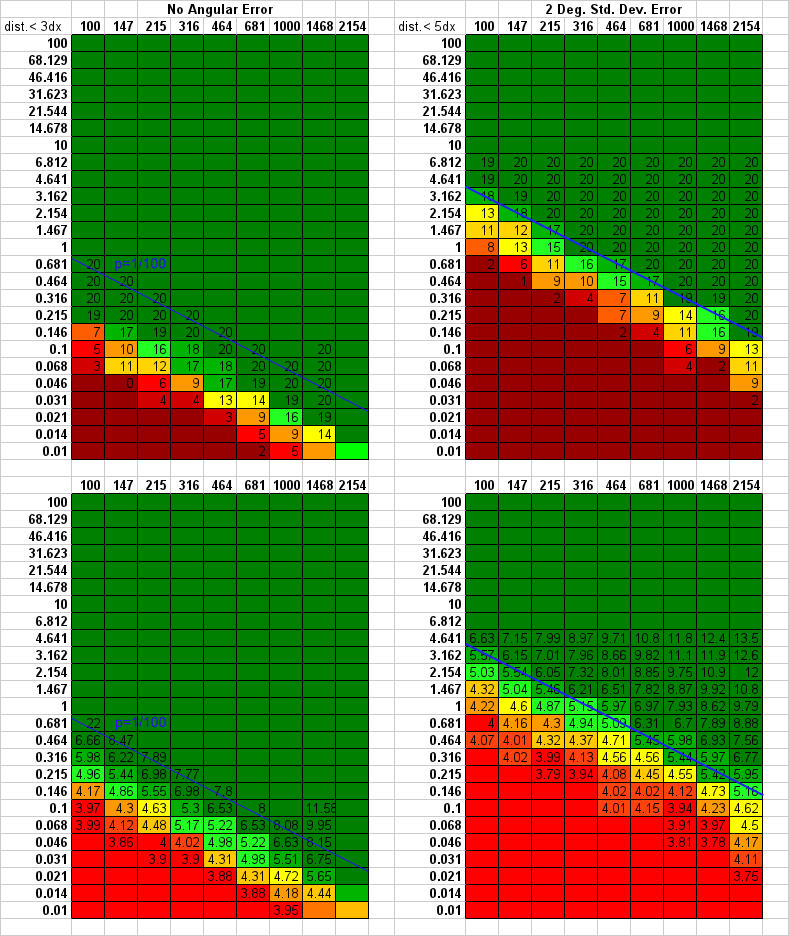}
\end{center}
\caption{
Vertical axis - SNR in $\%$. Horizontal axis - the number of detected ballistic particles from the source. Both axes are in logarithmic scale. From left to right the angular standard error is $\alpha=0$ and $2$ degrees.
Top row, in each box - numbers of successful detections out of $20$ random realizations. Bottom row, in each box - numbers of standard deviations above the mean at the detected source.
Blue line - the theoretical estimate (\ref{E:rule}) and (\ref{E:rule_error}) of the detectability region.}\label{DetTables0-2}
\end{figure}

In the tables in the top rows of Fig. \ref{DetTables0-2} $20$ realizations for each box were used. If a box contains a number, say $18$, it indicates in how many out of $20$ simulations the source was detected.
Detection was considered to be confirmed, if the confidence probability was at least $95\%$ that the location's high concentration cannot be explained by random fluctuations of the background.

The color scheme was used that indicated completely successful detection in green (dark grey) in the upper right corner and then changed to the red (black) part of the spectrum in the lower left with decreasing detectability.

The theoretical relations (\ref{E:rule}) and  (\ref{E:rule_error}) provide curves in the $N$-SNR parameter plane that are supposed to roughly separate
detectability and non-detectability regions. The blue straight lines (the logarithmic scale was used) were drawn that correspond to these theoretical boundaries of the detectability regions.

The variable along the vertical axis in all tables was the SNR value $s$ in percents, varying from $0.01\%$ to $100\%$. The horizontal axis indicates the number of detected ballistic source particles
(so $1000$ ballistic source particles with SNR $= 1\%$ mean that the total number of detected particles was $10^5$).

Additional parameter of interest (especially important for neutron Compton type detectors, where good angular precision is hard and costly to obtain), is the error of the detection of directions.
We assumed that the error in the recorded directions is normally distributed centered at zero with a standard deviation of $\alpha$ degrees. The tables from the left to right in Fig.
\ref{DetTables0-2} show the results for $\alpha = 0$ and $2$ degrees correspondingly. In presence of angular errors, instead of formula  (\ref{E:rule}), its analog  (\ref{E:rule_error}) is used.

The bottom row of tables shows, rather the number of successful detection, the average over the $20$ runs of the hight above the mean of the peak in the reconstruction at the detection location,
measured in standard deviations.

Some boxes do not have numbers in them, which indicates that no test runs were done for those values. However, all such boxes are either above (better) than others with perfect detection or lower (worse)
than those with bad detection. Thus the colors (shades) were just extrapolated there.

\begin{figure}[H]
\begin{center}
 \includegraphics[width=6in]{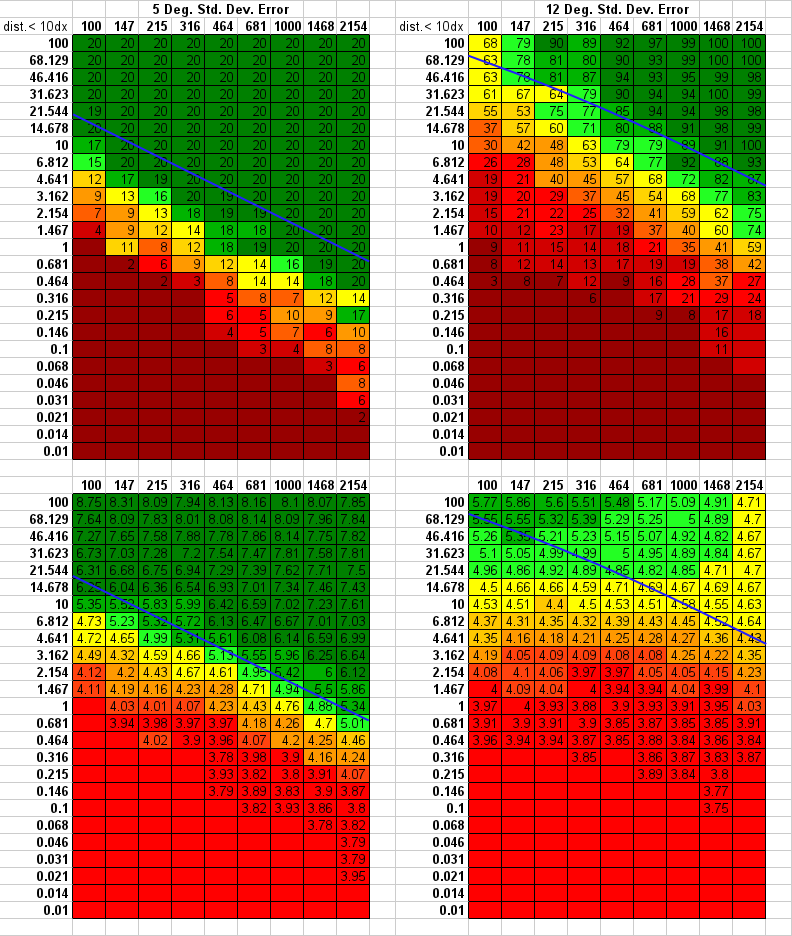}
\end{center}
\caption{
Vertical axis - SNR in $\%$. Horizontal axis - the number of detected ballistic particles from the source. Both axes are in logarithmic scale. From left to right the angular standard error is $\alpha=5$ and $12$ degrees.
Top row, in each box - numbers of successful detections out of $20$ (left column) and $100$ (right column) random realizations. Bottom row, in each box - numbers of standard deviations above the mean at the detected source.
Blue line - the theoretical estimate (\ref{E:rule_error}) of the detectability region.}\label{DetTables5-12}
\end{figure}

The tables in Fig. \ref{DetTables5-12} are similar, except that the angular error was $\alpha=5$ (left column) and $\alpha=12$ (right column) degrees correspondingly.
Also, in the right column, $100$, rather than $20$, random realizations for each box were used.

The tables show that the theoretically predicted in \cite{AHKK} detectability results do hold in numerical experiments.
The theoretical lines, in spite of simplifying assumptions used, amazingly well indicate the detectability regions of parameters. We also notice that they do reach the desired SNR level,
if the observation time (and thus the number of particles detected) increases. The angular errors certainly lead to deterioration of the detection (still corresponding well to the theoretical lines of Section \ref{S:remarks}).
The deviation of around $5\sigma$ leads to detection, also in accordance with analysis done in \cite{AHKK}.

%%%%%%%%%%%%%%%%%%%%%%%%%%%
\subsection{Cargo Scenarios}
%%%%%%%%%%%%%%%%%%%%%%%%%%%%

In the scenarios below, it is assumed that an HEU source (shown as $\boxbox$ in figures) is of size $5 \mathrm{cm} \times 5 \mathrm{cm}$, while each out of eight rectangular material blocks is
$60\mathrm{cm} \times 120\mathrm{cm}$, forming the cargo hold of the size $2.4\times 2.4\, meter^2$. The materials in each box are shown in figures as follows:
plastic (striped texture), cotton (plaid texture), wood (wood grain texture), concrete (concrete texture), iron (metal tread texture).
The background is assumed to come from the concrete base below the cargo as described in Section \ref{S:physics}, while the background radiation is shielded from the other three sides.

Four arrays (in some simulations three, when indicated, with the bottom one missing) of equally spaced $48$ detectors, each with $180$ angular bins were used in each simulation.
Both reconstructions with the complete direction sensitive ``detectors,'' as well as with ``Compton detectors'' were used, as indicated in captions.

The forward data were simulated according to physical parameters described in Section \ref{S:physics} and numerical techniques indicated in Section \ref{S:forward}.

The reconstructed images were thresholded at the level of $4.3\sigma$ above the (local) mean and results are shown both as density and surface plots.

%%%%%%%%%%%%%%%%%%
\subsubsection{Cargo Scenario \#1}
\begin{figure}[H]
\begin{center}
 \includegraphics[width=6.5cm]{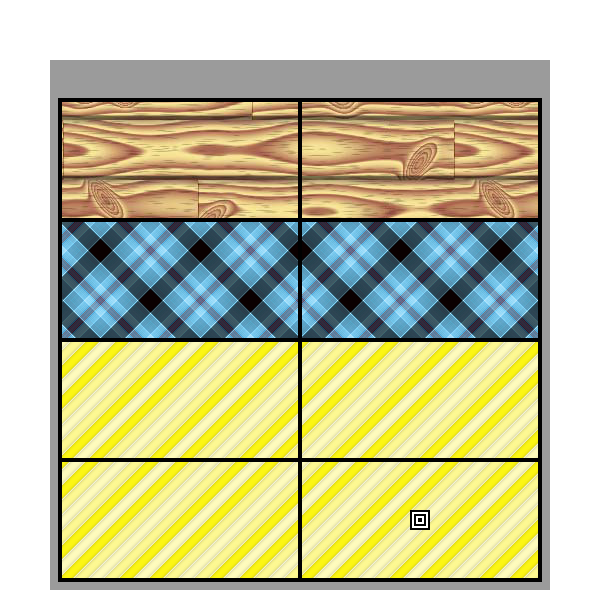}
\end{center}
\caption{Scenario \#1 Material Arrangement: plastic (striped texture), cotton (plaid texture), wood (wood grain texture), and an HEU source ($\boxbox$).}\label{SCN01}
\end{figure}

%%%%%%%%%%%%%%%%%%%%%%%%
Here the cargo contains plastic, cotton, wood, and an HEU source ($\boxbox$), as shown in Fig. \ref{SCN01}.
%four detector arrays were used with 48 equally spaced detectors per array and 180 angular bins per detector.

\begin{figure}[H]
\begin{center}
\includegraphics[width=7cm]{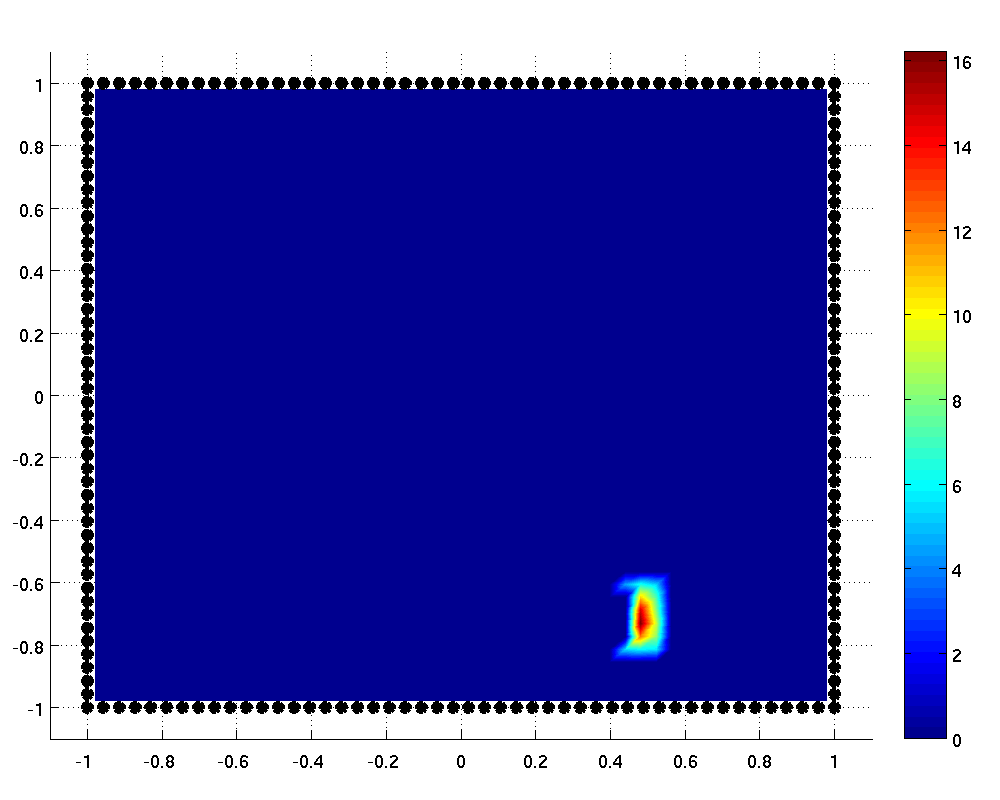}
\includegraphics[width=7cm]{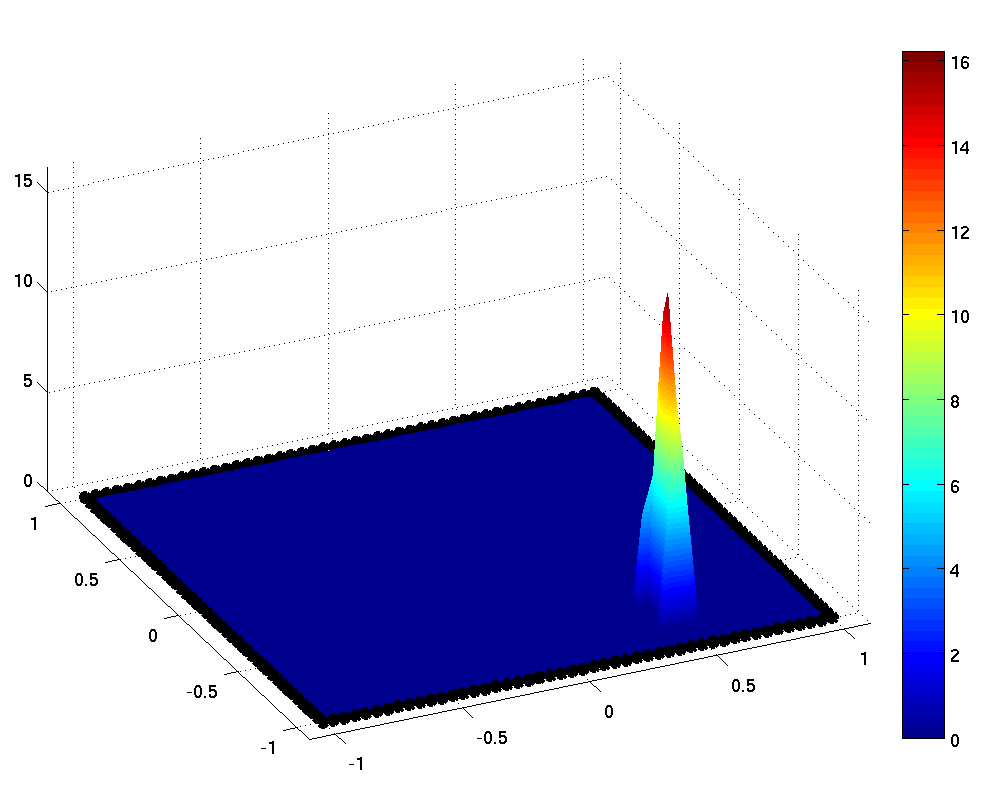}
\caption{Scenario \#1 with 4 detector arrays. Exposure Time = 2 seconds. Density and surface plots of backprojection with the cut off at $4.3\sigma$. The source is clearly found.}\label{4DS1T2}
\end{center}
\end{figure}

Figure \ref{4DS1T2} shows the density and surface plots of backprojection with the cut off at $4.3\sigma$. The exposure time is 2 seconds. $124$ particles at 1 MeV were detected, $119$ of which are ballistic,
which gives a signal to noise ratio of 23.8.

In the next figure we show the same results obtained by using Compton data and converting them to the synthetically collimated ones.
\begin{figure}[H]
\begin{center}
\includegraphics[width=7cm]{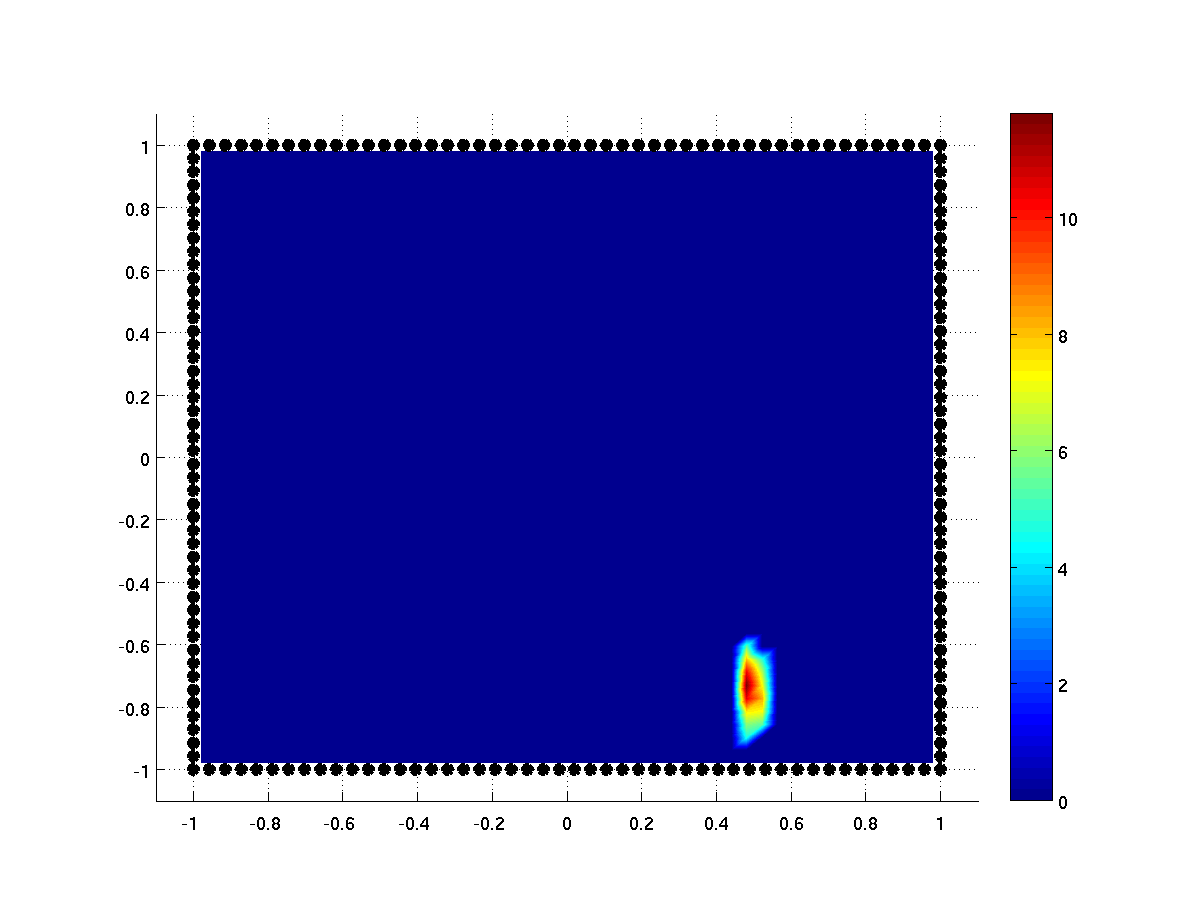}
\includegraphics[width=7cm]{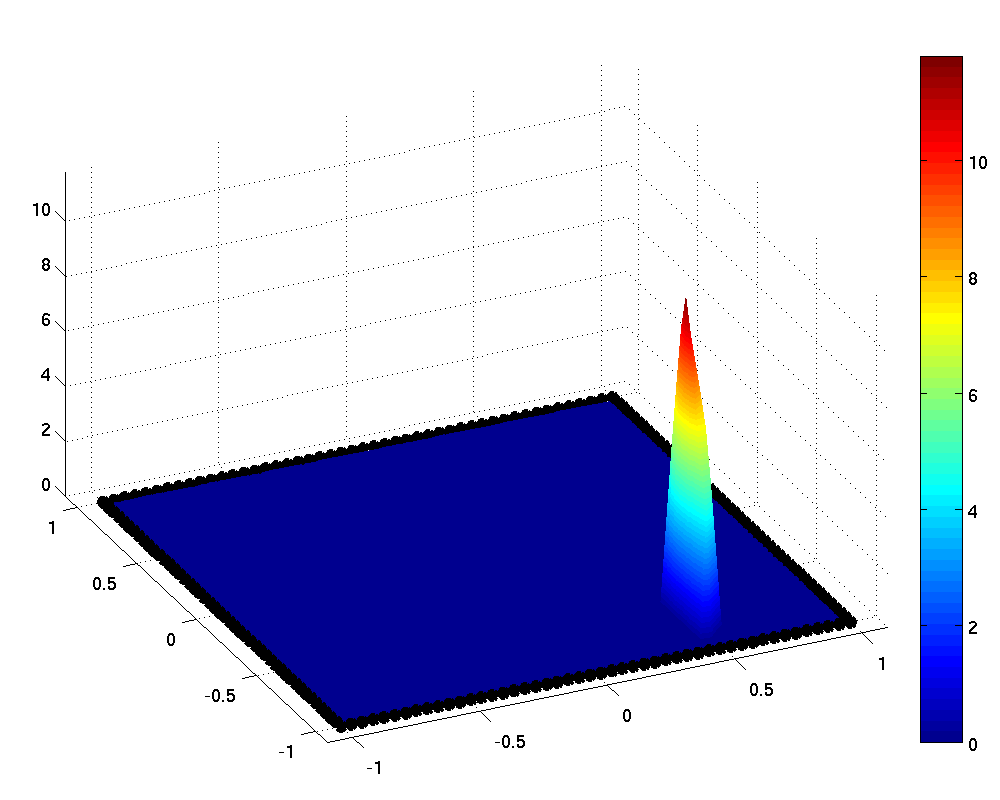}
\caption{Scenario \#1 with 4 Compton detector arrays. Exposure Time = 2 seconds}\label{4DS1T2Compton}
\end{center}
\end{figure}

%\pagebreak

\begin{figure}[H]
\begin{center}
\includegraphics[width=7cm]{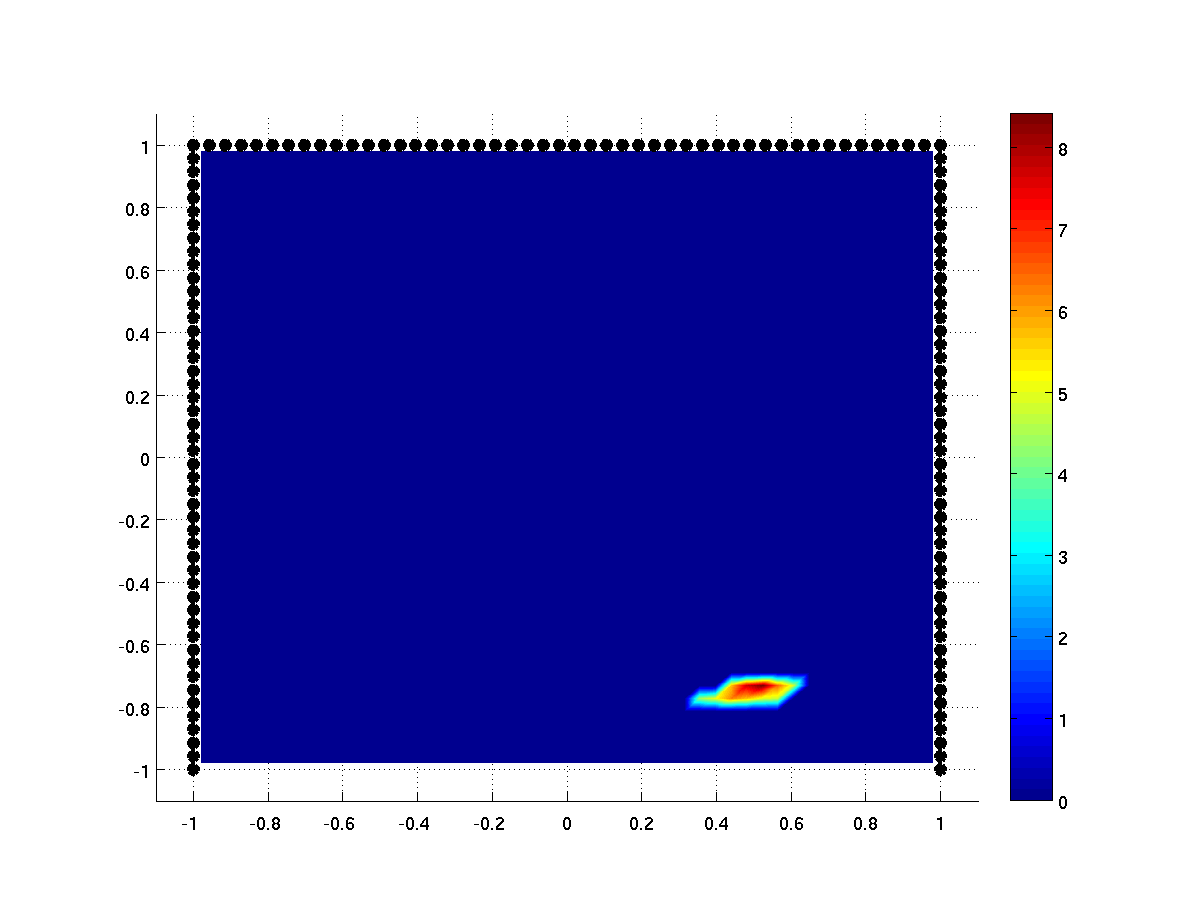}
\includegraphics[width=7cm]{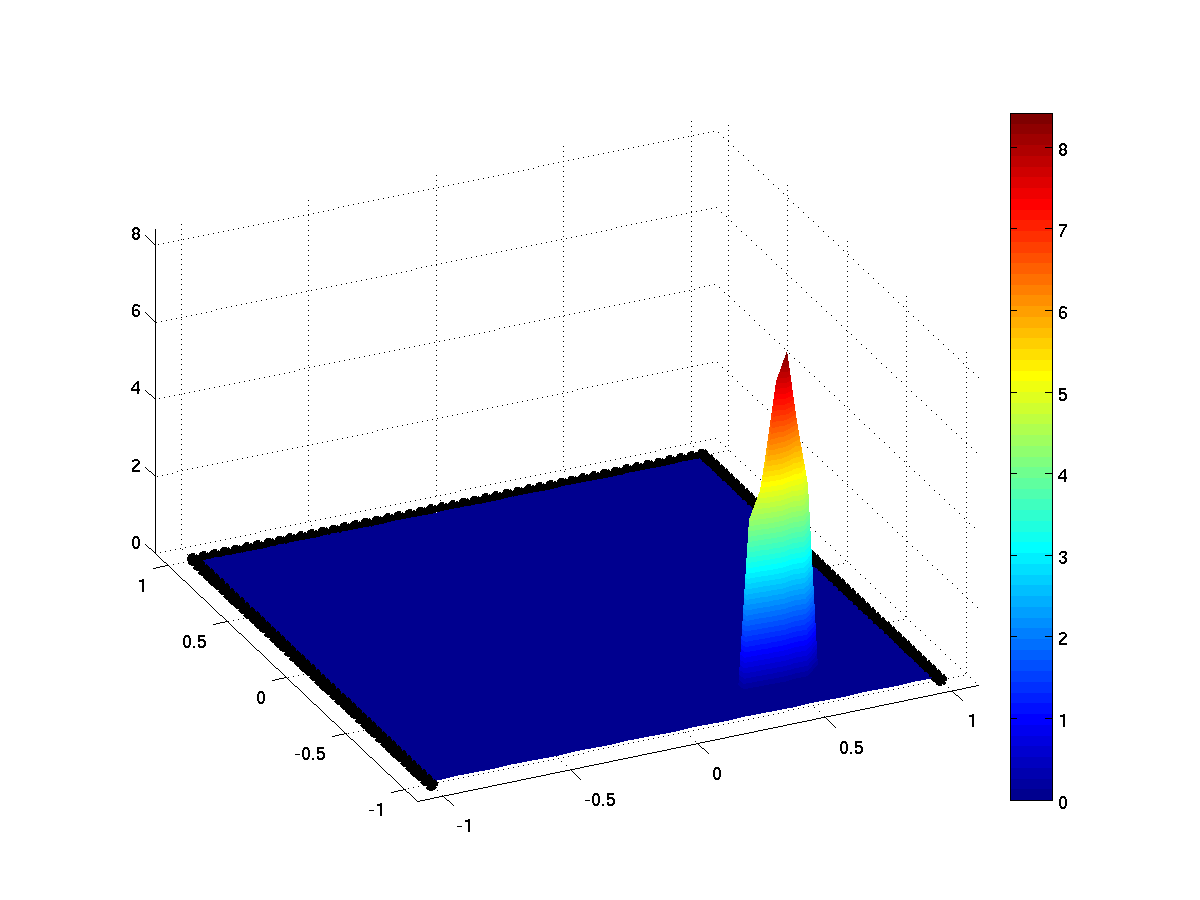}
\caption{Scenario \#1 with 3 detector arrays (the bottom one missing). Exposure Time = 10 seconds}\label{3DS1T10}
\end{center}
\end{figure}

In Figure \ref{3DS1T10}, we use three detector arrays.% with 48 equally spaced detectors per array and 180 angular bins.
The bottom array (closest to the source) is missing. %In 10 seconds, we detect 35 particles at 1 MeV, 34 of which are ballistic, which gives a signal to noise ratio of 34.

Due to the absence of the detector array closest to the source, many ballistic particles from the source are not accounted for, which results in a longer observation time for detection.

\begin{figure}[H]
\begin{center}
\includegraphics[width=7cm]{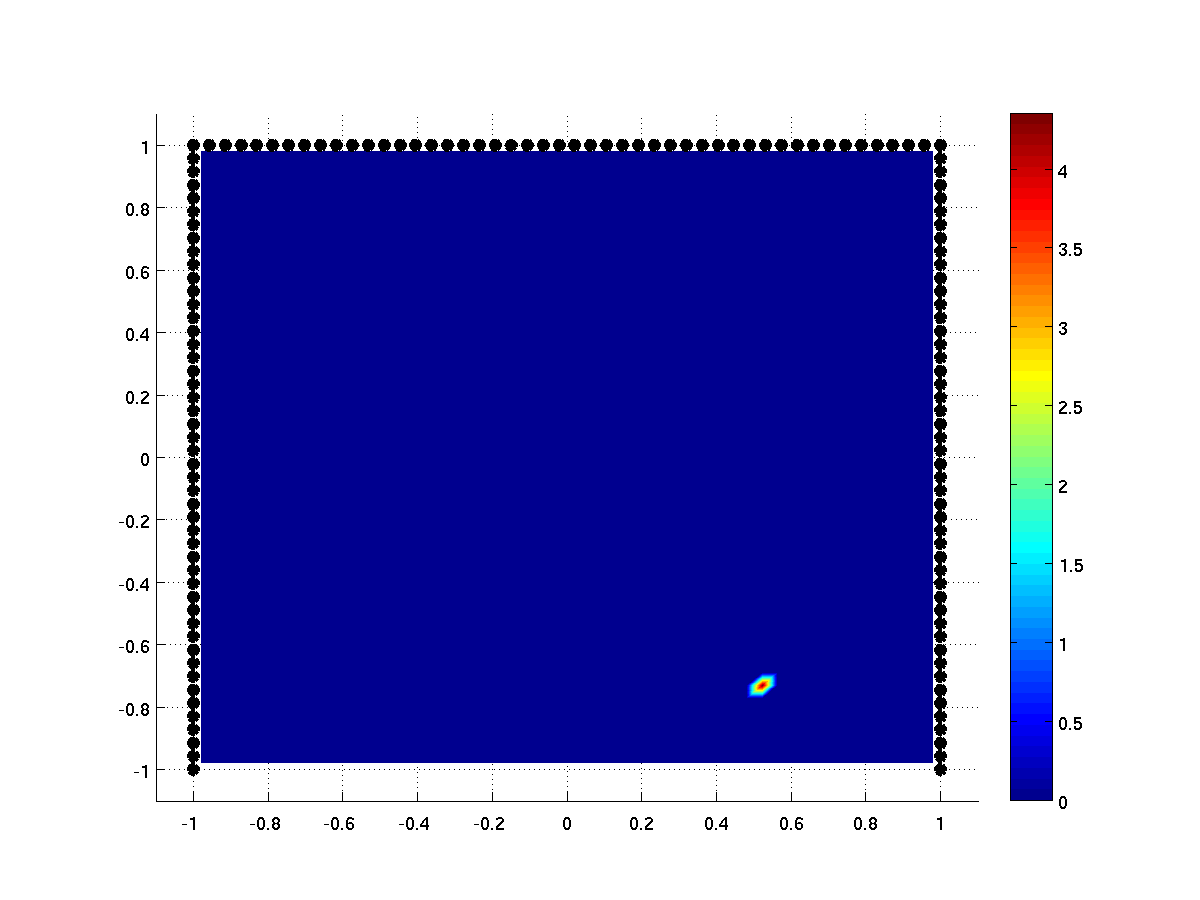}
\includegraphics[width=7cm]{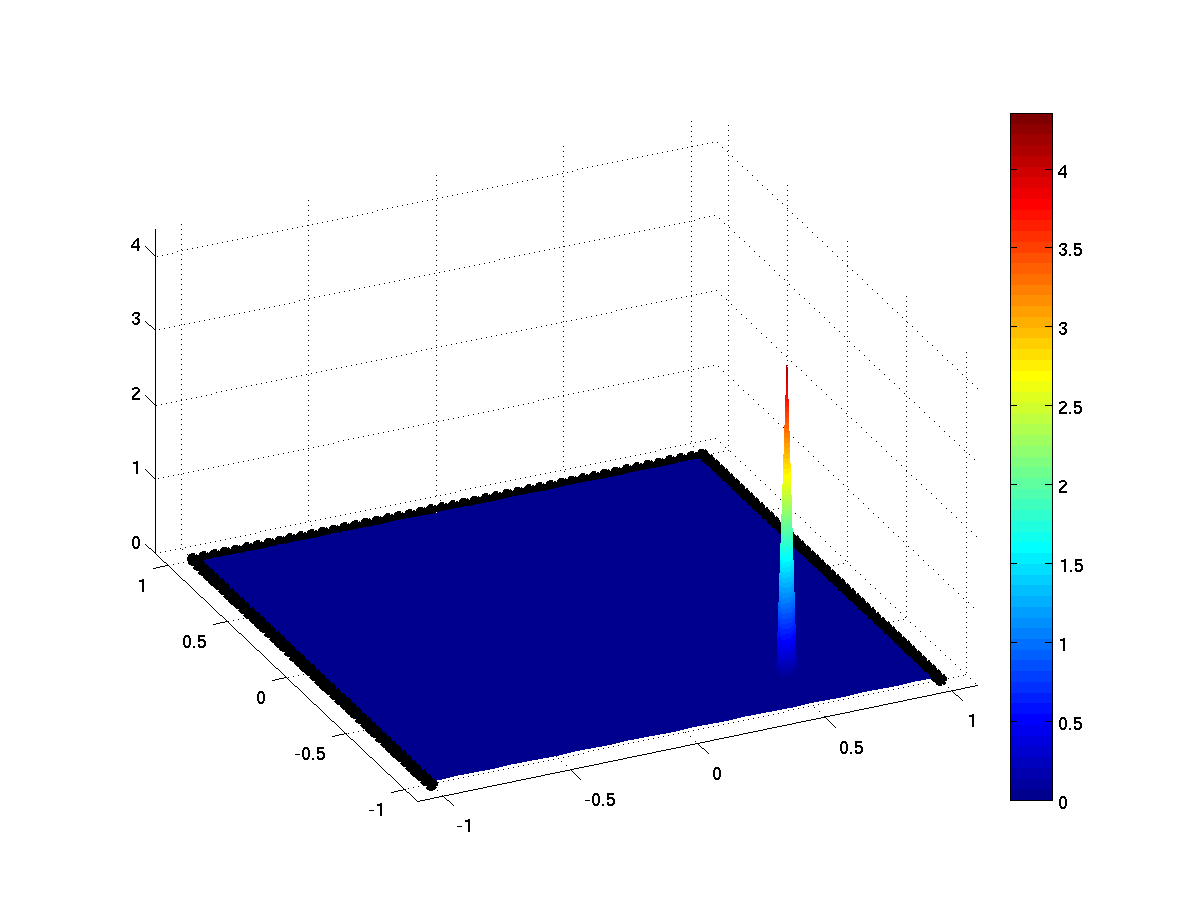}
\caption{Scenario \#1 with 3 Compton detector arrays (the bottom one missing. Exposure Time = 10 seconds}\label{3DS1T10Compton}
\end{center}
\end{figure}

Fig. \ref{3DS1T10Compton} shows the similar set-up, but with three Compton detector arrays.

\begin{figure}[H]
\begin{center}
\includegraphics[width=7cm]{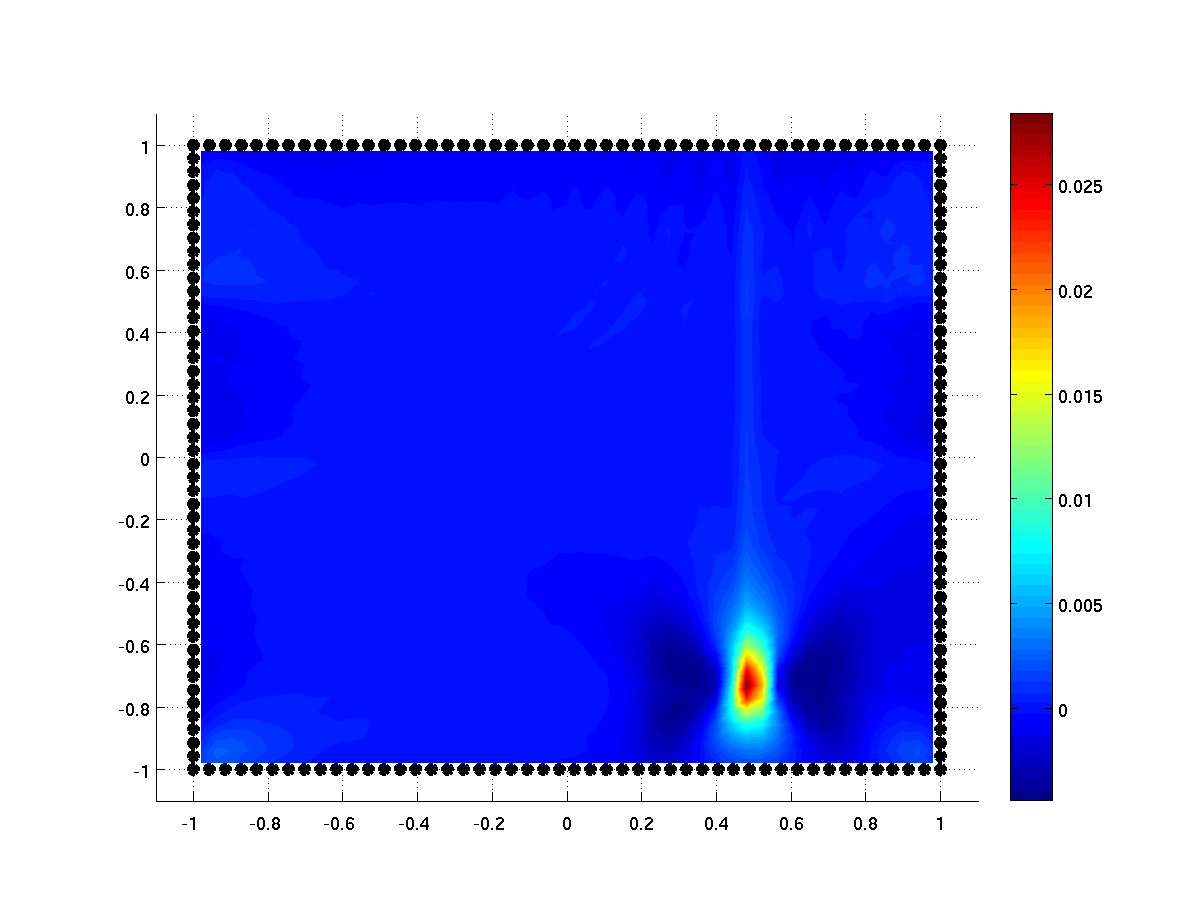}
\includegraphics[width=7cm]{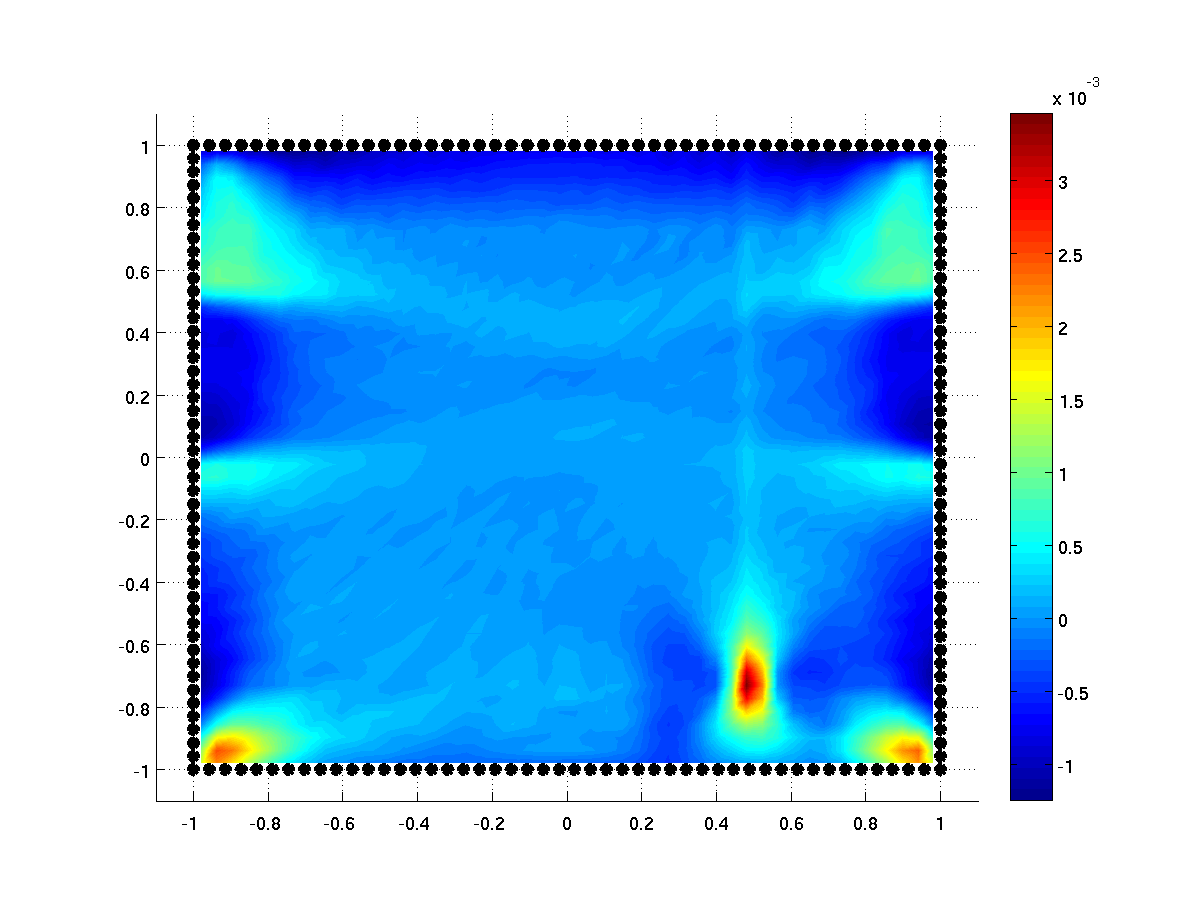}
\includegraphics[width=7cm]{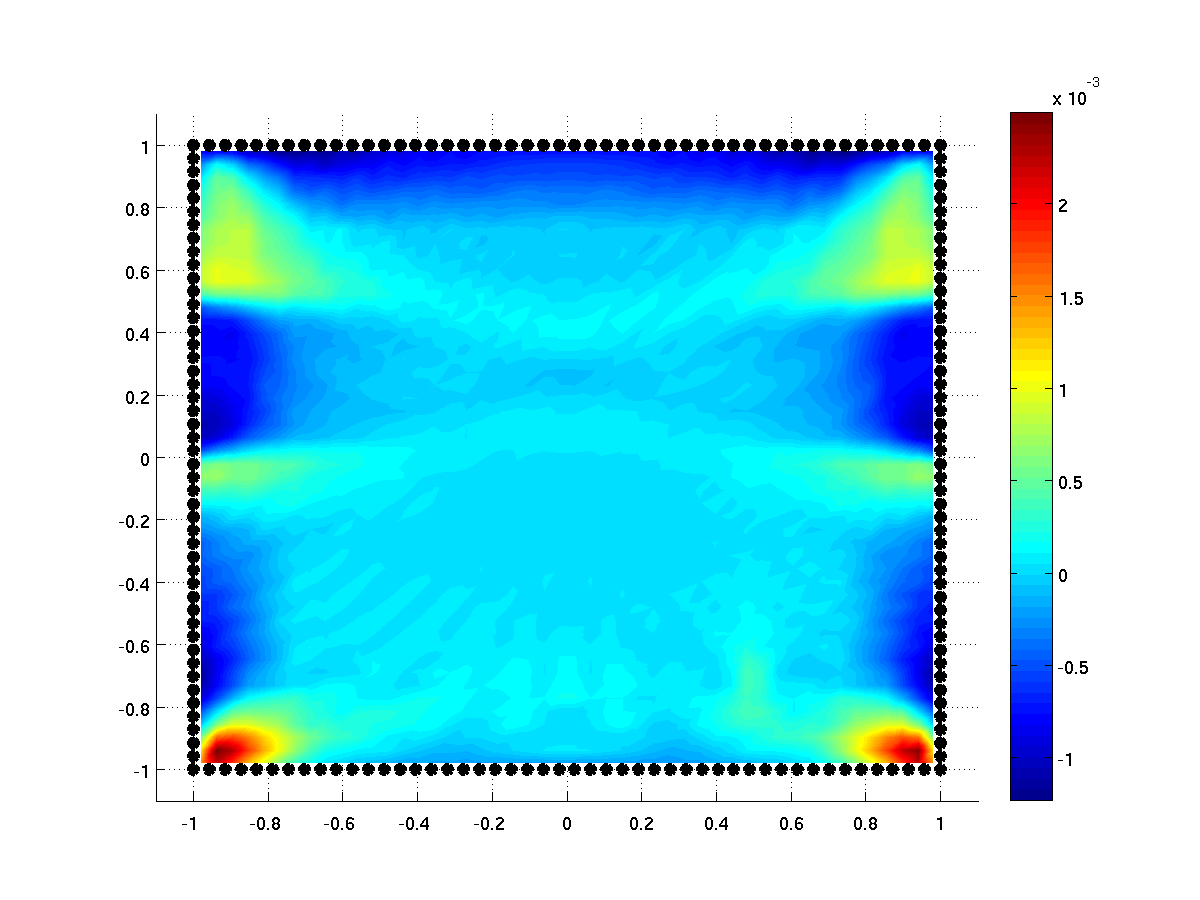}
\caption{Scenario \#1 with 4 detector arrays at SNR equal to $0.1, \, 0.01$, and $0.001$.}\label{4DS1SNR}
\end{center}
\end{figure}

Fig. \ref{4DS1SNR} shows the effect of dropping SNR to 0.1, 0.01, and 0.001 respectively, on the reconstruction
And now, the similar situation with three detectors at SNR of 0.1 and 0.01:
\begin{figure}[H]
\begin{center}
\includegraphics[width=7cm]{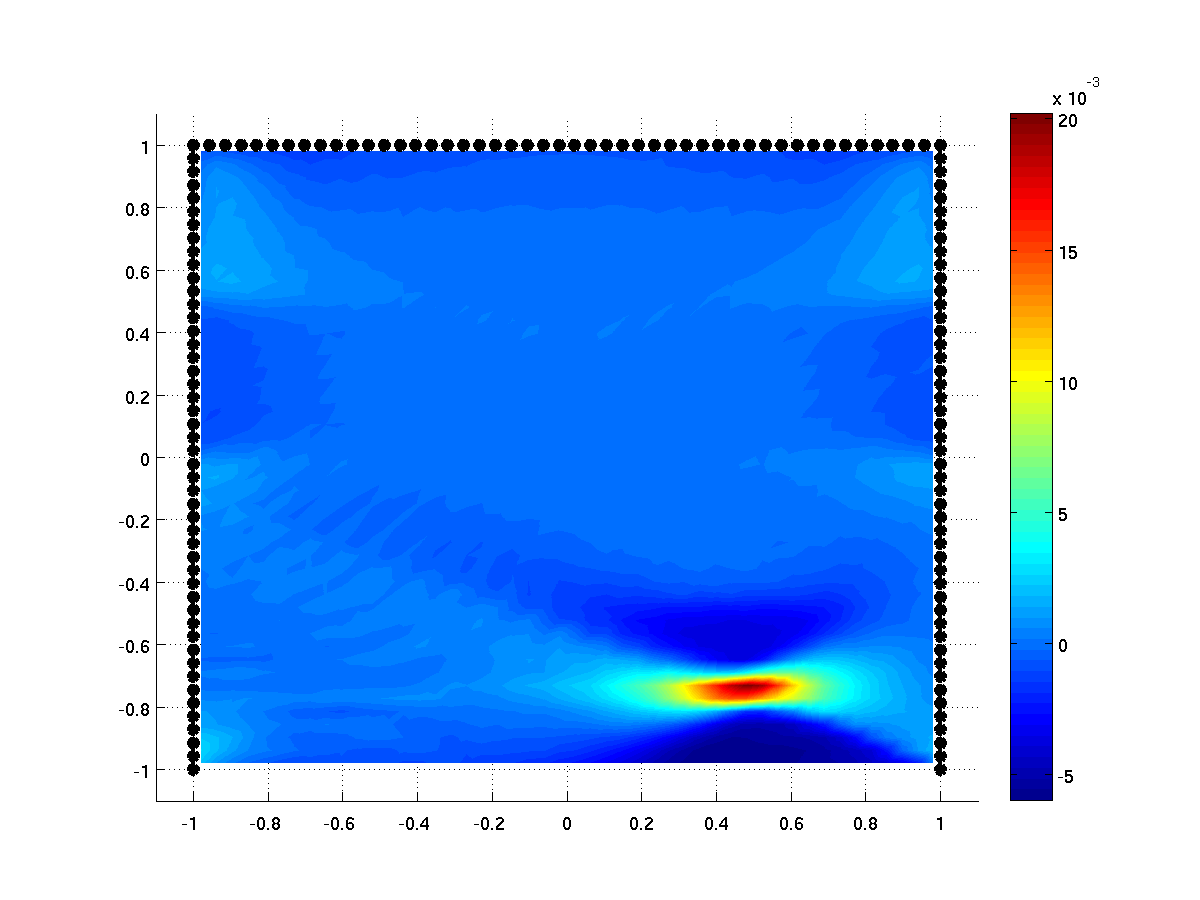}
\includegraphics[width=7cm]{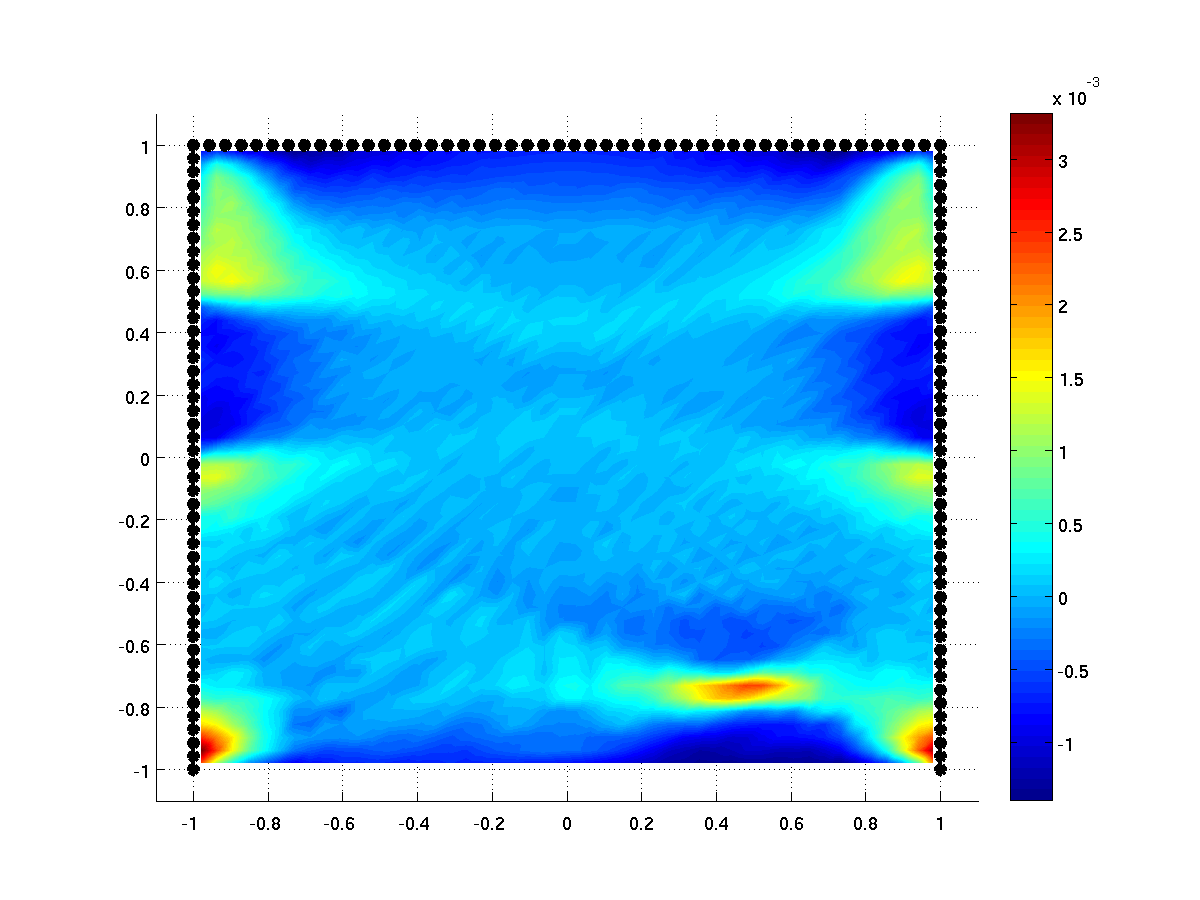}
\caption{Scenario \#1 with 3 detectors at various SNR}\label{3DS1SNR}
\end{center}
\end{figure}
Since, the most valuable detector array is missing, one observes clear deterioration.

One can see some corner artifacts, which arise due to the square geometry of the observation (detector) surface. See also Section \ref{S:remarks} for this discussion.

%%%%%%%%%%%%%%%%%%%
\subsubsection{Cargo Scenario \#2}
%%%%%%%%%%%%%%%%

\begin{figure}[H]
\begin{center}
 \includegraphics[width=6.5cm]{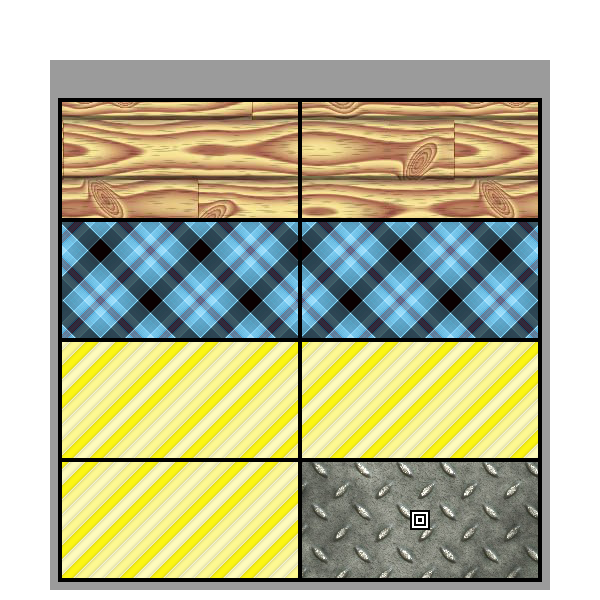}
\end{center}
\caption{Scenario \#2 Material Arrangement: plastic (striped texture), cotton (plaid texture), wood (wood grain texture), iron
(metal tread texture) at $50\%$ density, and an HEU source ($\boxbox$).}\label{SCN02}
\end{figure}

Here we try to make the detection harder, so the Cargo Scenario \#2 contains plastic, cotton, wood texture, iron at $50\%$ density, and an HEU source ($\boxbox$) inside the iron box, as shown in Figure \ref{SCN02}.
The HEU source is $5 \mathrm{cm} \times 5 \mathrm{cm}$ and each rectangular material block is $60\mathrm{cm} \times 120\mathrm{cm}$.
The particle count is clearly lower here, so longer observation time is required.

In Fig. \ref{4DS2T300} below, the same as before set-up of four detector arrays is used.
Figure shows the detection with the same cut off at $4.3\sigma$ as before and 5 min observation.
\begin{figure}[H]
\begin{center}
\includegraphics[width=7cm]{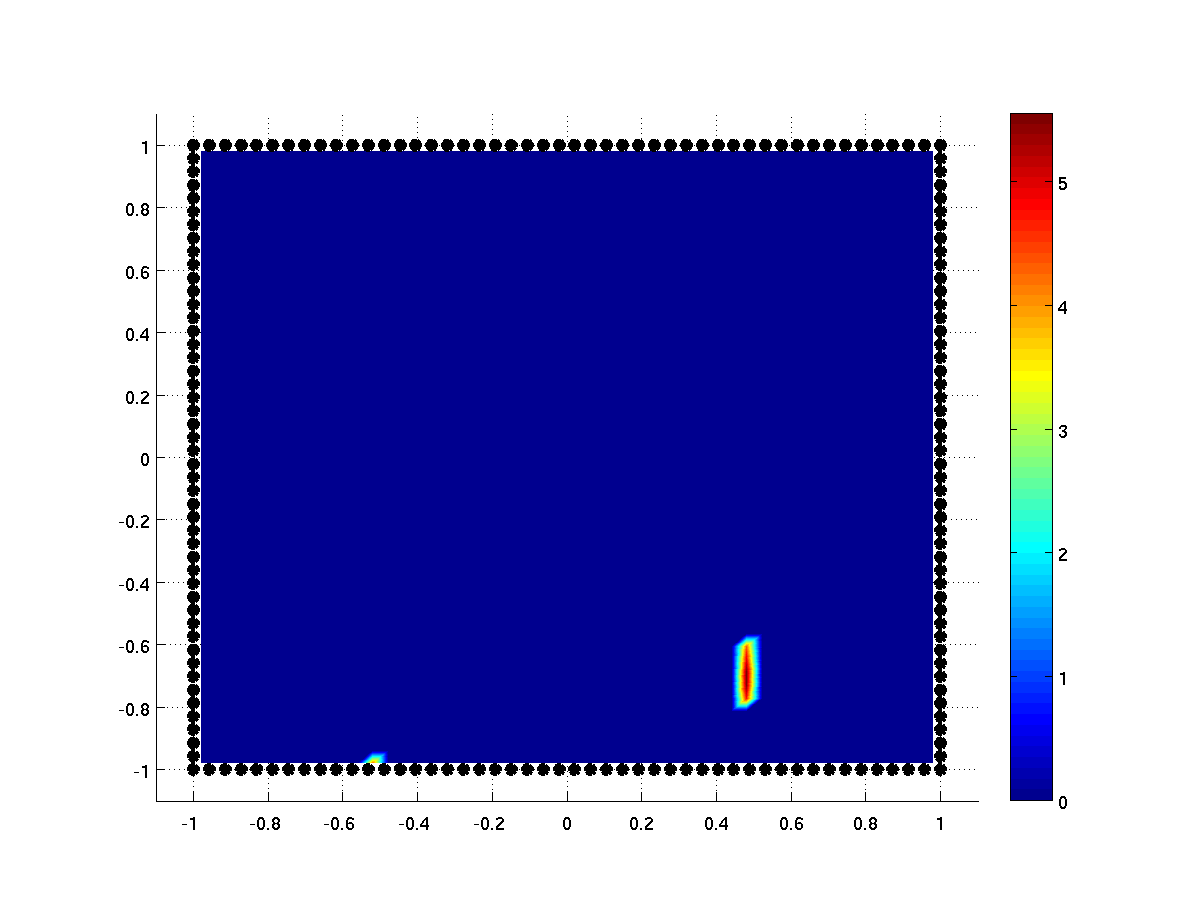}
\includegraphics[width=7cm]{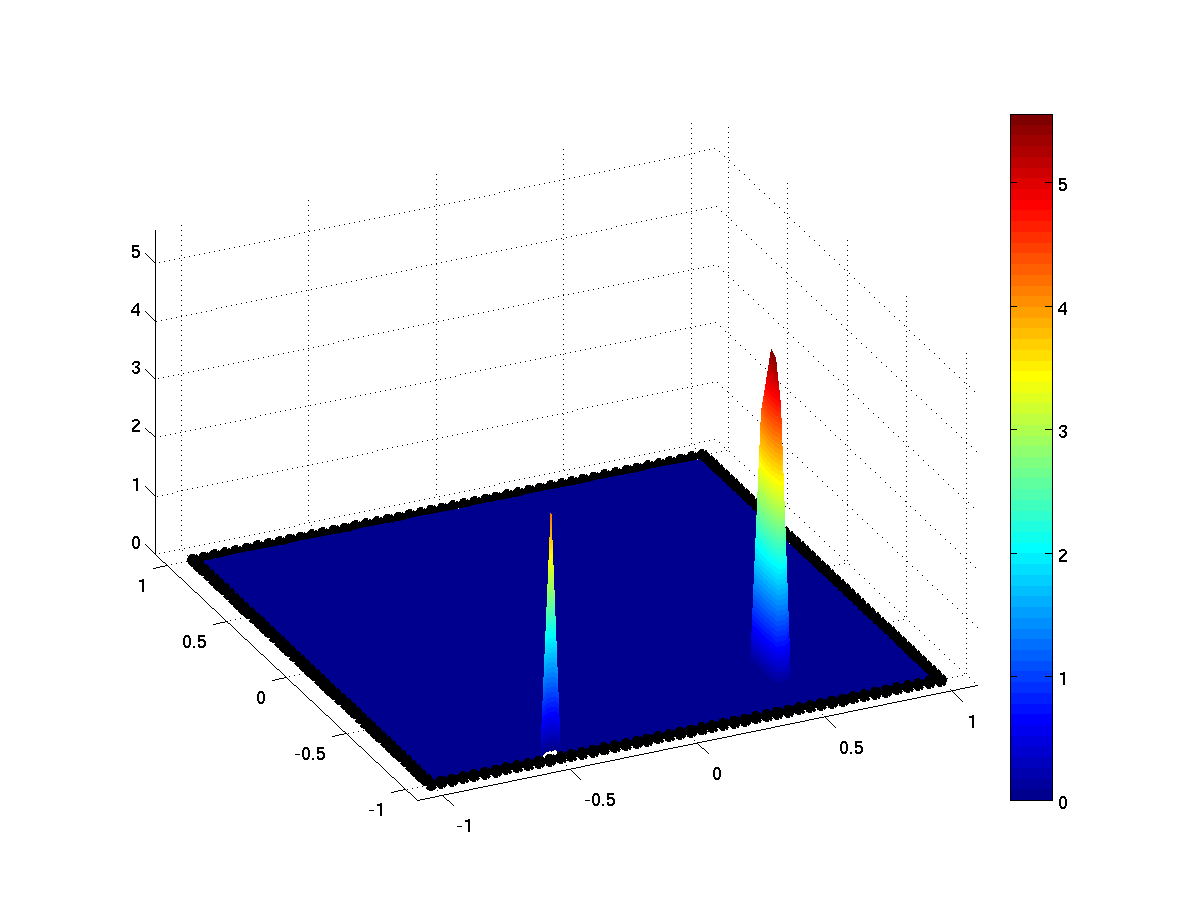}
\caption{Scenario \#2 with 4 detector arrays. Exposure Time = 5 minutes}\label{4DS2T300}
\end{center}
\end{figure}
And now the same using Compton data:
\begin{figure}[H]
\begin{center}
\includegraphics[width=7cm]{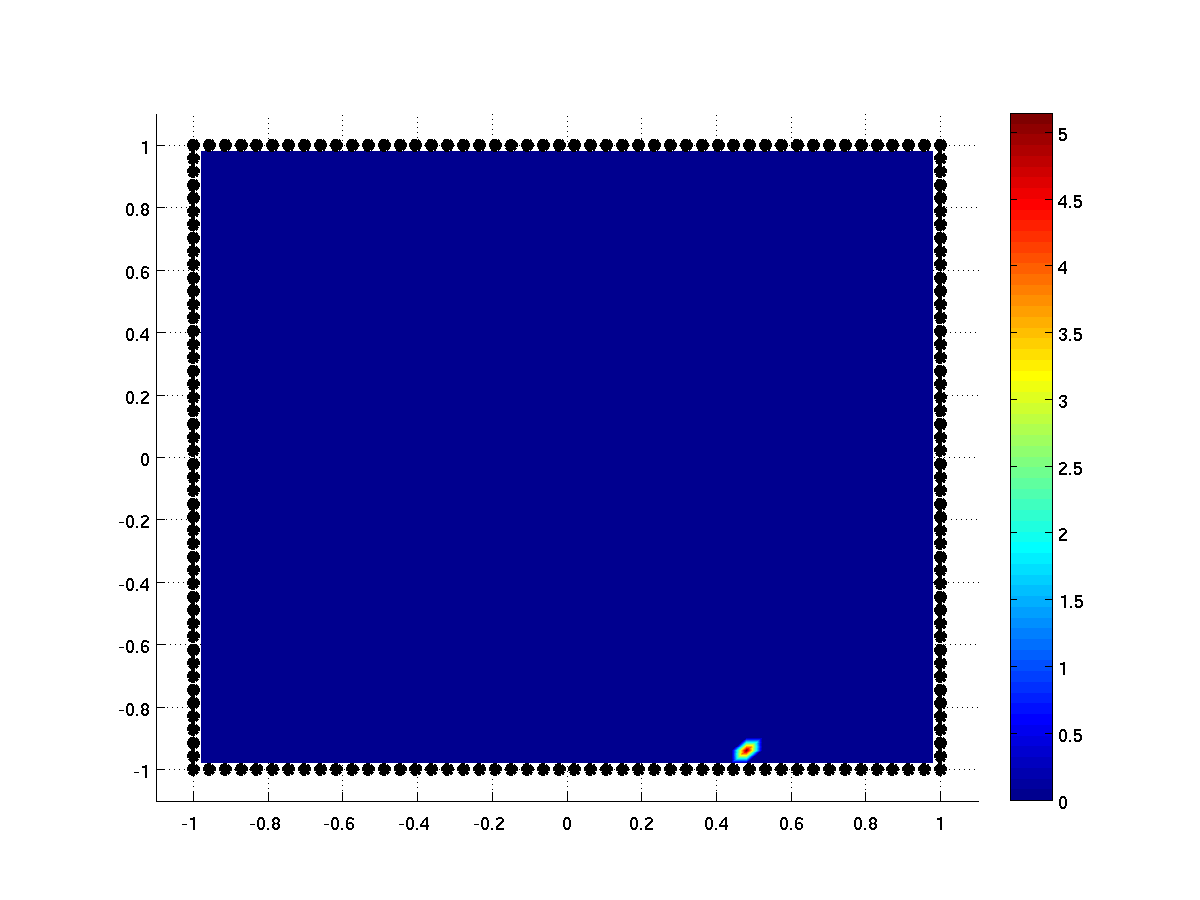}
\includegraphics[width=7cm]{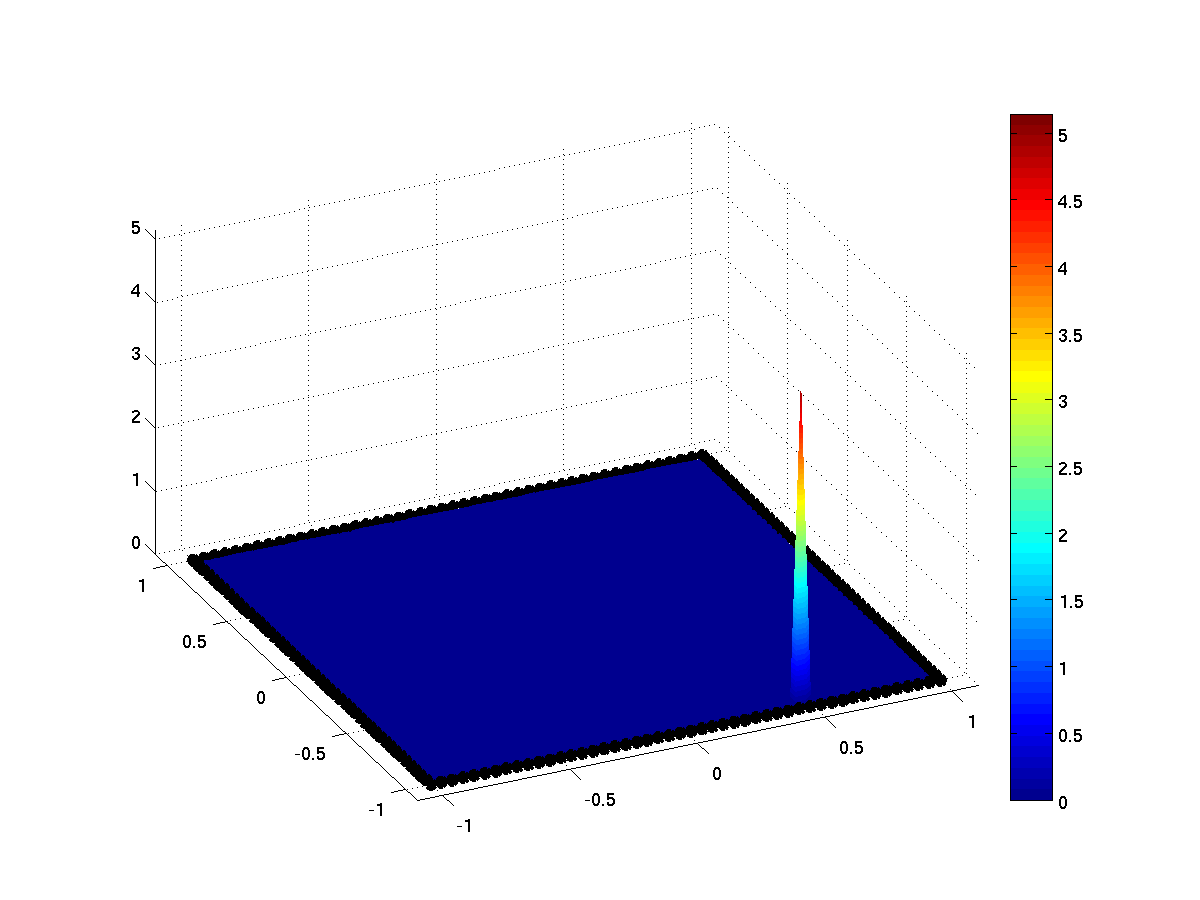}
\caption{Scenario \#2 with 4 Compton detector arrays. Exposure Time = 5 minutes}
\end{center}
\label{4DS2T300Compton}
\end{figure}

In the next Fig. \ref{3DS2T604800}, three detector arrays are used (with the most relevant bottom one missing). Here a much longer observation time of 7 days is required to reach the same level of detection.
% with 48 equally spaced detectors per array and 180 angular bins.
%In 7 days, we detect 3308 particles at 1 MeV, 120 of which are ballistic, which gives a signal to noise ratio of 0.0376.

\begin{figure}[H]
\begin{center}
\includegraphics[width=7cm]{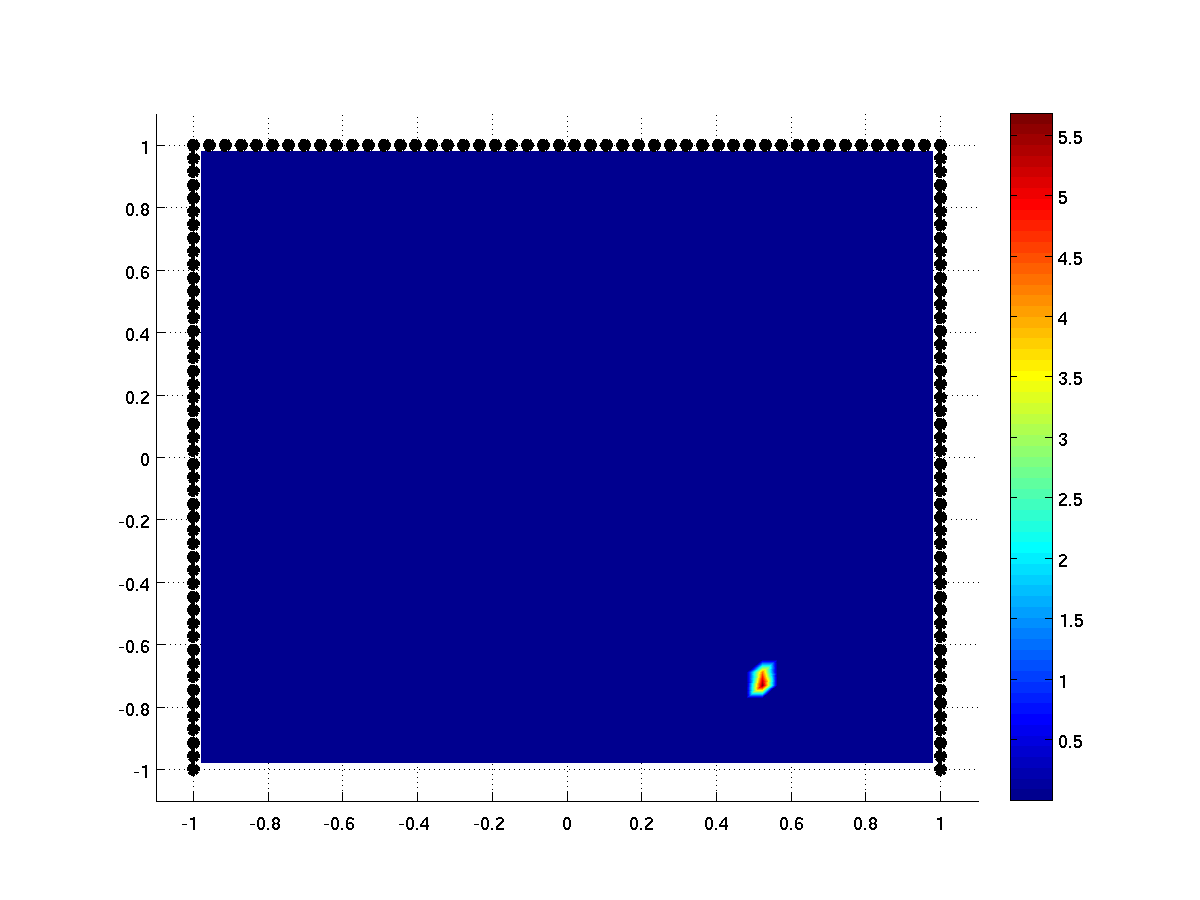}
\includegraphics[width=7cm]{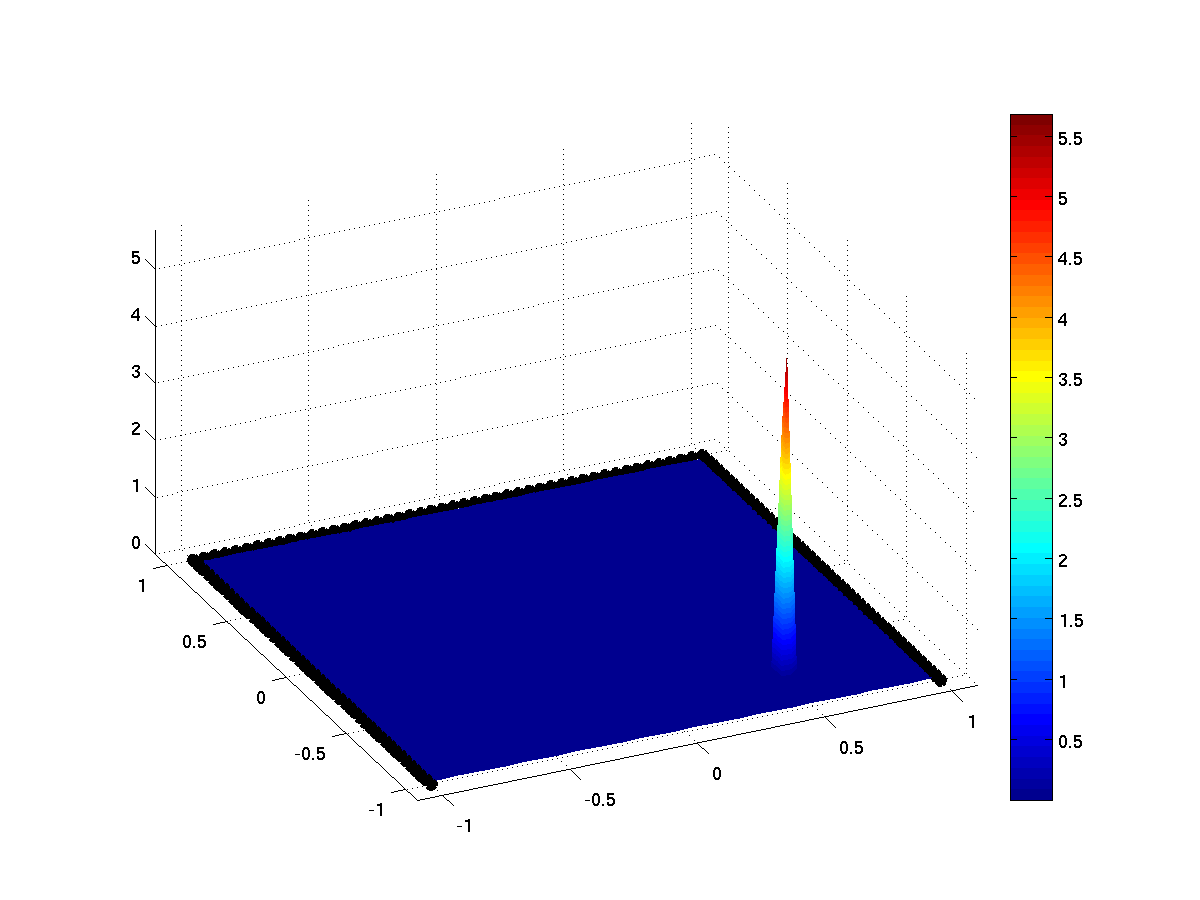}
\caption{Scenario \#2 with 3 detector arrays: Exposure Time = 7 days}\label{3DS2T604800}
\end{center}
\end{figure}
About the same situation is with the Compton data, Fig. \ref{3DS2T604800Compton}:
\begin{figure}[H]
\begin{center}
\includegraphics[width=7cm]{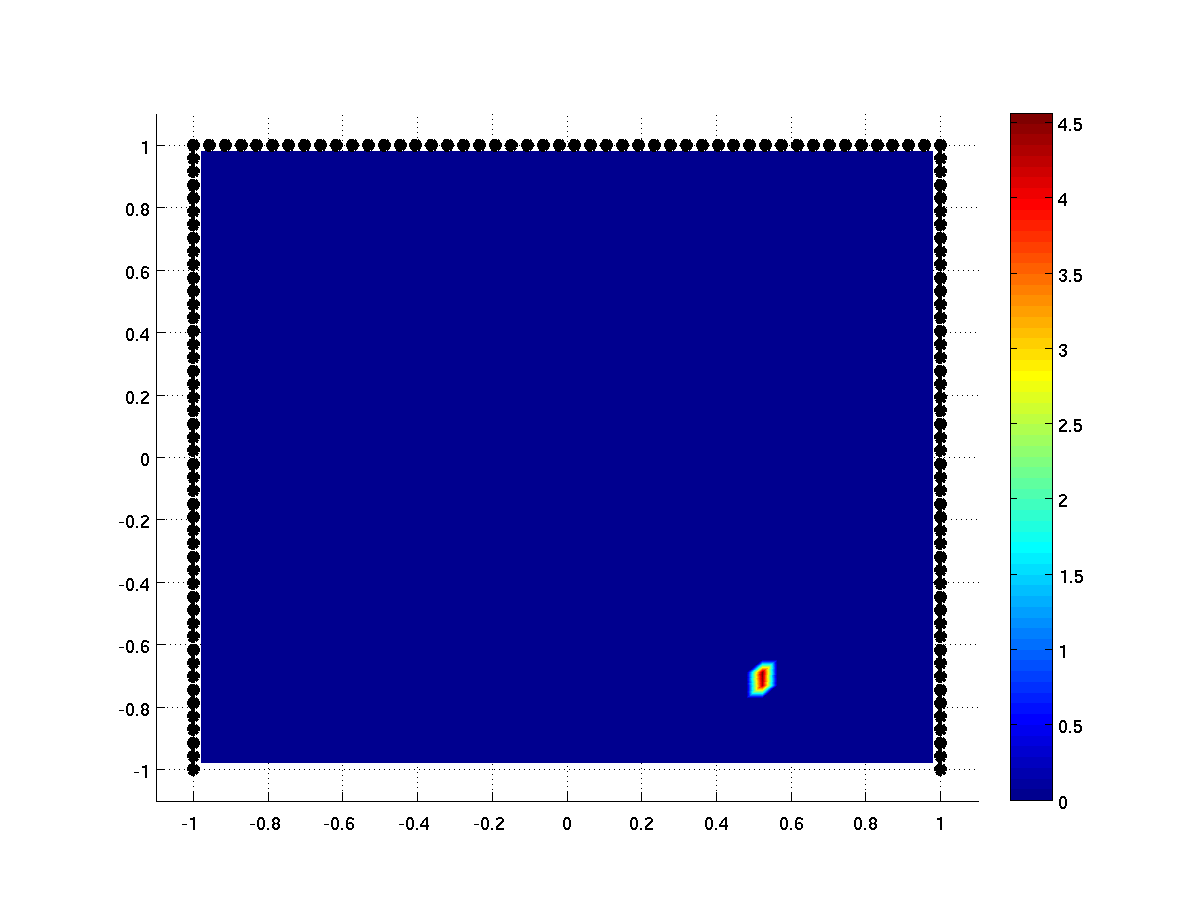}
\includegraphics[width=7cm]{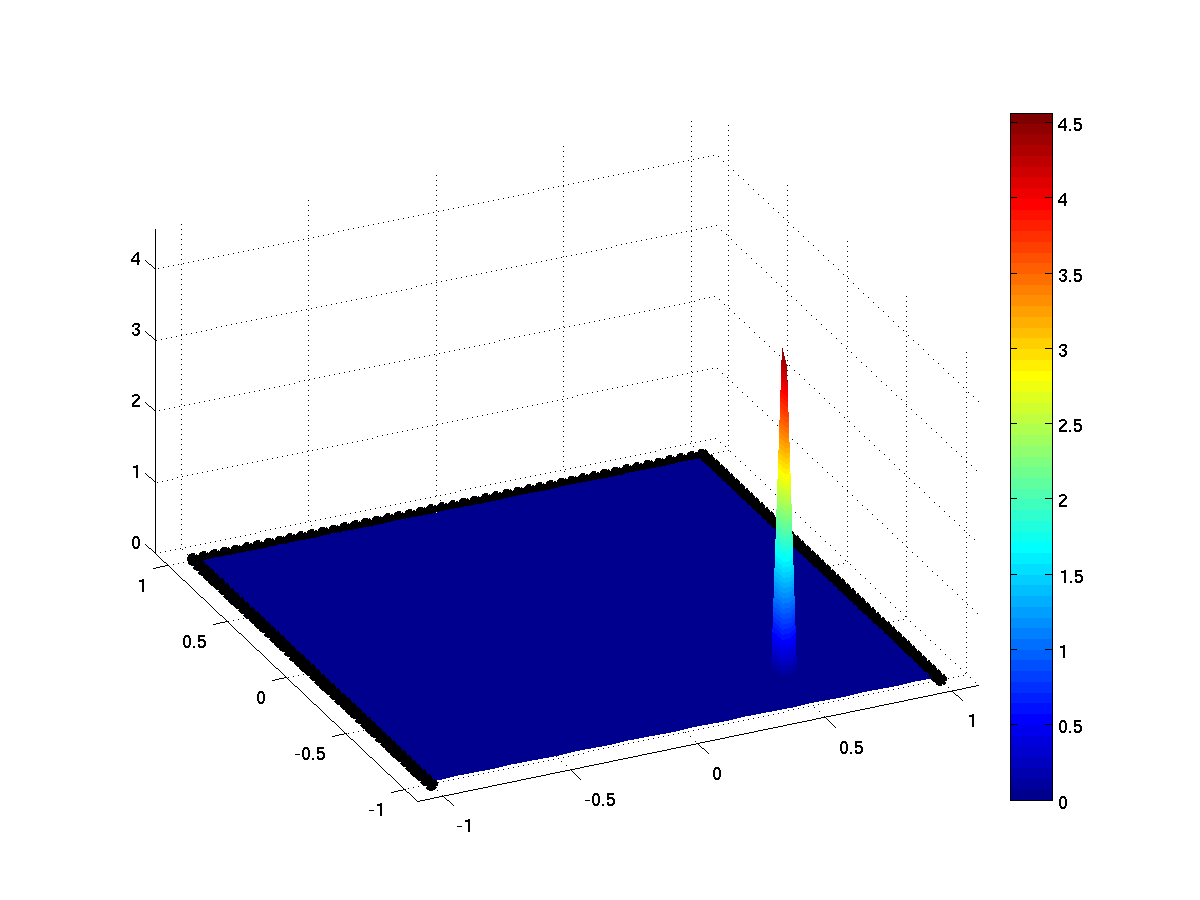}
\caption{Scenario \#2 with 3 Compton detector arrays. Exposure Time = 7 days}\label{3DS2T604800Compton}
\end{center}
\end{figure}
Here is the detection as a function of SNR$=0.1, 0.01, 0.001$:
\begin{figure}[H]
\begin{center}
\includegraphics[width=7cm]{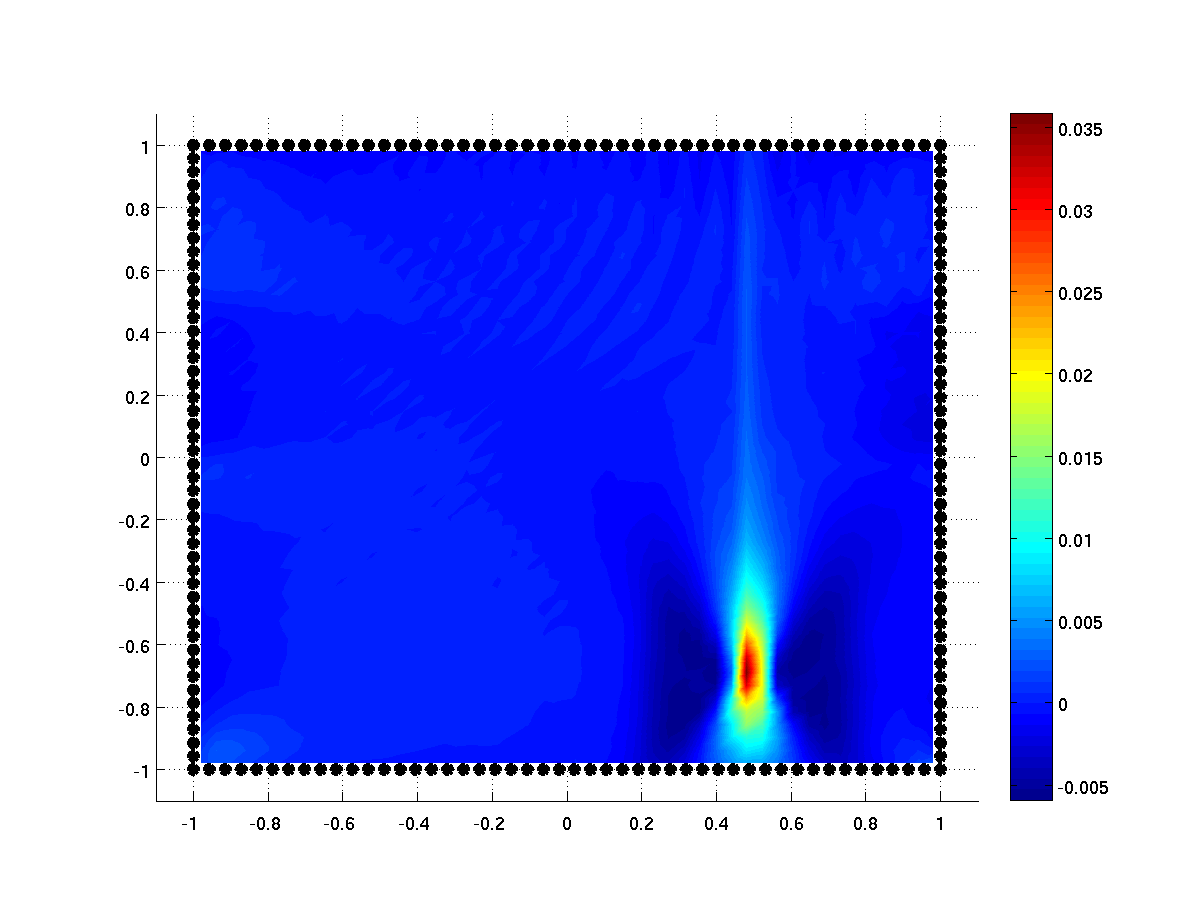}
\includegraphics[width=7cm]{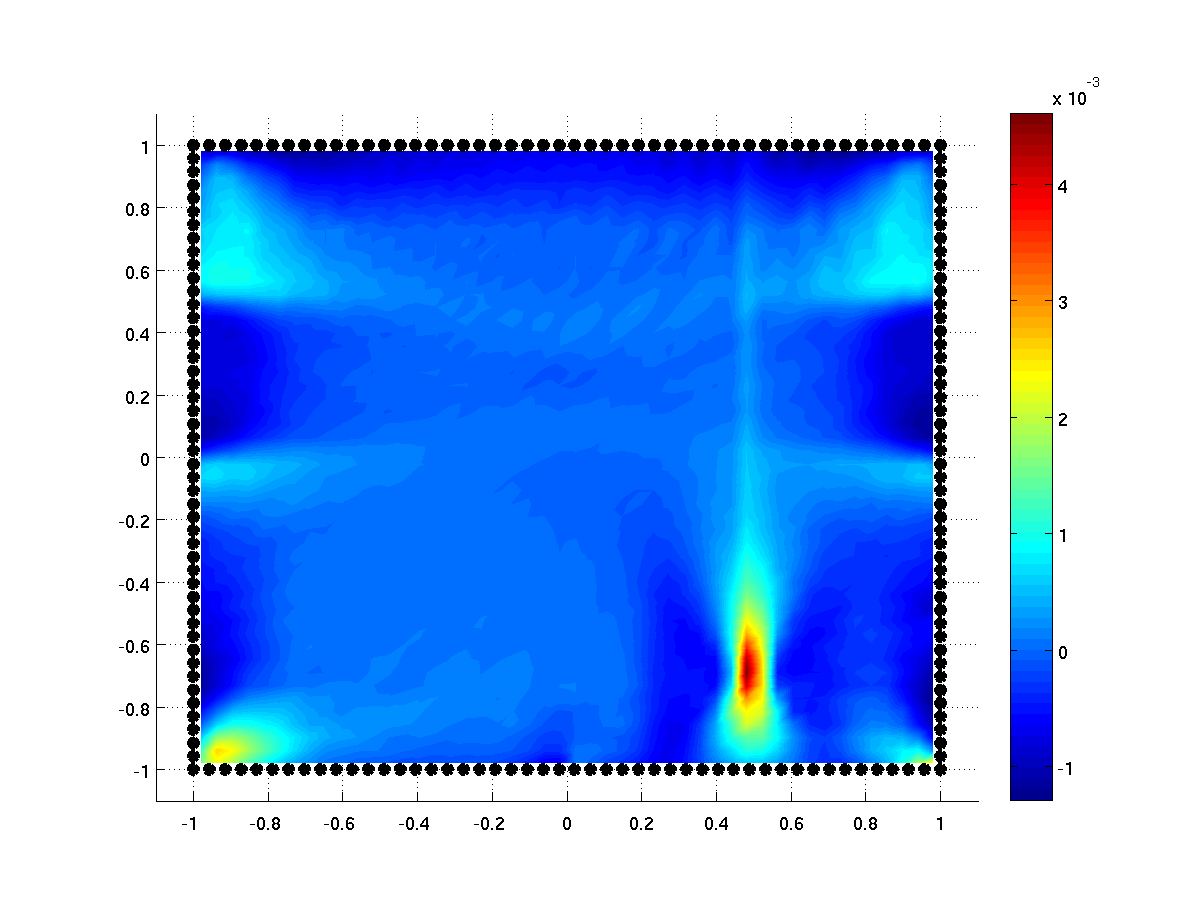}
\includegraphics[width=7cm]{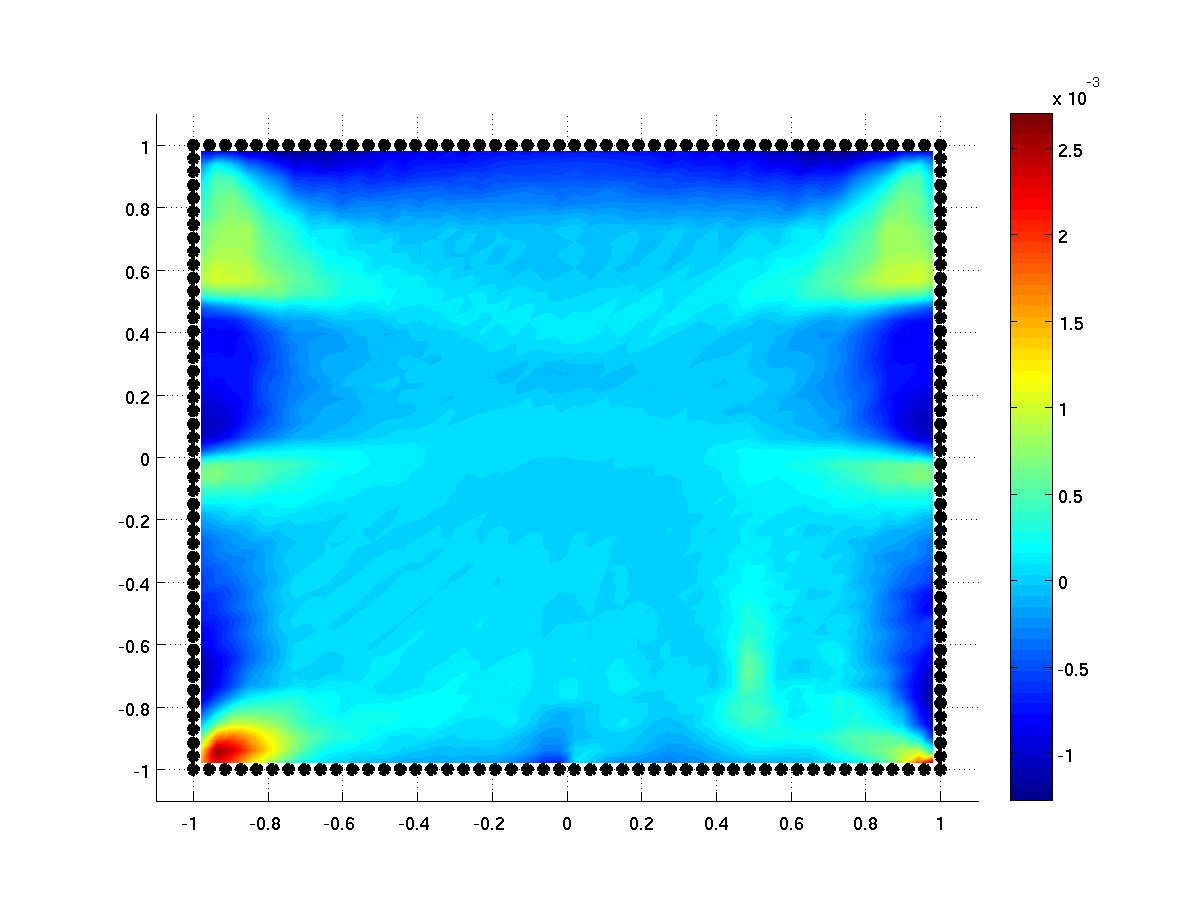}
\caption{Scenario \#2 with 4 detector arrays at various levels of SNR: $0.1, 0.01, 0.001$.}\label{4DS2SNR}
\end{center}
\end{figure}
And finally, the same with three detector arrays at SNR of $0.1$ and $0.01$:
\begin{figure}[H]
\begin{center}
\includegraphics[width=7cm]{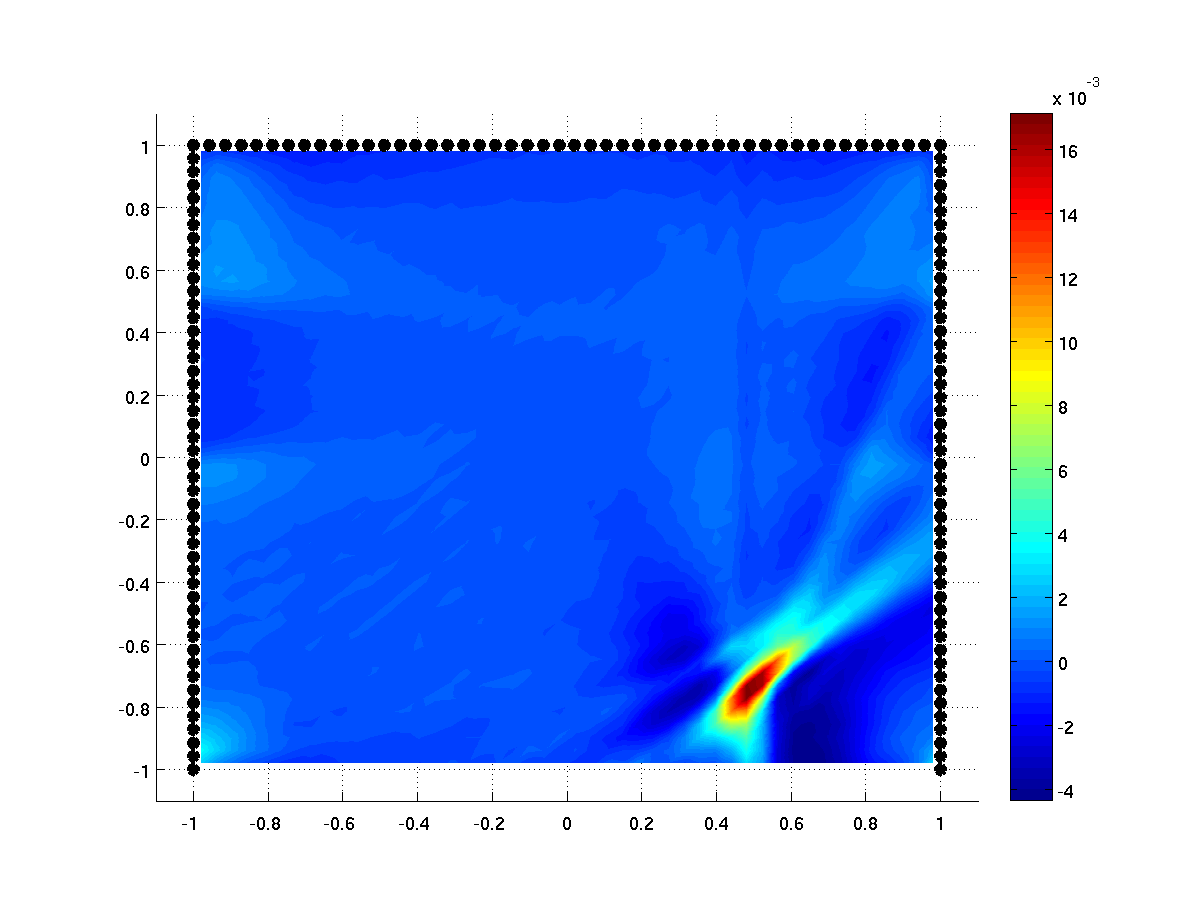}
\includegraphics[width=7cm]{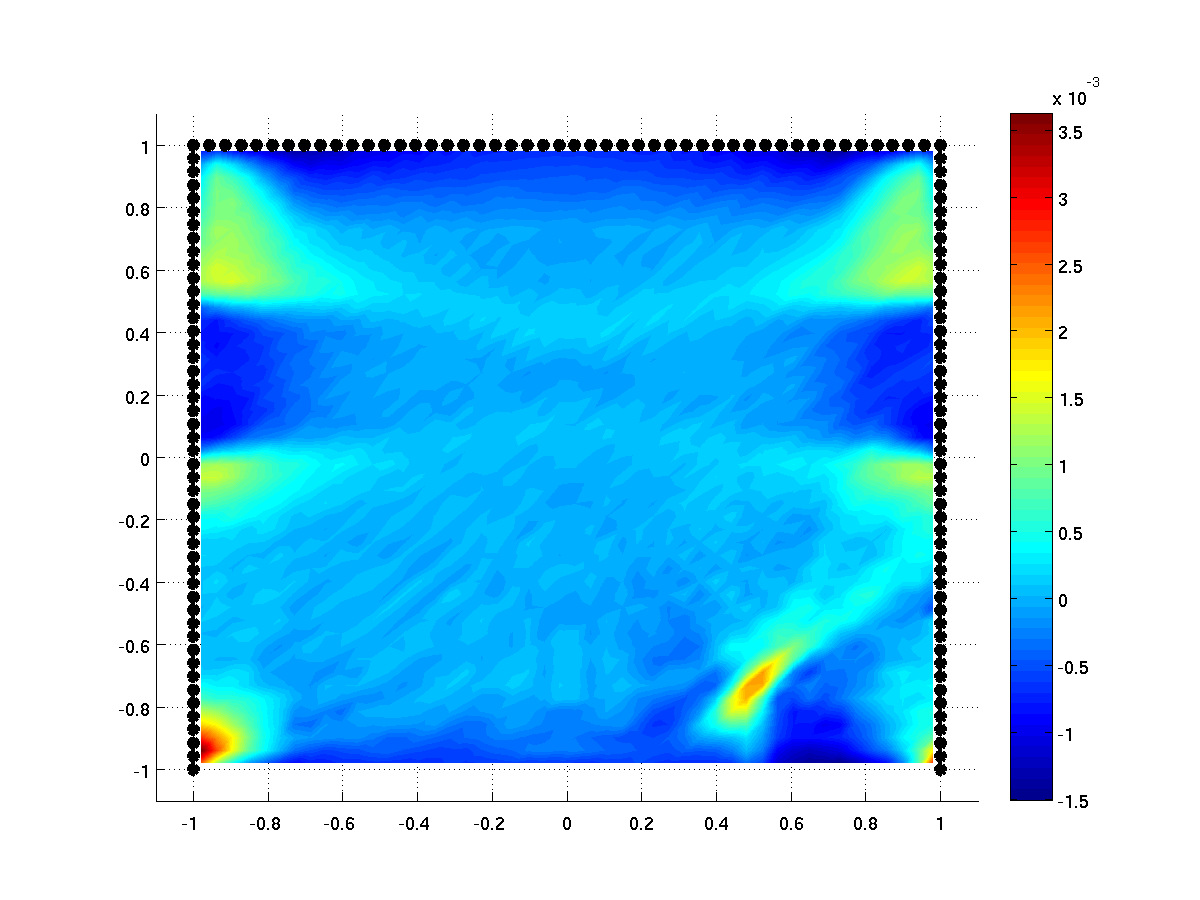}
\caption{Scenario \#2 with 3 detector arrays at SNR=$0.1, 0.01$.}\label{3DS2SNR}
\end{center}
\end{figure}

%%%%%%%%%%%%%%%%%%
\subsubsection{Cargo Scenario \#3}
\begin{figure}[H]
\begin{center}
 \includegraphics[width=6.5cm]{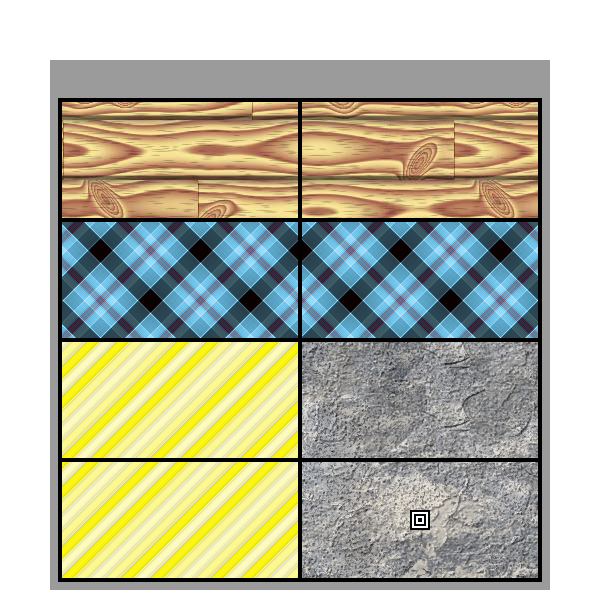}
\end{center}
\caption{Scenario \#3 Material Arrangement: plastic (striped texture), cotton (plaid texture), wood (wood grain texture), concrete (concrete texture), and an HEU source ($\boxbox$.}\label{SCN03}
\end{figure}

%%%%%%%%%%%%%%%%%%%
The next Cargo Scenario \#3 contains plastic, cotton, wood, concrete, and an HEU source as shown in Figure \ref{SCN03}. The sizes of the boxes and source are the same as before.
%The HEU source is $5 \mathrm{cm} \times 5 \mathrm{cm}$ and each rectangular material block is $60\mathrm{cm} \times 120\mathrm{cm}$.
It is clear that the source is shielded much better now, especially if the bottom detector array is missing.

The detection with four and three detector arrays are shown in the next two figures:
%In the next figure, we use four detector arrays with 48 equally spaced detectors per array and 180 angular bins.
%Figure \ref{4DS3T60} shows the back projection, back projection with the local mean subtracted, and the back projection minus the %local mean cut off at 4.3 sigmas for an exposure time of 60 seconds.
%In 1 minute, we detect 34 particles at 1 MeV, 32 of which are ballistic, which gives a signal to noise ratio of 16.
\begin{figure}[H]
\begin{center}
\includegraphics[width=7cm]{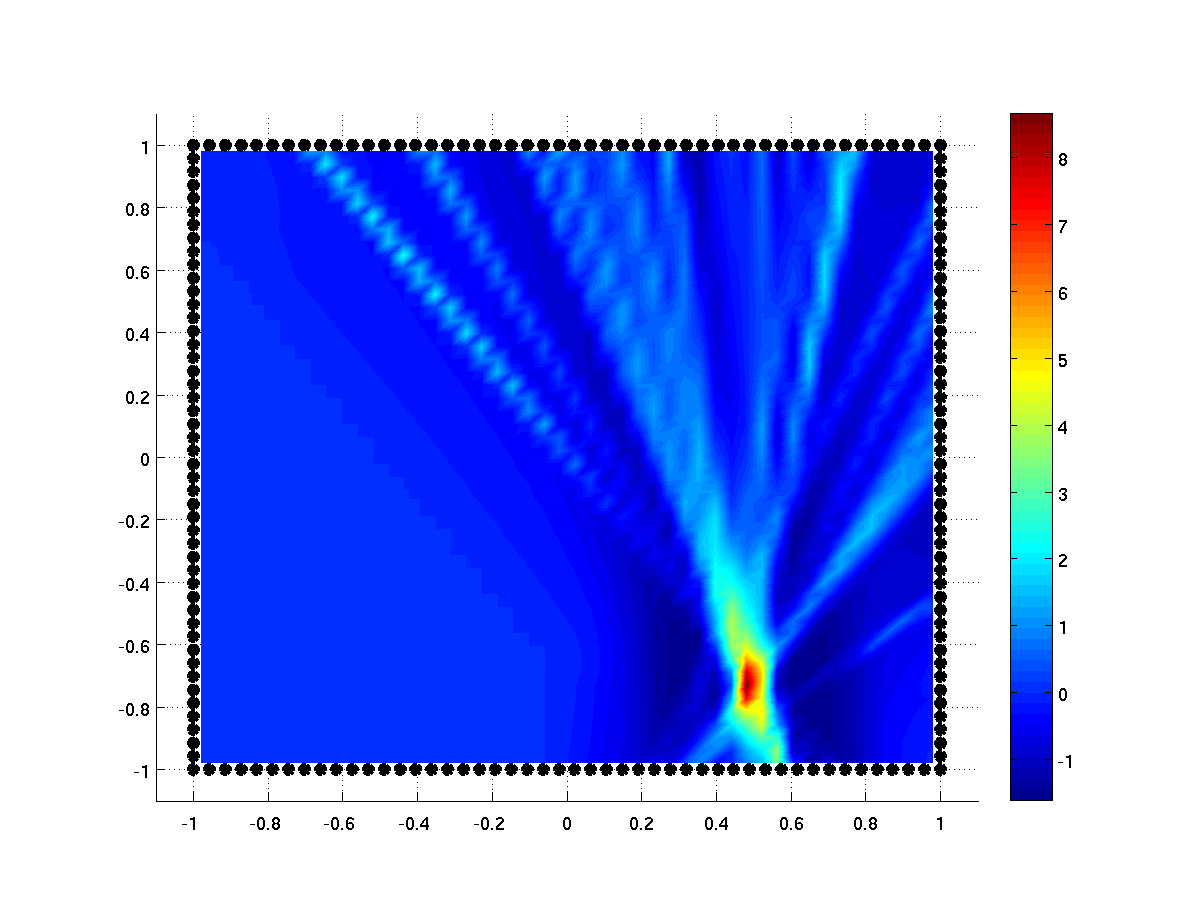}
\includegraphics[width=7cm]{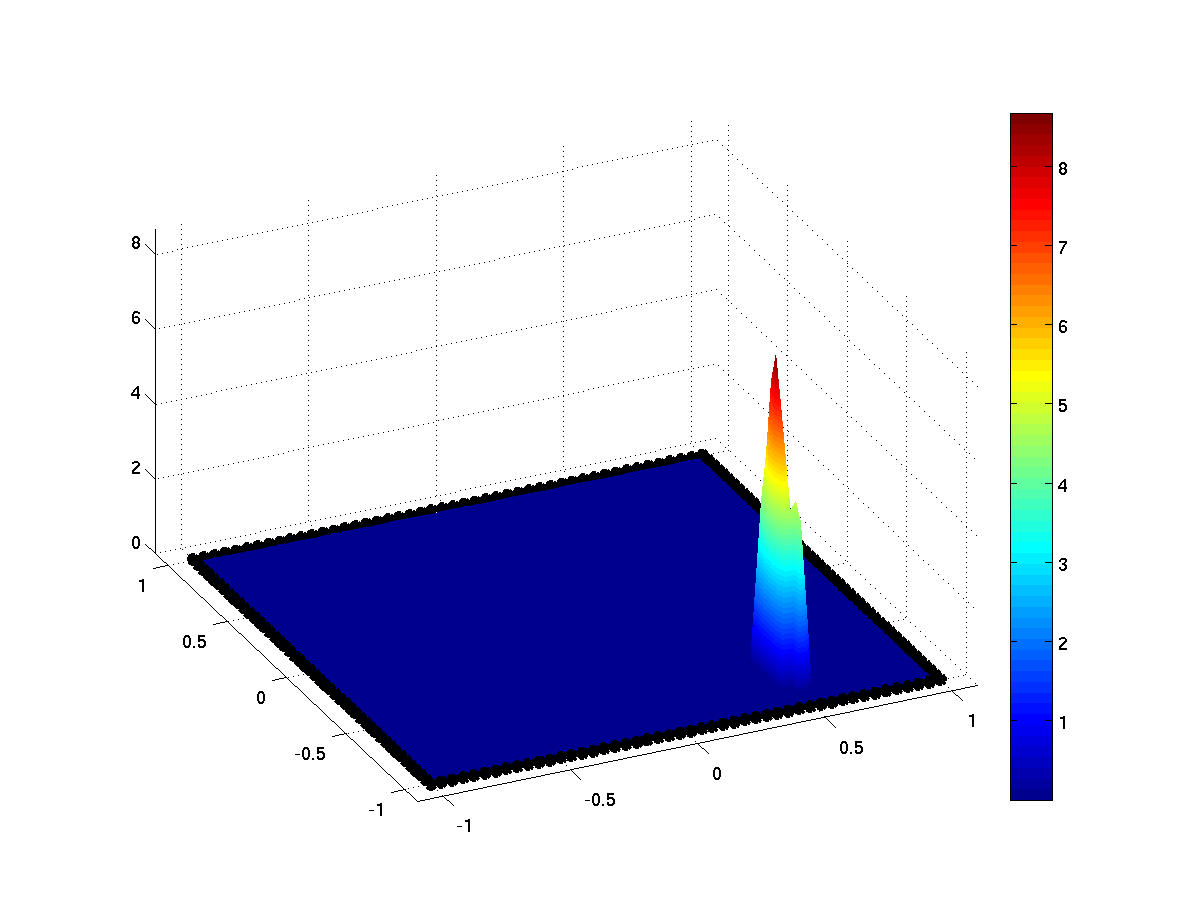}
\caption{Scenario \#3 with 4 detector arrays. Exposure Time = 1 minute}\label{4DS3T60}
\end{center}
\end{figure}
As expected, in this scenario it takes a significantly longer time to detect with three detector arrays:
%
%In Figure \ref{3DS3T86400}, we use three detector arrays with 48 equally spaced detectors per array and 180 angular bins.
%In 24 hours, we detect 482 particles at 1 MeV, 29 of which are ballistic, which gives a signal to noise ratio of 0.064.

\begin{figure}[H]
\begin{center}
\includegraphics[width=7cm]{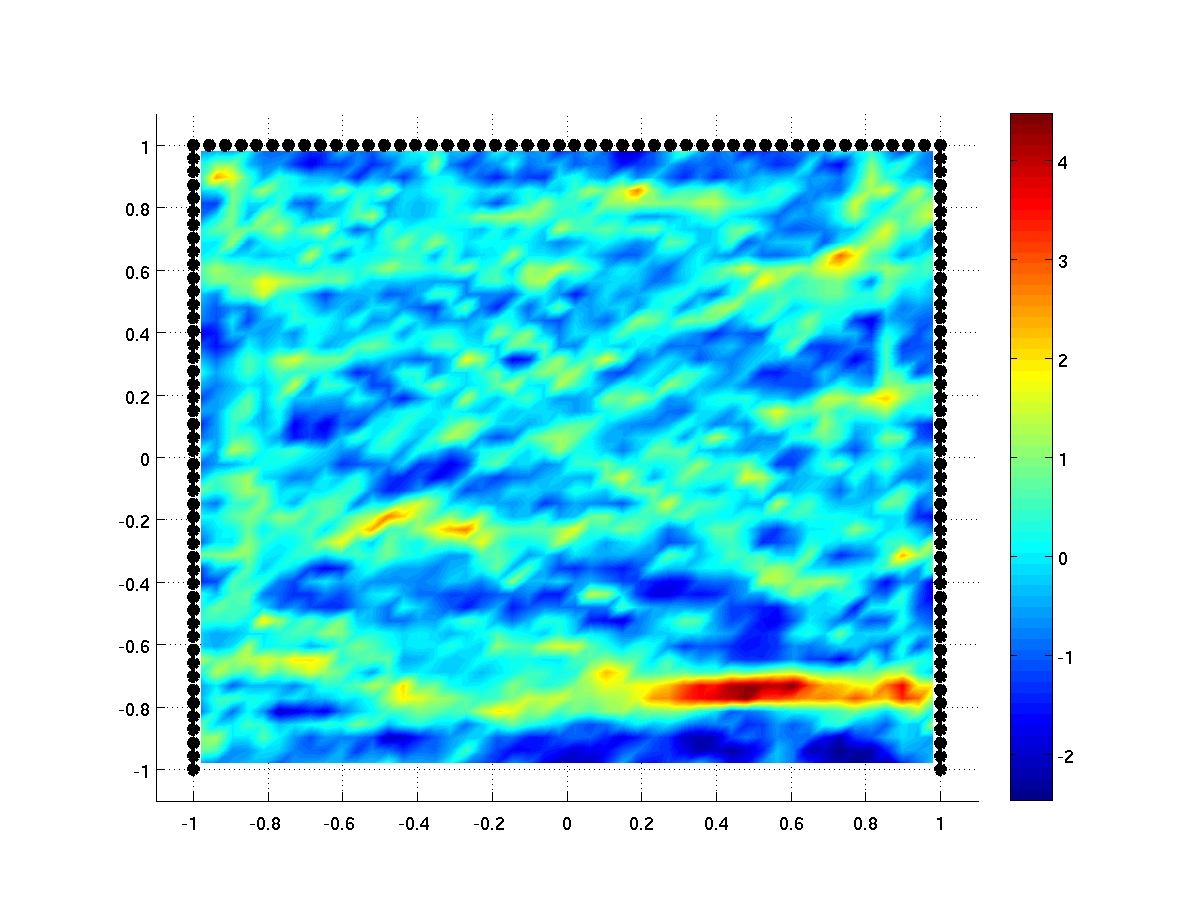}
\includegraphics[width=7cm]{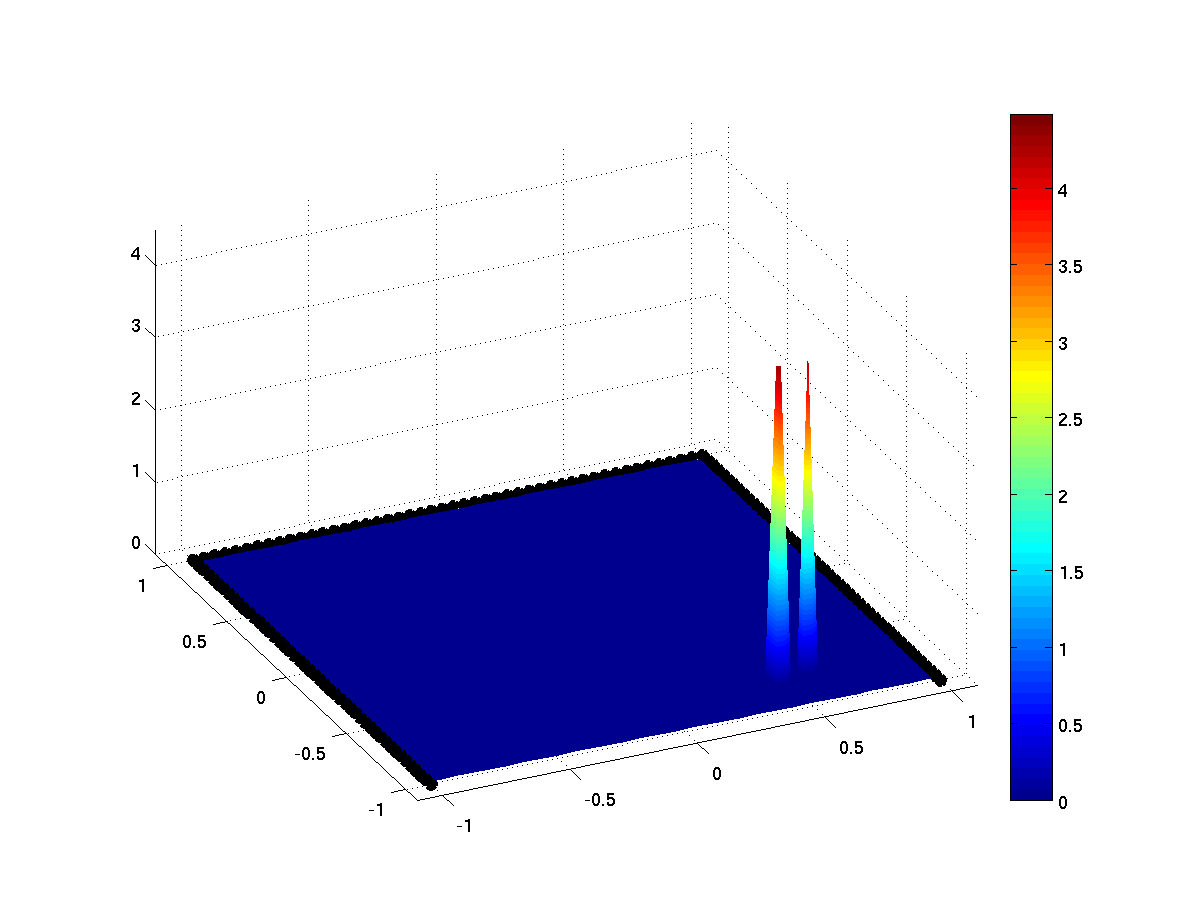}
\caption{Scenario \#3 with 3 detectors: Exposure Time = 24 hours}\label{3DS3T86400}
\end{center}
\end{figure}
As we did before, we show in the next two figures the SNR sensitivity of the detection for four and three detector arrays:
\begin{figure}[H]
\begin{center}
\includegraphics[width=7cm]{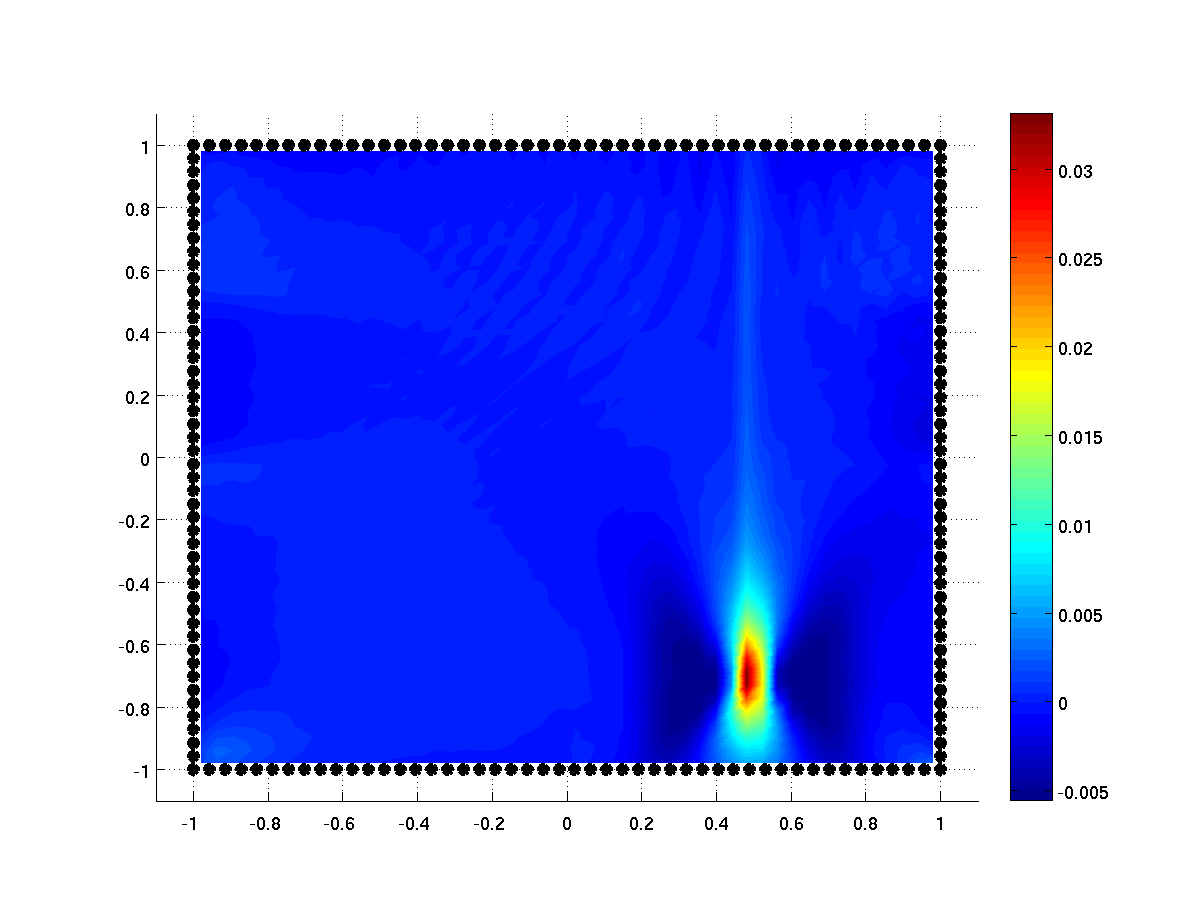}
\includegraphics[width=7cm]{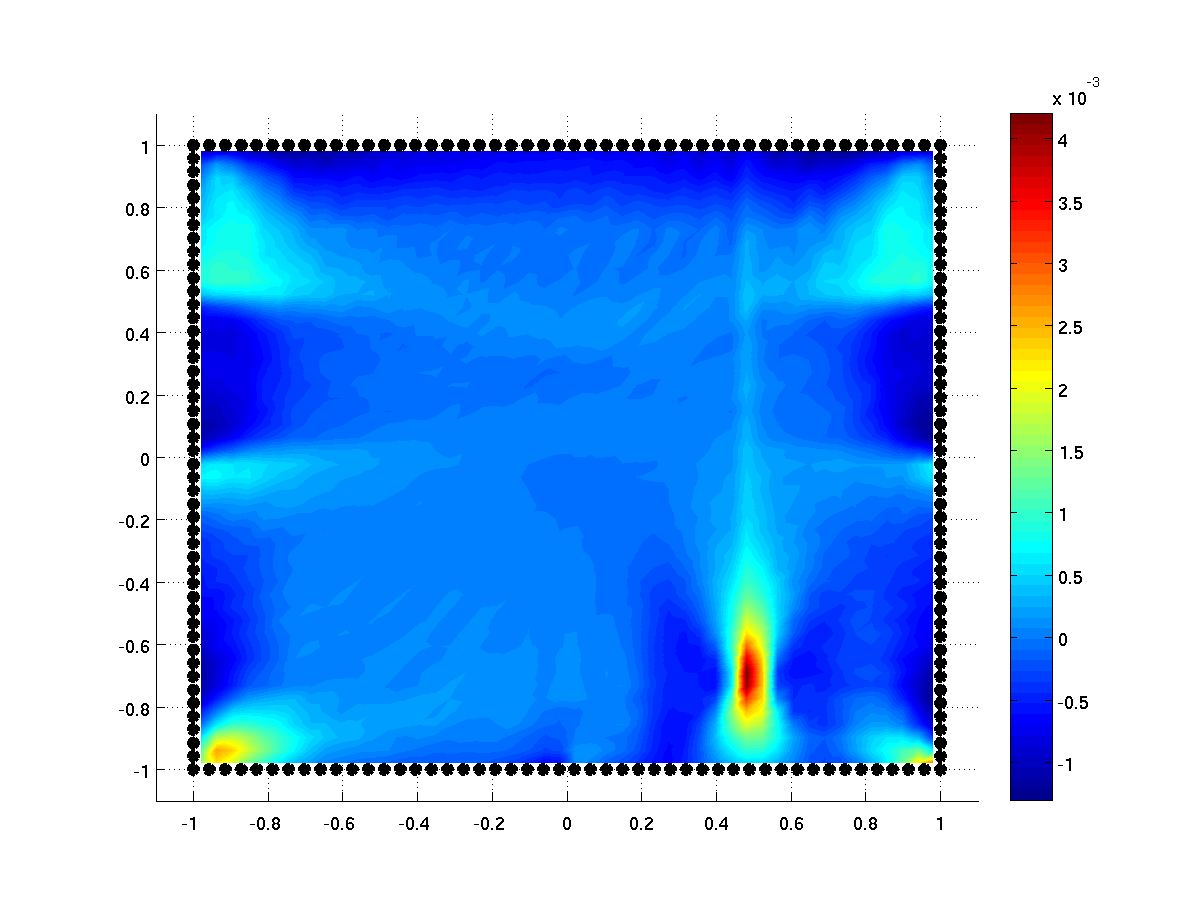}
\includegraphics[width=7cm]{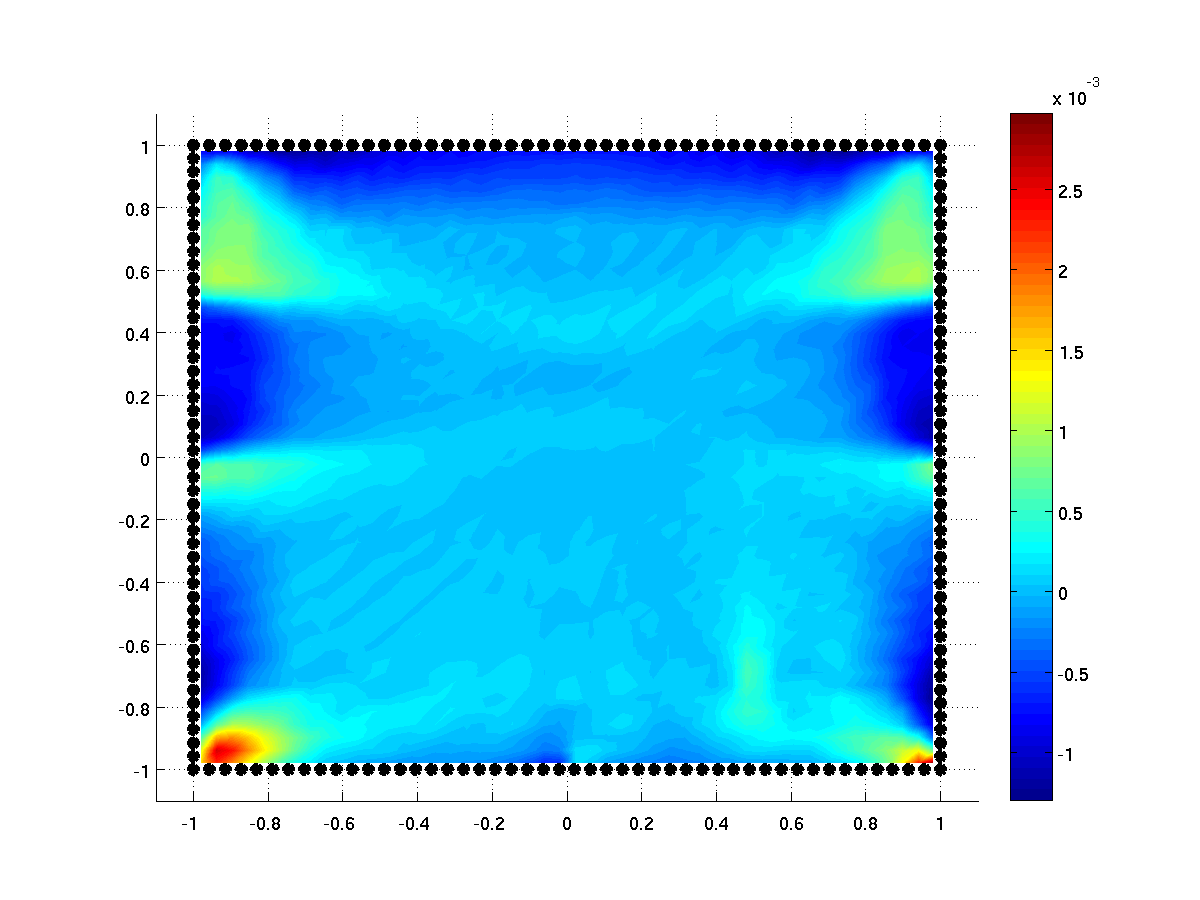}
\caption{Scenario \#3 with 4 detector arrays at SNR levels $0.1, 0.01, 0.001$.}\label{4DS3SNR}
\end{center}
\end{figure}

\begin{figure}[H]
\begin{center}
\includegraphics[width=7cm]{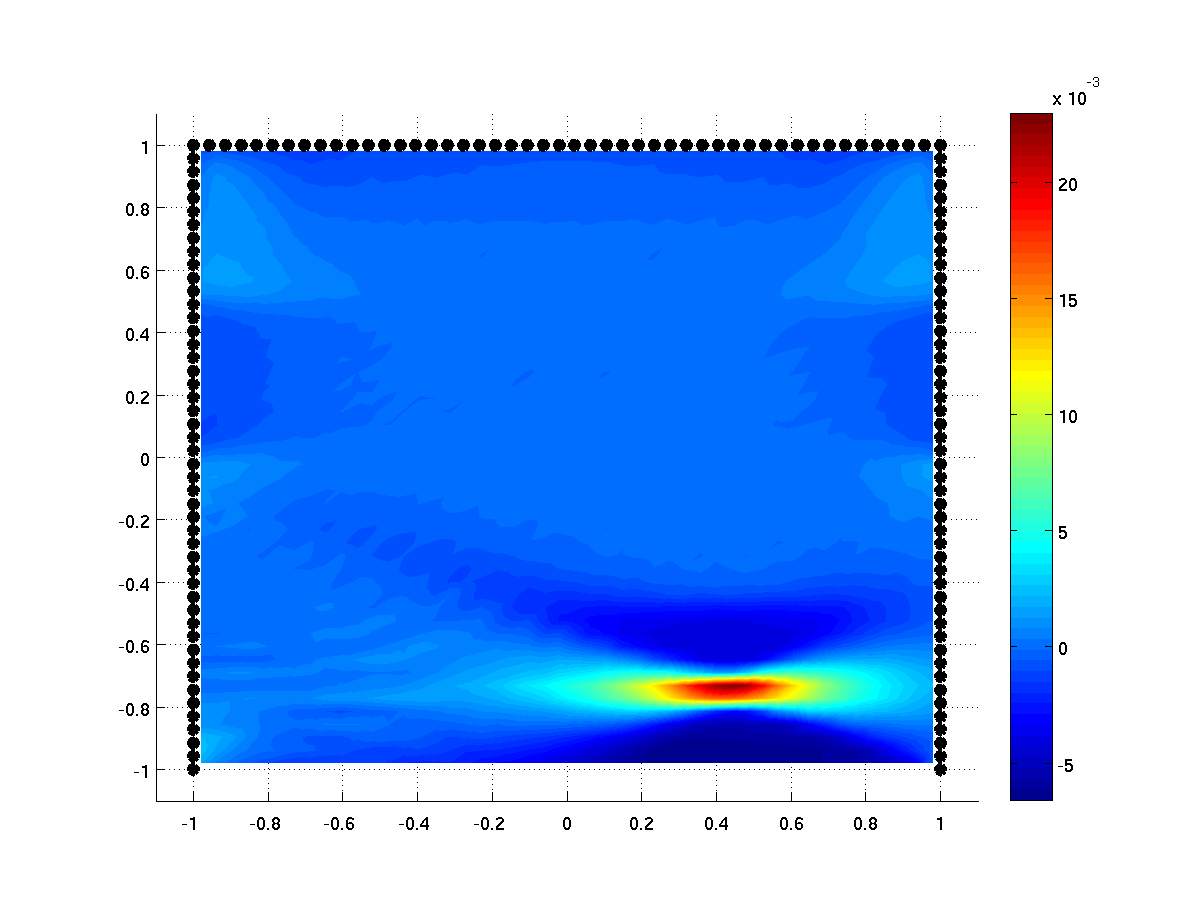}
\includegraphics[width=7cm]{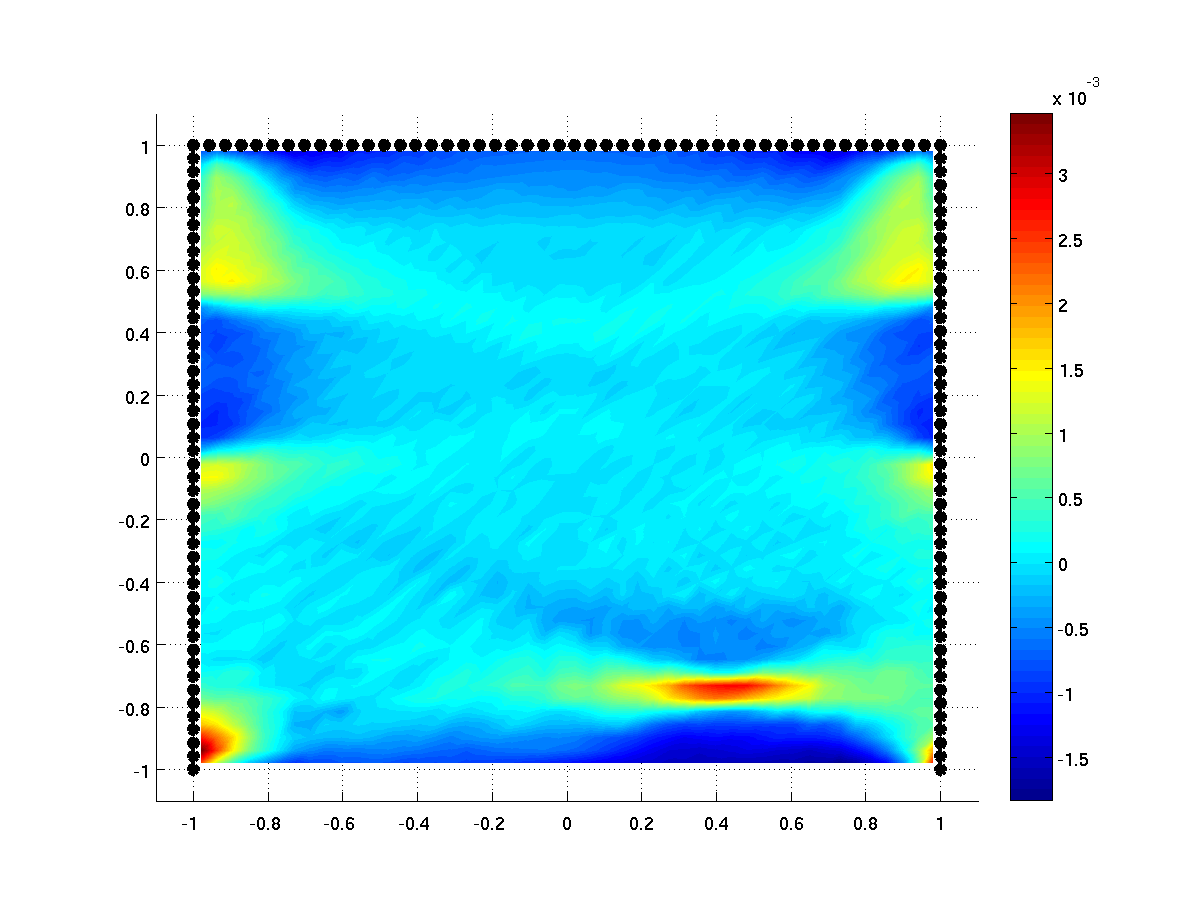}
\caption{Scenario \#3 with 3 detector arrays at SNR levels $0.1, 0.01$}\label{3DS3SNR}
\end{center}
\end{figure}

We now modify the scenario by moving deeper the location of the source.

%%%%%%%%%%%%%%%%%%%%%%%%%%%%%%%%
\subsubsection{A modified Cargo Scenario \#3 (Source Moved)}
%%%%%%%%%%%%%%%%%%%%%%%%%%%%%%

%Description: Cargo Scenario \#3 contains plastic (yellow), cotton (blue), wood (brown), iron at 50\% density (dark gray), %concrete (tan), and an HEU source (red) as shown in Figure \ref{SCN03m}.
%The HEU source is $5 \mathrm{cm} \times 5 \mathrm{cm}$ and each rectangular material block is $60\mathrm{cm} \times 120\mathrm{cm}$.
\begin{figure}[H]
\begin{center}
 \includegraphics[width=6.5cm]{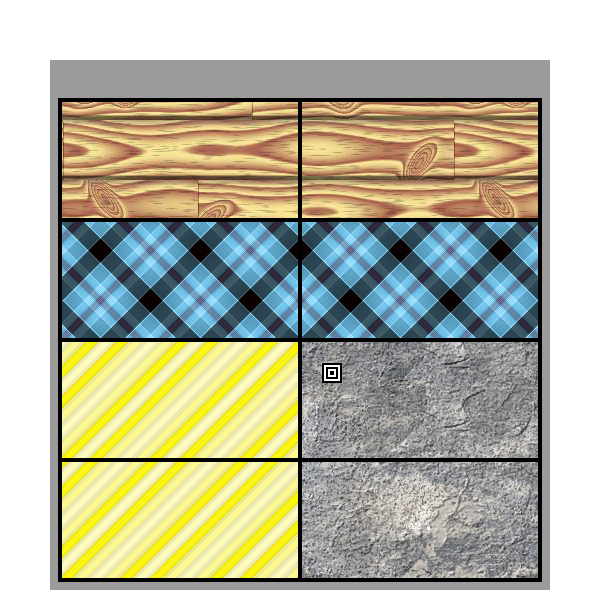}
\end{center}
\caption{Modified Scenario \#3. Material Arrangement}\label{SCN03m}
\end{figure}
One discovers much better detection here:
%In the next figure, we use four detector arrays with 48 equally spaced detectors per array and 180 angular bins.
%Figure \ref{4DS3mT1} shows the back projection, back projection with the local mean subtracted, and the back projection minus the %local mean cut off at 4.3 sigmas for an exposure time of 1 seconds.
%In 1 second, we detect 24 particles at 1 MeV, 23 of which are ballistic, which gives a signal to noise ratio of 23.
\begin{figure}[H]
\begin{center}
\includegraphics[width=7cm]{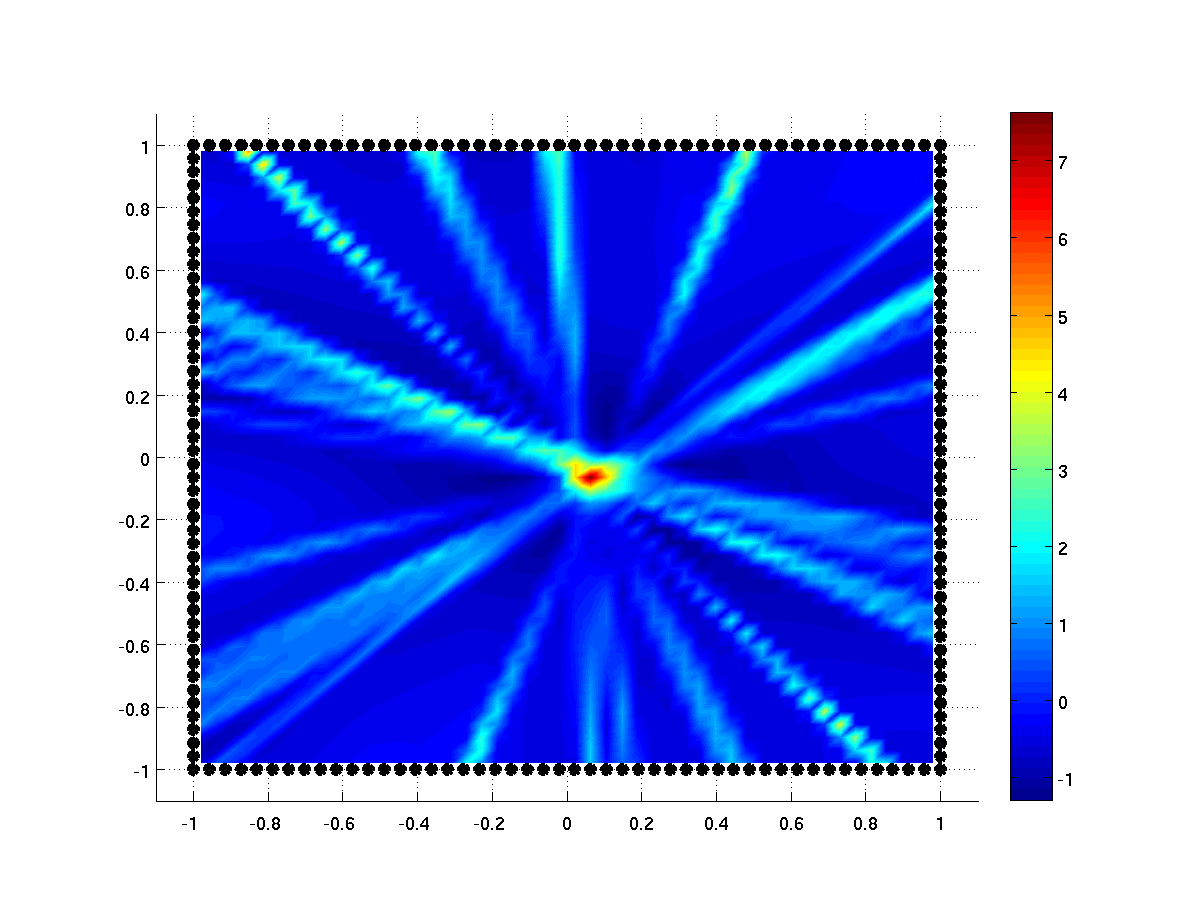}
\includegraphics[width=7cm]{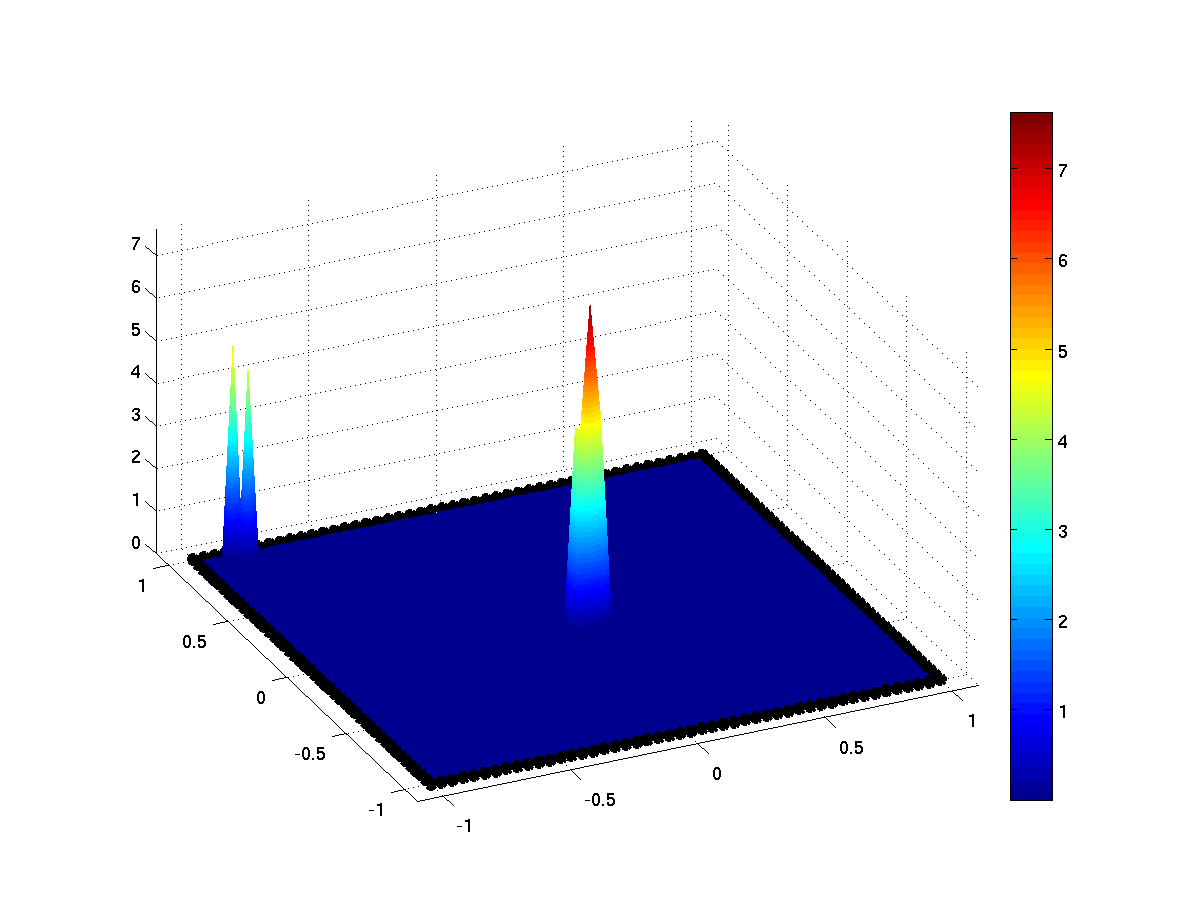}
\caption{Modified Scenario \#3 with 4 detector arrays. Exposure Time = 1 second}
\end{center}
\label{4DS3mT1}
\end{figure}
%\pagebreak
%Using just 3 detector arrays we get essentially the same thing since nearly all particles are detected by the left and top %detector arrays due to the location of the source.
\begin{figure}[H]
\begin{center}
\includegraphics[width=7cm]{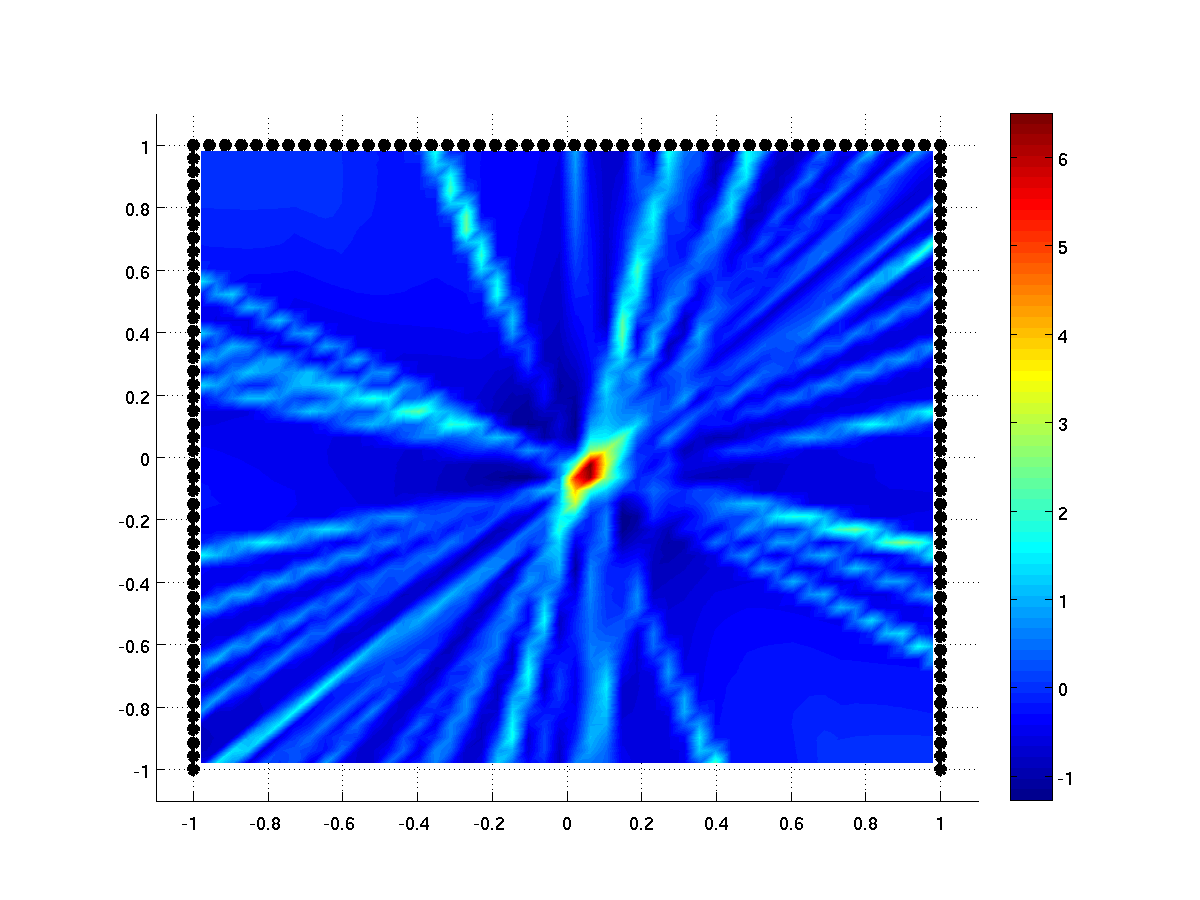}
\includegraphics[width=7cm]{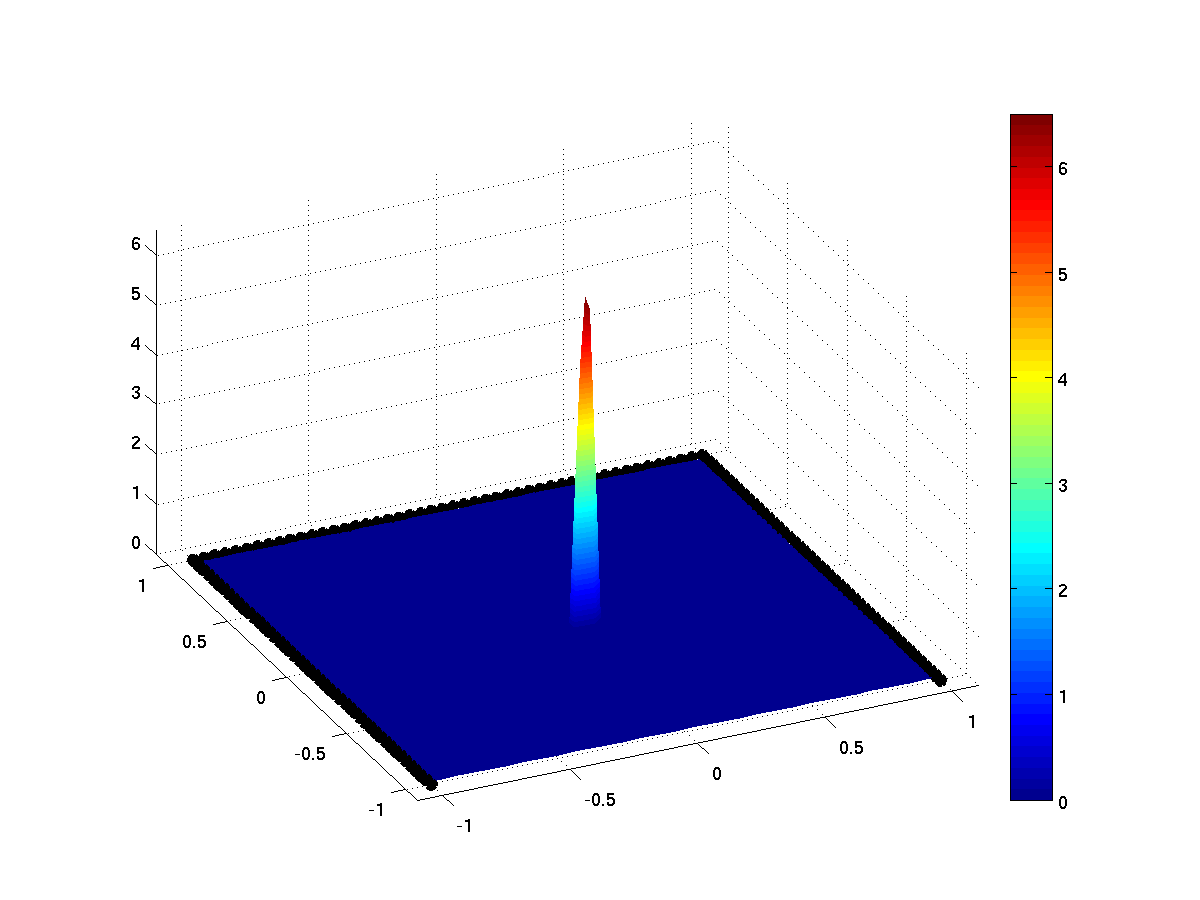}
\caption{Modified Scenario \#3 with 3 detector arrays (the bottom one missing). Exposure Time = 1 second}\label{3DS3mT1}
\end{center}
\end{figure}
One notices no deterioration due to the missing bottom detector array, since for the current location of the source the bottom array does not play such a major role anymore.
%\begin{figure}[H]
%\includegraphics[width=7cm]{BP_S03mod-NAng180-4detectors-S0-Tinfinite-SNR01.png}
%\includegraphics[width=7cm]{LocalMean_S03mod-NAng180-4detectors-S0-Tinfinite-SNR01.png}\\
%\includegraphics[width=7cm]{BP_S03mod-NAng180-4detectors-S0-Tinfinite-SNR001.png}
%\includegraphics[width=7cm]{LocalMean_S03mod-NAng180-4detectors-S0-Tinfinite-SNR001.png}\\
%\includegraphics[width=7cm]{BP_S03mod-NAng180-4detectors-S0-Tinfinite-SNR0001.png}
%\includegraphics[width=7cm]{LocalMean_S03mod-NAng180-4detectors-S0-Tinfinite-SNR0001.png}
%\caption{Scenario \#3mod with 4 detectors at various SNR}\label{4DS3mSNR}
%\end{figure}
The sensitivity for low SNR also improves:
\begin{figure}[H]
\begin{center}
\includegraphics[width=7cm]{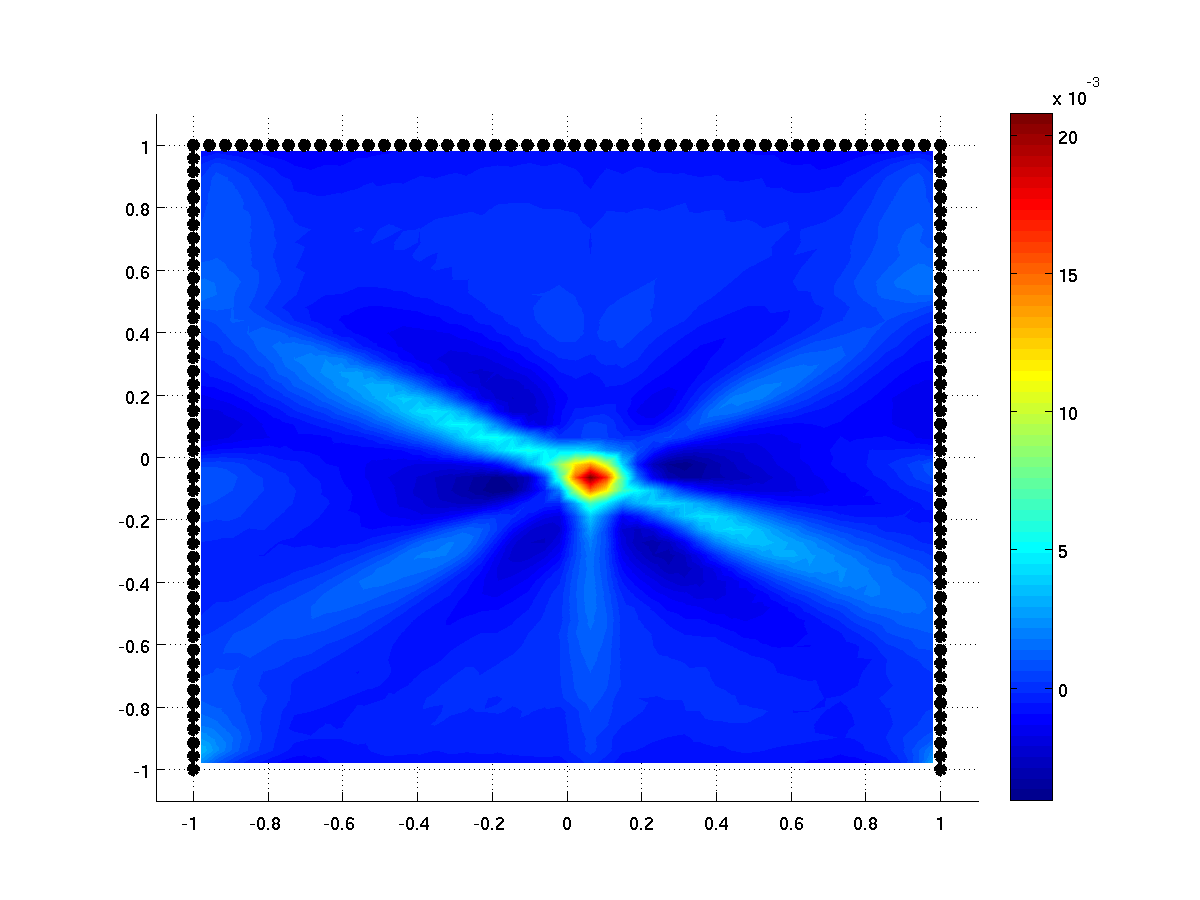}
\includegraphics[width=7cm]{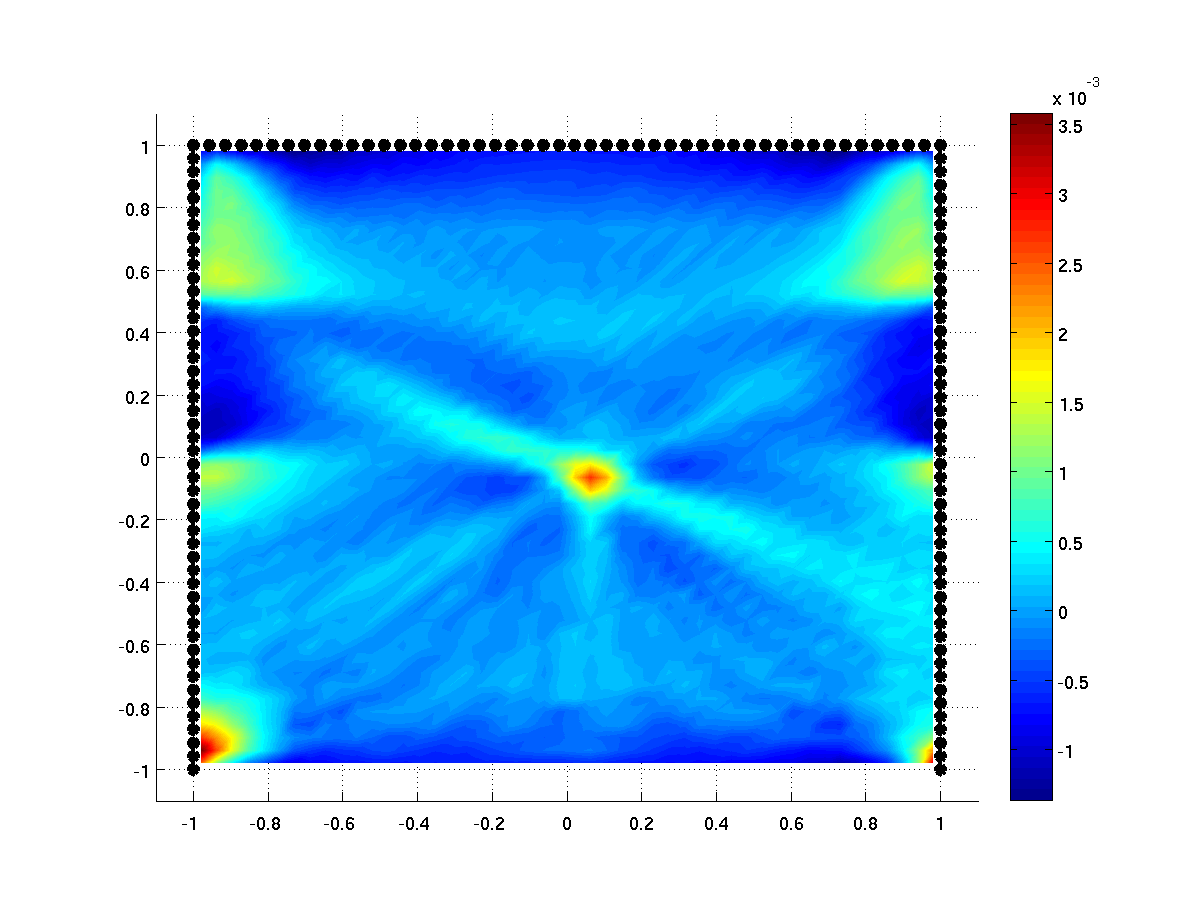}
\caption{Modified Scenario \#3 with 3 detectors at SNR=$0.1, 0.01$.}\label{3DS3mSNR}
\end{center}
\end{figure}

%%%%%%%%%%%%%%%%%%%%%%%%
\subsubsection{Cargo Scenario \#4}
%%%%%%%%%%%%%%%%%%%%%%%

\begin{figure}[H]
\begin{center}
 \includegraphics[width=6.5cm]{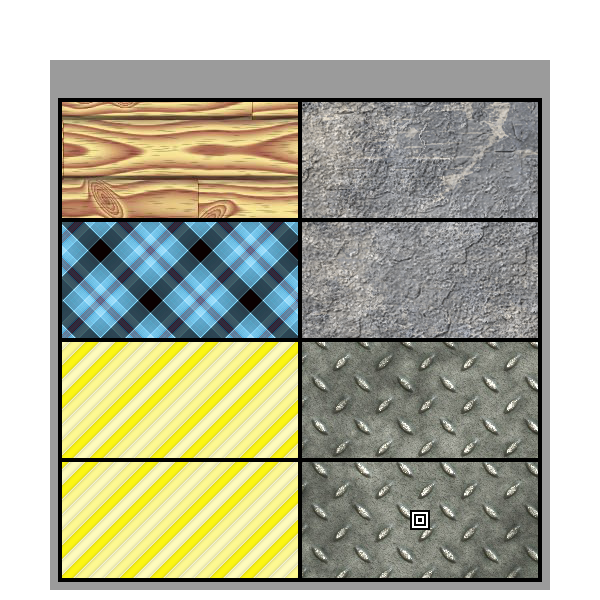}
\end{center}
\caption{Scenario \#4. Material Arrangement: plastic (striped texture), cotton (plaid texture), wood (wood grain texture), iron at 50\% density (metal tread texture), concrete (concrete texture), and an HEU source ($\boxbox$).}\label{SCN04}
\end{figure}

The Cargo Scenario \#4 contains plastic, cotton, wood, iron at $50\%$ density, concrete, and an HEU source
as shown in Figure \ref{SCN04}.
%The HEU source is $5 \mathrm{cm} \times 5 \mathrm{cm}$ and each rectangular material block is $60\mathrm{cm} \times %120\mathrm{cm}$.
This scenario is significantly more difficult, due to the thick iron and concrete shielding. Nevertheless, with complete enclosure by detector arrays detection is possible in a reasonable amount of time:
%In the next figure, we use four detector arrays with 48 equally spaced detectors per array and 180 angular bins.
%Figure \ref{4DS4T120} shows the back projection, back projection with the local mean subtracted, and the back projection minus the local mean cut off at 4.3 sigmas for an exposure time of 120 seconds.
%In 2 minutes, we detect 9 particles at 1 MeV, 7 of which are ballistic, which gives a signal to noise ratio of 3.5.
\begin{figure}[H]
\begin{center}
\includegraphics[width=7cm]{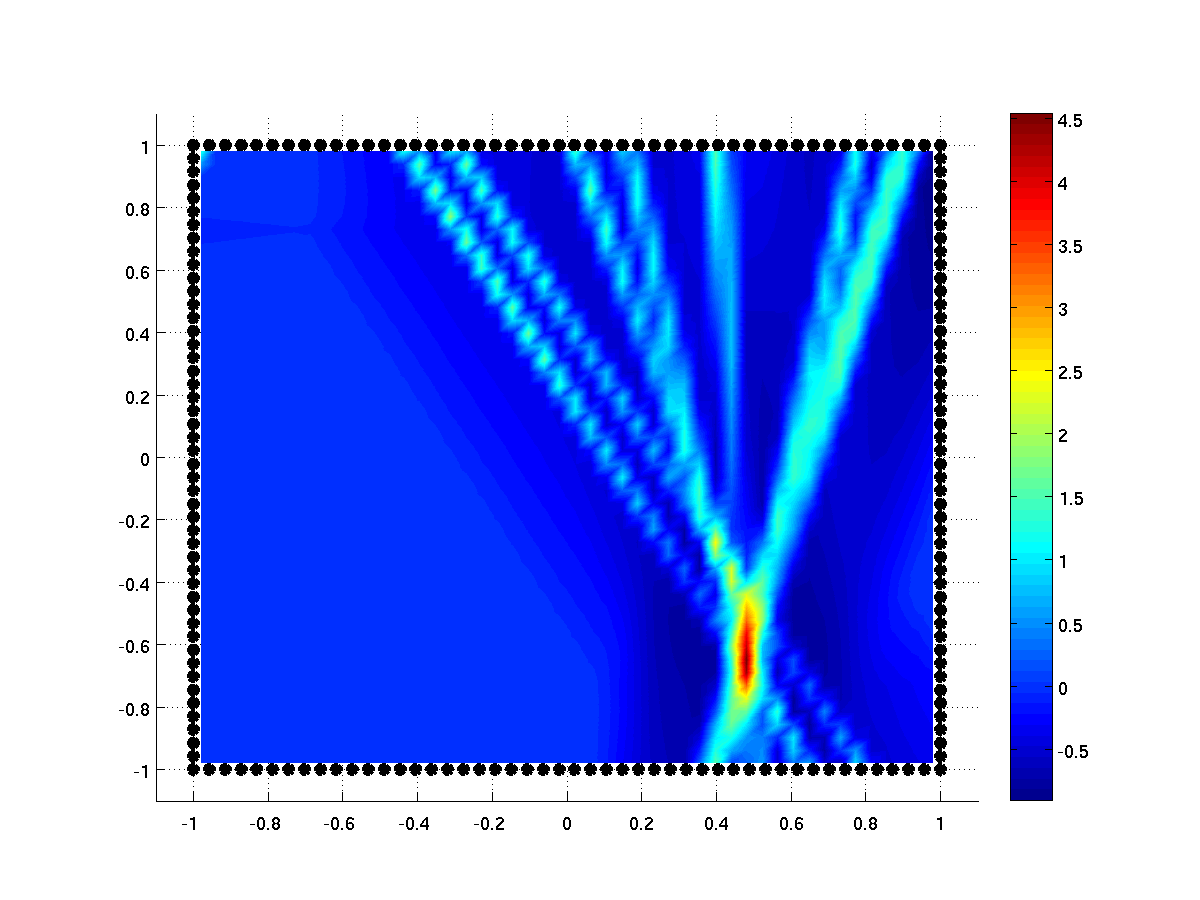}
\includegraphics[width=7cm]{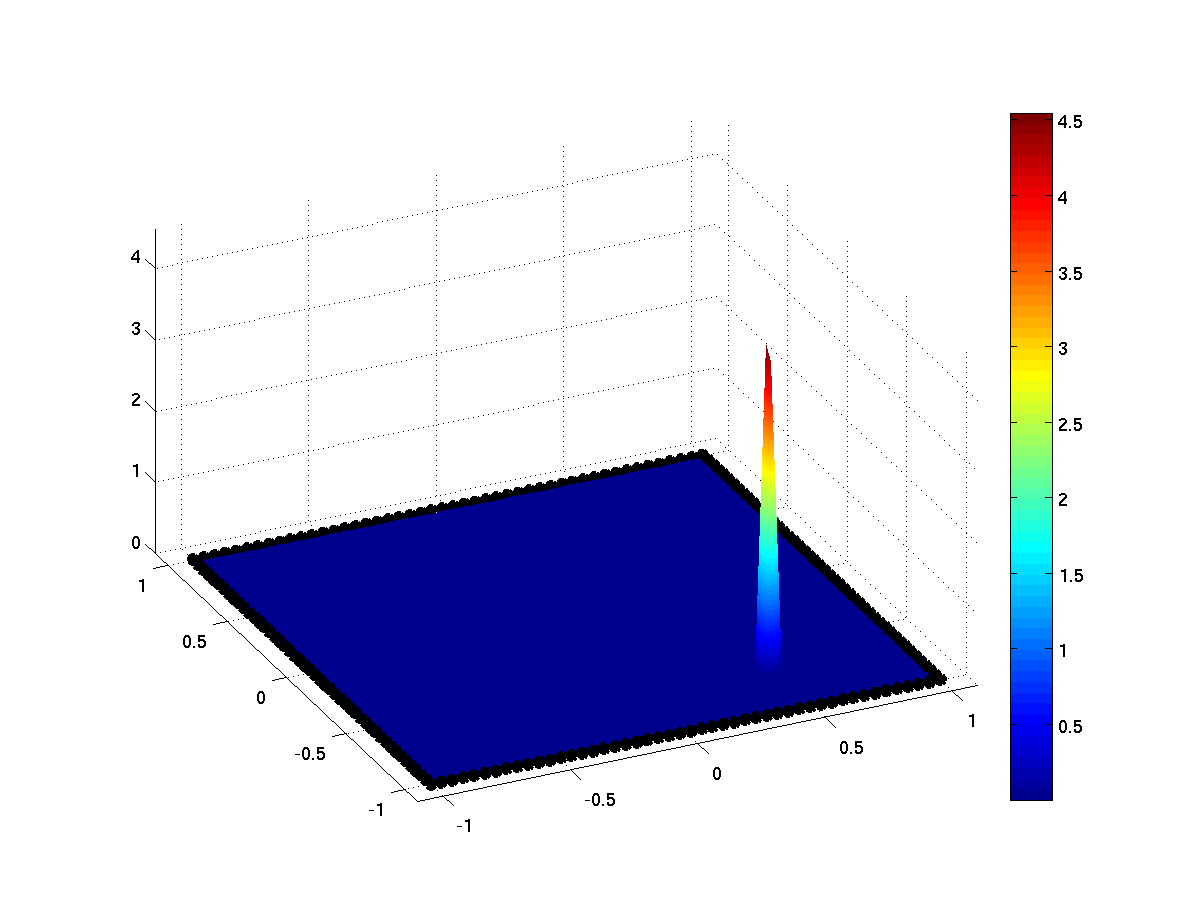}
\caption{Scenario \#4 with 4 detector arrays. Exposure Time = 2 minutes}\label{4DS4T120}
\end{center}
\end{figure}
However, using three detector arrays, with the bottom one missing, reliable detection becomes impossible even for long observation times.
%the source is not detectable.
%Figure \ref{4DS4Tinf} shows the back projection and back projection with the local mean subtracted for an infinite exposure time (i.e. using unsampled data).
%
%\begin{figure}[H]
%\includegraphics[width=7cm]{BP_S04m-XS-180-NAng180-3detectors-S0-Tinfinite.png}
%\includegraphics[width=7cm]{LocalMean_S04m-XS-180-NAng180-3detectors-S0-Tinfinite.png}
%\caption{Scenario \#4 with 3 detectors: Exposure Time = infinite}\label{4DS4Tinf}
%\end{figure}
The SNR dependence is shown below:
\begin{figure}[H]
\begin{center}
\includegraphics[width=7cm]{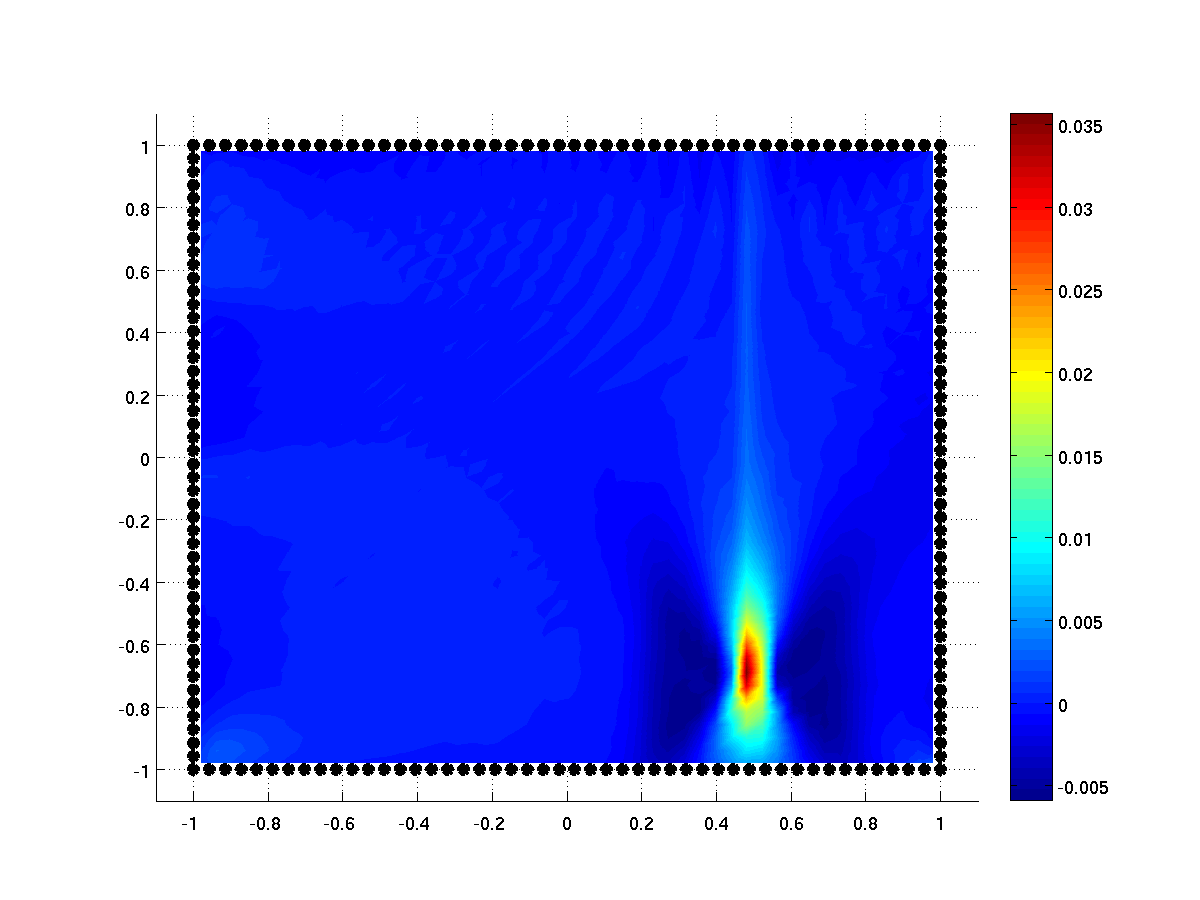}
\includegraphics[width=7cm]{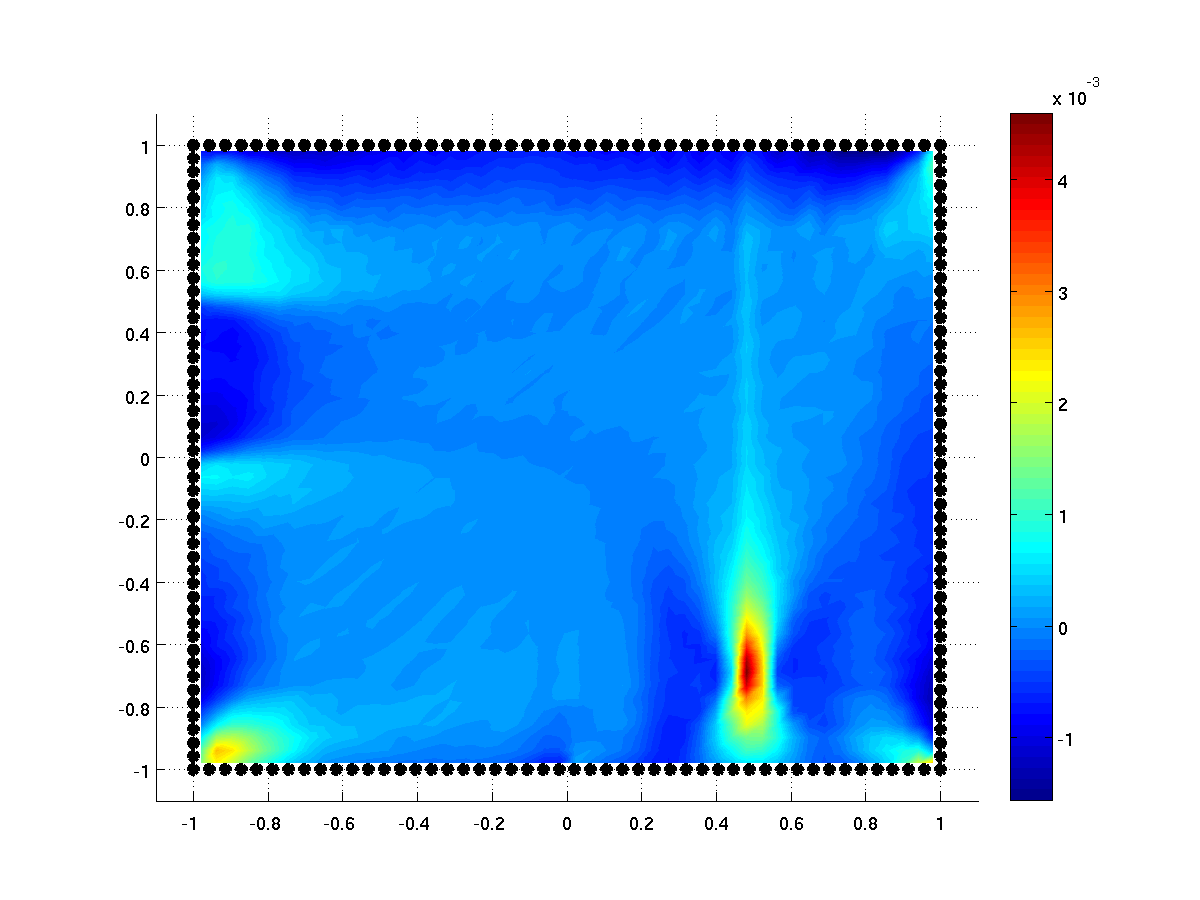}
\includegraphics[width=7cm]{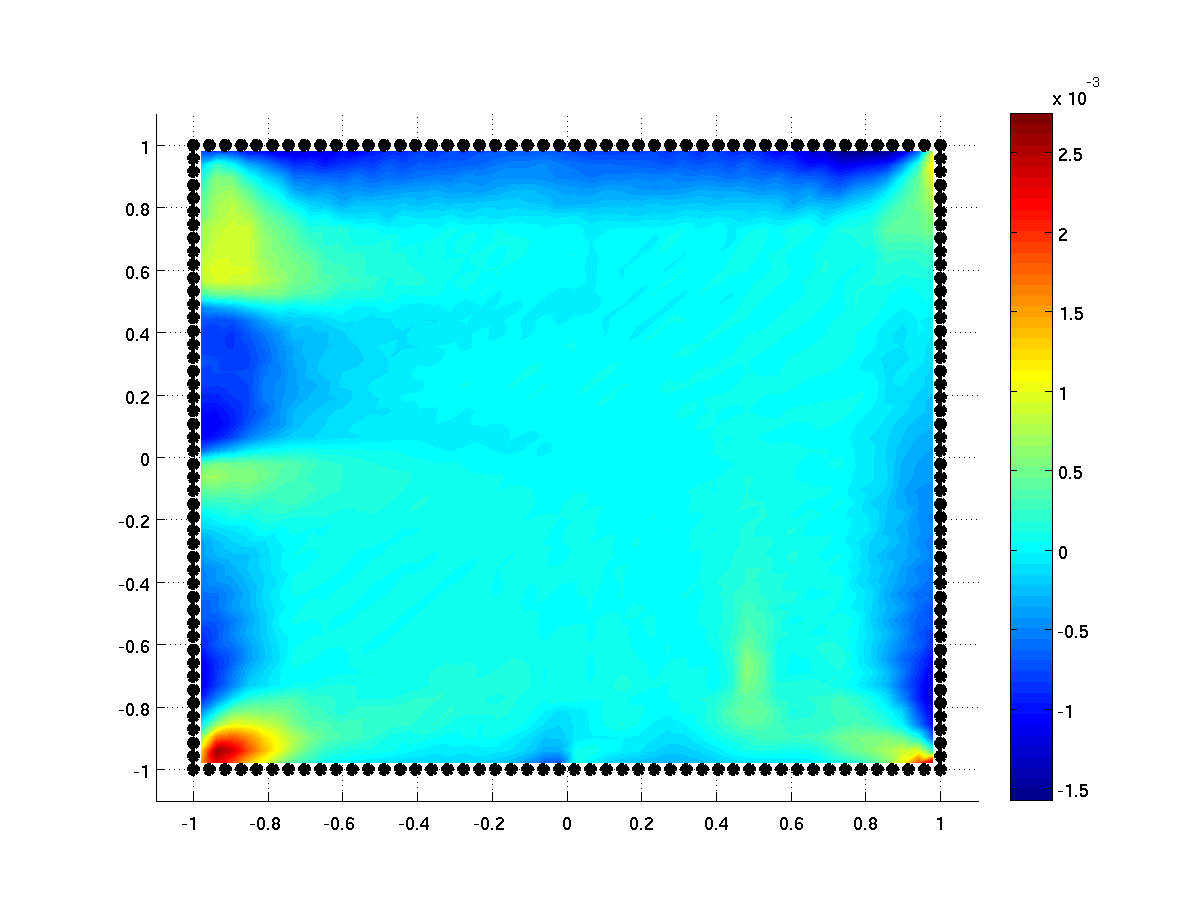}
\caption{Scenario \#4 with 4 detector arrays at SNR = $0.1, 0.01, 0.001$.}\label{4DS4SNR}
\end{center}
\end{figure}
%
%\begin{figure}[H]
%\includegraphics[width=7cm]{BP_S04m-NAng180-3detectors-S0-Tinfinite-SNR01.png}
%\includegraphics[width=7cm]{LocalMean_S04m-NAng180-3detectors-S0-Tinfinite-SNR01.png}\\
%\includegraphics[width=7cm]{BP_S04m-NAng180-3detectors-S0-Tinfinite-SNR001.png}
%\includegraphics[width=7cm]{LocalMean_S04m-NAng180-3detectors-S0-Tinfinite-SNR001.png}
%\caption{Scenario \#4 with 3 detectors at various SNR}\label{3DS4SNR}
%\end{figure}

%%%%%%%%%%%%%%%%%%%%%%%%%%%%%%%%%%%%%%%%%%
\subsubsection{Cargo Scenario \#5 (Fertilizer)}
\begin{figure}[H]
\begin{center}
\includegraphics[width=7cm]{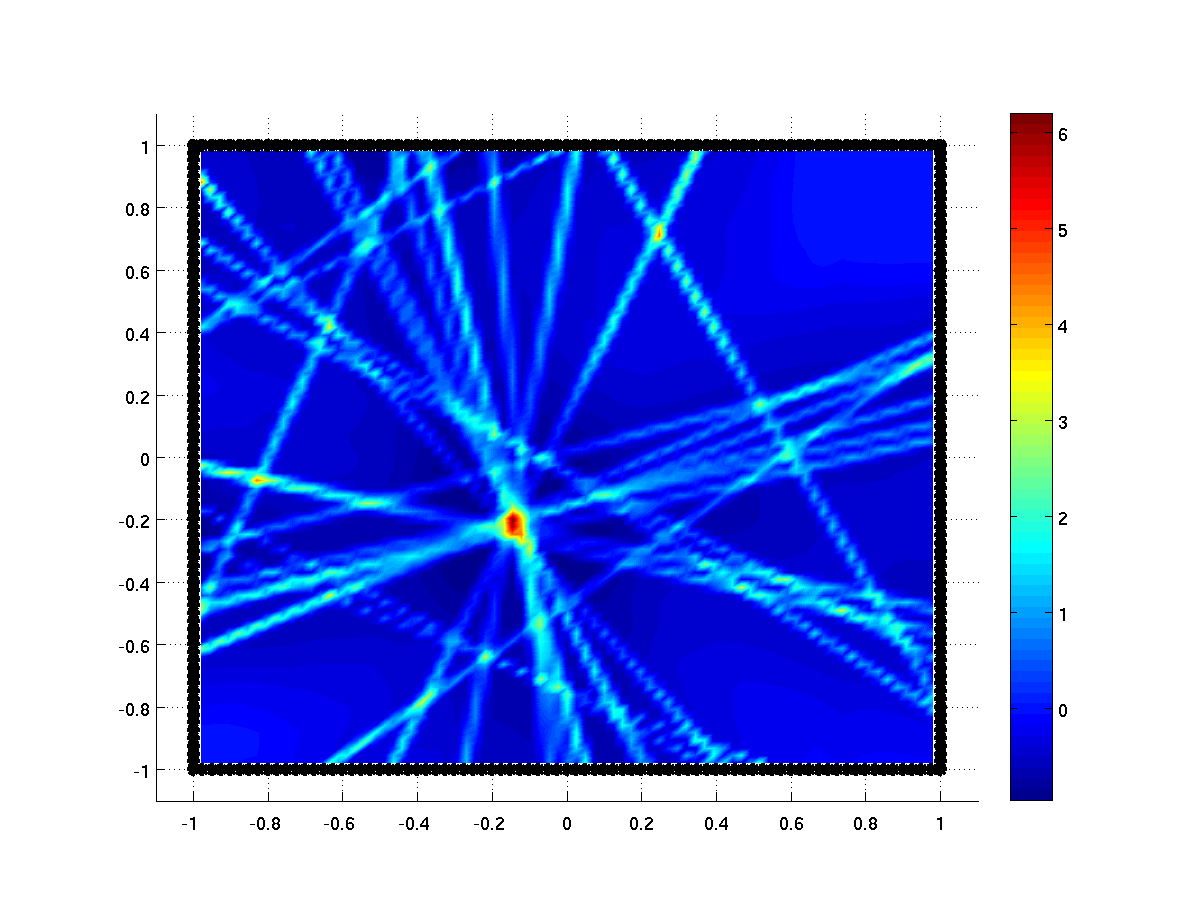}
\includegraphics[width=7cm]{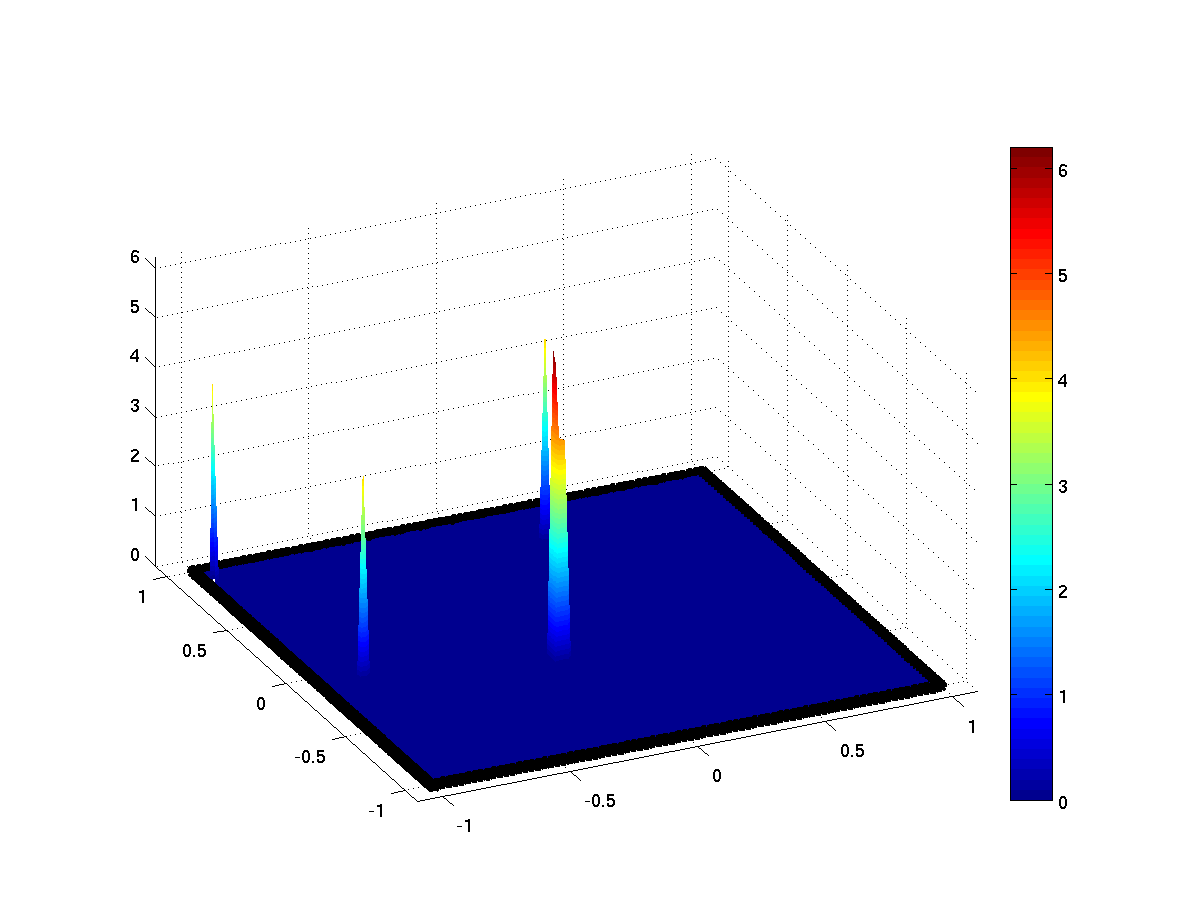}
\caption{Scenario \#5 with 4 detector arrays. Exposure Time = 60 seconds}\label{4DS5T60}
\end{center}
\end{figure}

%%%%%%%%%%%%%%%%%%%%%%%%%%%%%%%%
In Cargo Scenario \#5, the container is filled with self-radiating fertilizer (see its description in Section \ref{S:physics}) and contains an HEU source.
The situation is very much like the uniform random background case of \cite{AHKK} (see also Section \ref{S:math}).
One thus expects firm detection, which indeed happens to be the case, and detection probabilities computed in \cite{AHKK} being applicable.
%In the next figure, we use four detector arrays with 81 equally spaced detectors per array and 180 angular bins.
Figure \ref{4DS5T60} shows the back projection %, back projection with the local mean subtracted, and the back projection minus the local mean
with cut off at $4.3\sigma$ for an exposure time of 60 seconds.
%In 60 seconds, we detect 26 particles at 1 MeV, 16 of which are ballistic, which gives a signal to noise ratio of 1.45.
The confidence based on the threshold value of $4.3\sigma$ is 94.68, however the detection confidence based on the computed value of $6.2\sigma$ is 99.9998\%.
Elimination of the bottom detector does not spoil things much and gives detection with the confidence of 99.98\%.
%In Figure \ref{3DS5T120} we use three detector arrays with 81 equally spaced detectors per array and 180 angular bins.
%In 120 seconds, we detect 26 particles at 1 MeV, 11 of which are ballistic, which gives a signal to noise ratio of 0.733.
%The confidence based on the threshold value of 4.3 sigmas is 94.68\%.
%The confidence based on the maximum value of 6.2 sigmas is 99.98\%.
\begin{figure}[H]
\begin{center}
\includegraphics[width=7cm]{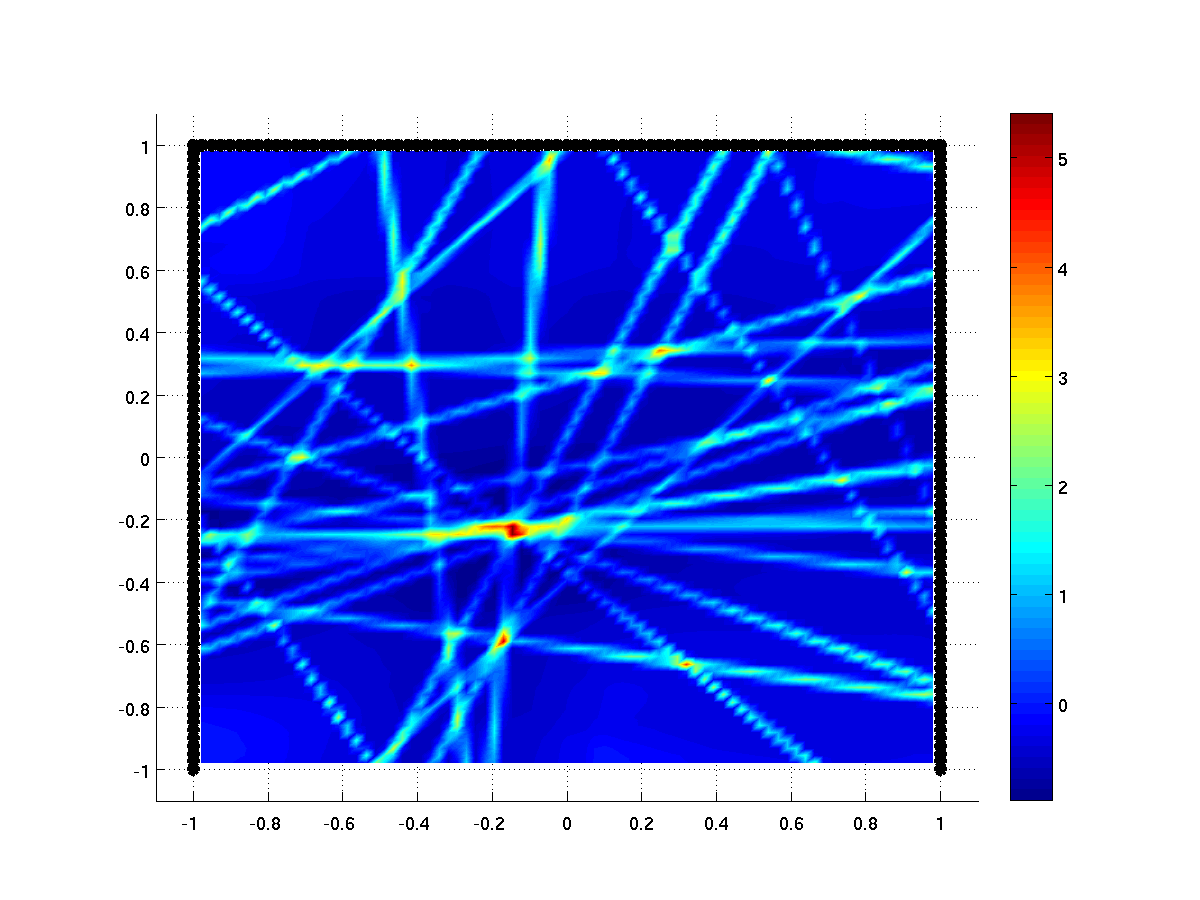}
\includegraphics[width=7cm]{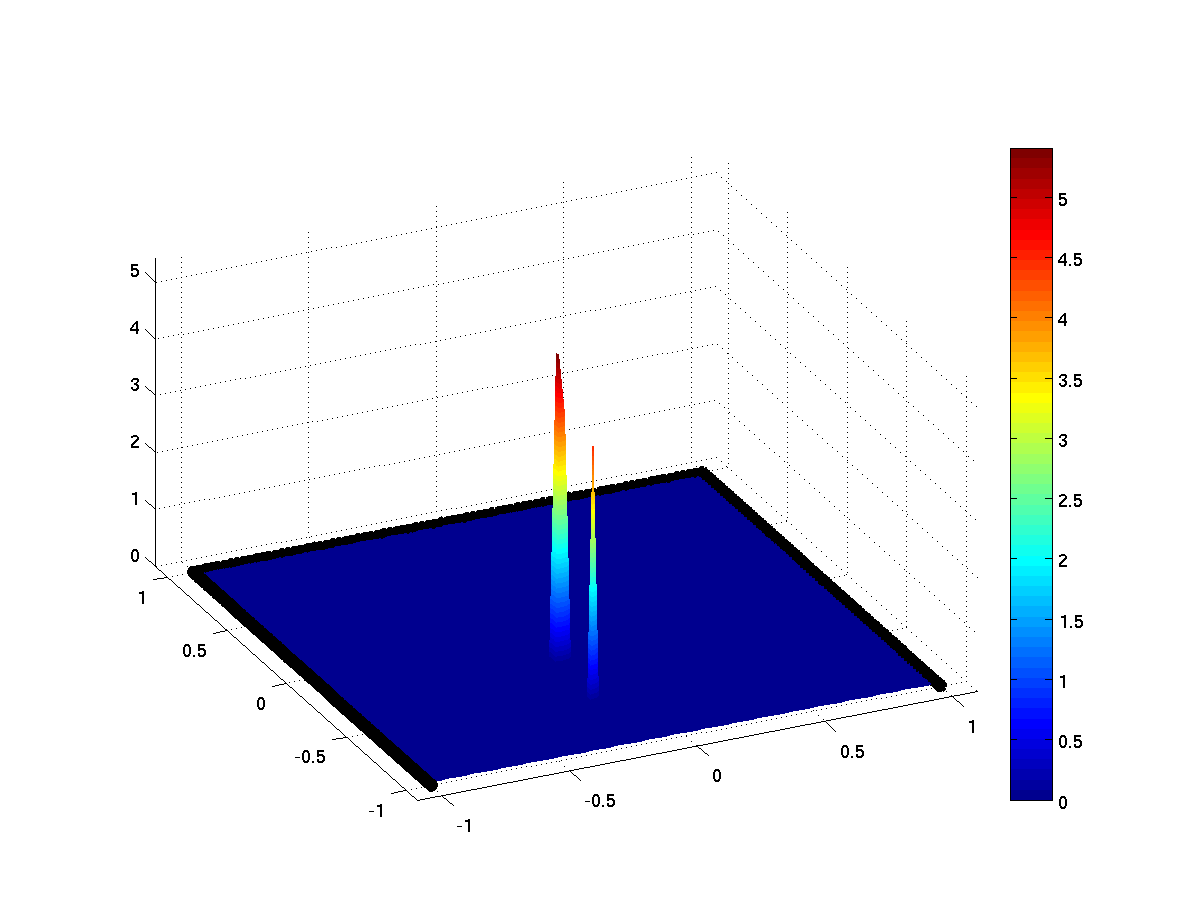}
\caption{Scenario \#5 with 3 detector arrays. Exposure Time = 2 minutes}\label{3DS5T120}
\end{center}
\end{figure}

%%%%%%%%%%%%%%%%%%%%%%%%
\section{Empty spaces scenarios}
%%%%%%%%%%%%%%%%%%

In this section, we make an interesting numerical observation, which needs to be studied further analytically and numerically.
Namely, numerical evidence that we provide below indicates that, unlike in previous scenarios, where detection of the presence of a source coincides with finding its exact location,
there are situations when one might have a strong evidence of presence of the source, without detecting its location. This seems to depend on existence of narrow unfilled spaces that usually exist even in well loaded containers.

%%%%%%%%%%%%%
\subsection{Single Energy Group}
%%%%%%%%%%%%%%

For the following figures, in the forward calculation all radiation (source and background) was considered to be in one large energy group. No energy changes are taken into account and all detected particles are used for the detection.

This scenario consists of an HEU source in the middle. The gray blocks are iron. The rest is void.
The background radiation comes from all sides isotropically.
\begin{figure}[H]
\begin{center}
 \includegraphics[width=5cm]{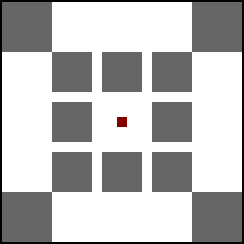}
\end{center}
\caption{Material Arrangement}\label{Grid}
\end{figure}
In this scenario, there is no direct (straight line) path between the source and the detector arrays that does not pass through the iron blocks, which practically eliminated the chance to rely upon ballistic particles coming from the source\footnote{Even tighter shielding has been tried with similar results.}. However, as the simulations below show, one can still reasonably infer the presence of an internal radiation source from differences between the images obtained.

\begin{figure}[H]
\begin{center}
\includegraphics[width=7cm]{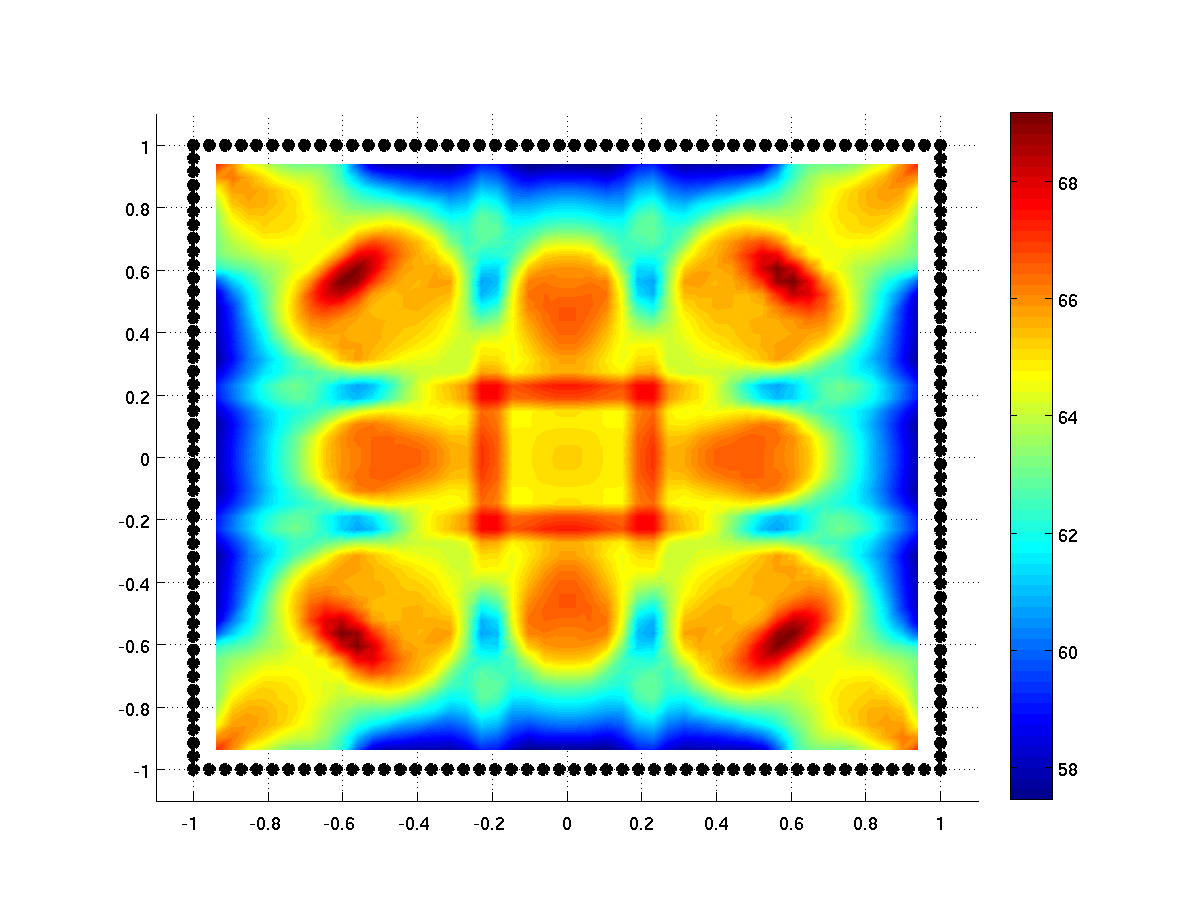}
\includegraphics[width=7cm]{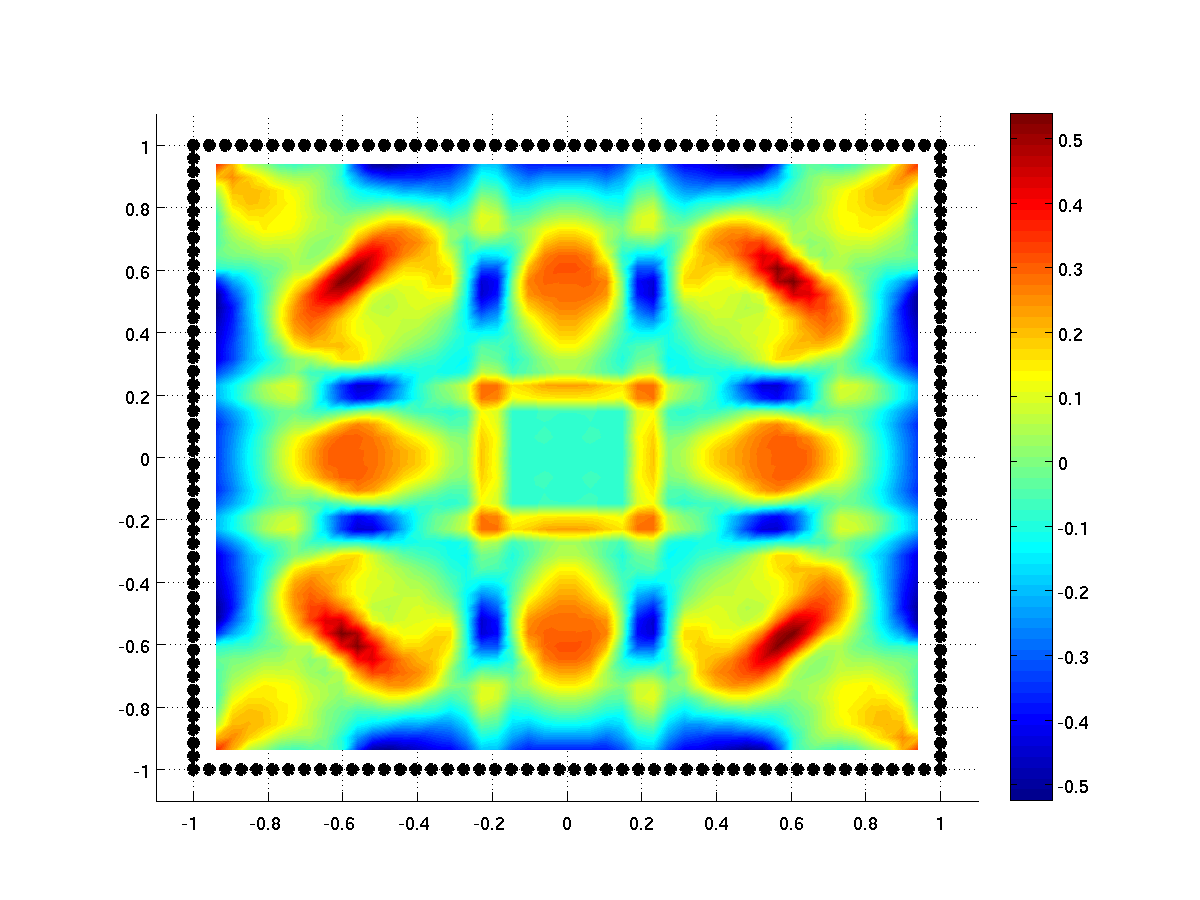}\\
\includegraphics[width=7cm]{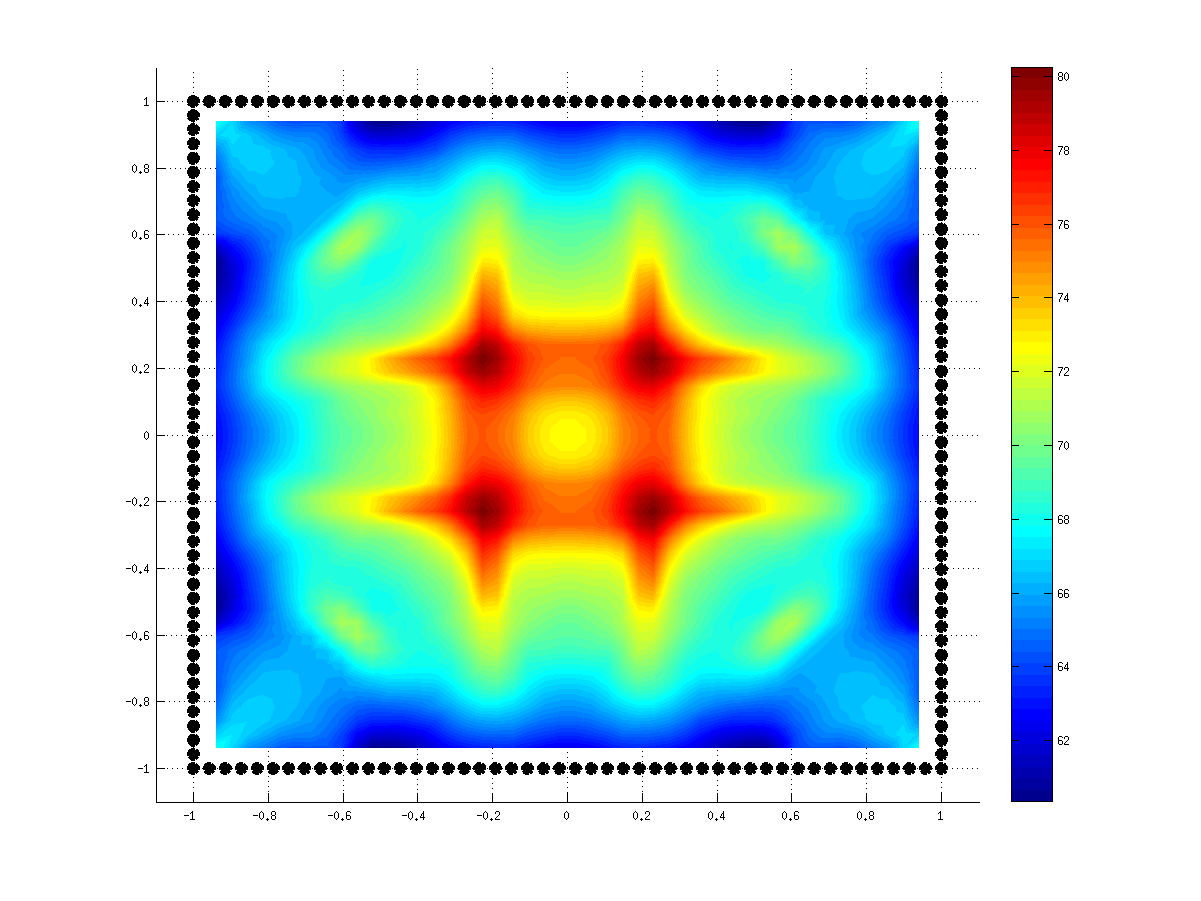}
\includegraphics[width=7cm]{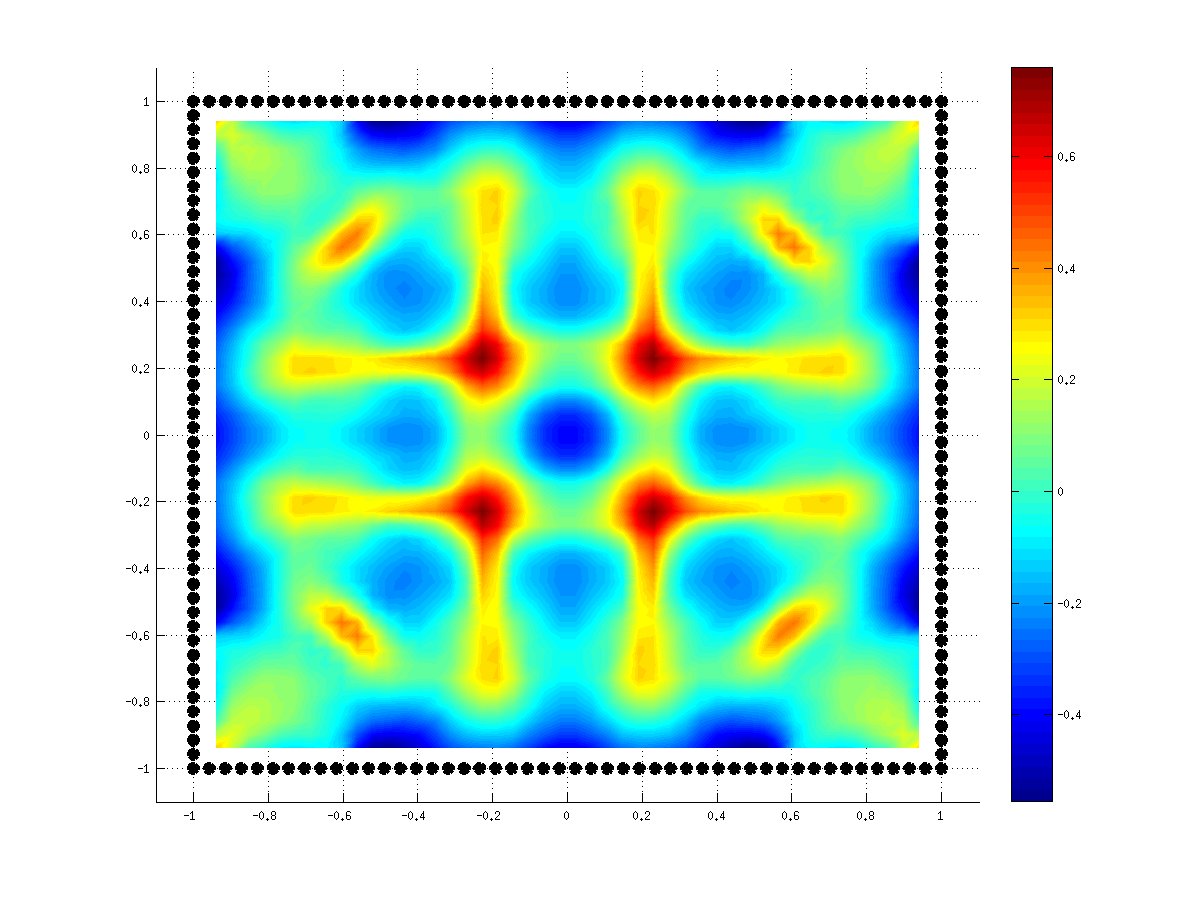}\\
\includegraphics[width=7cm]{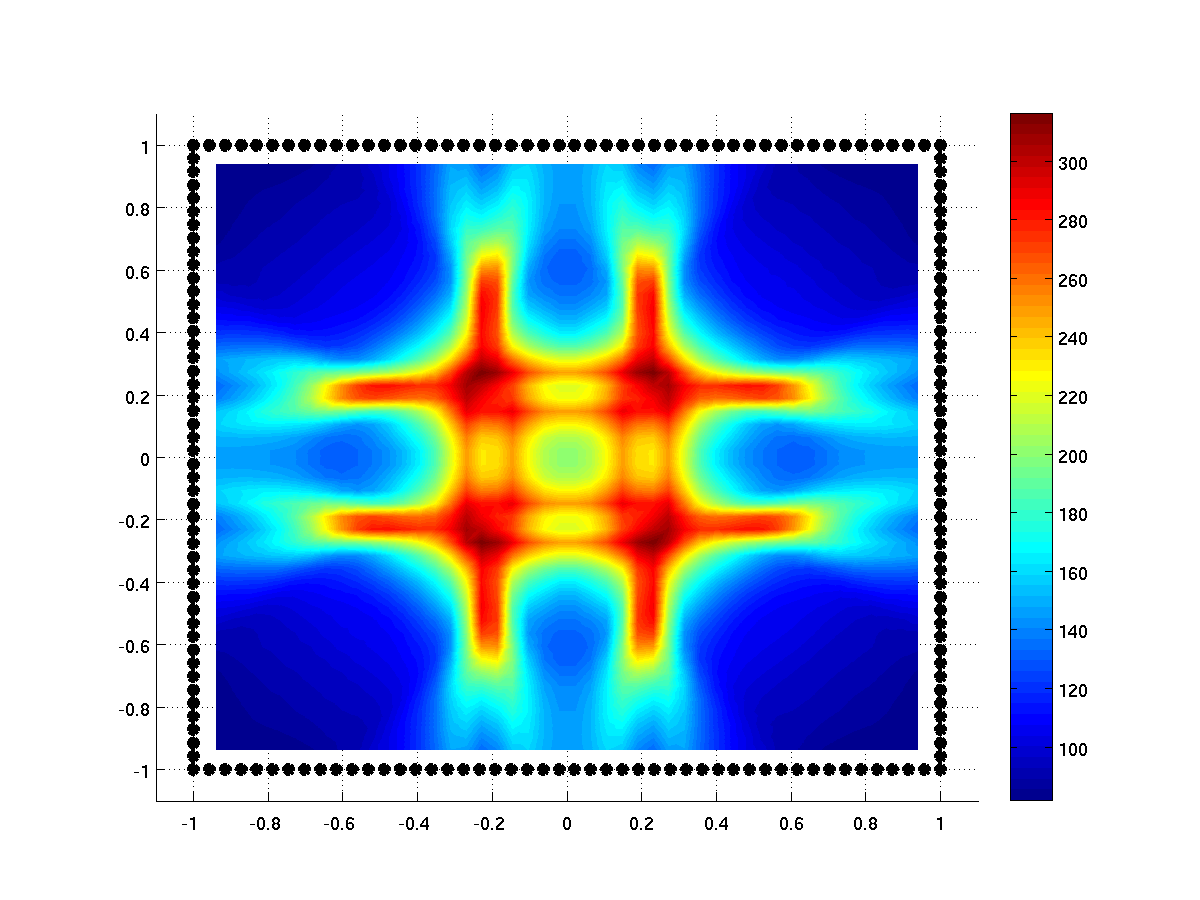}
\includegraphics[width=7cm]{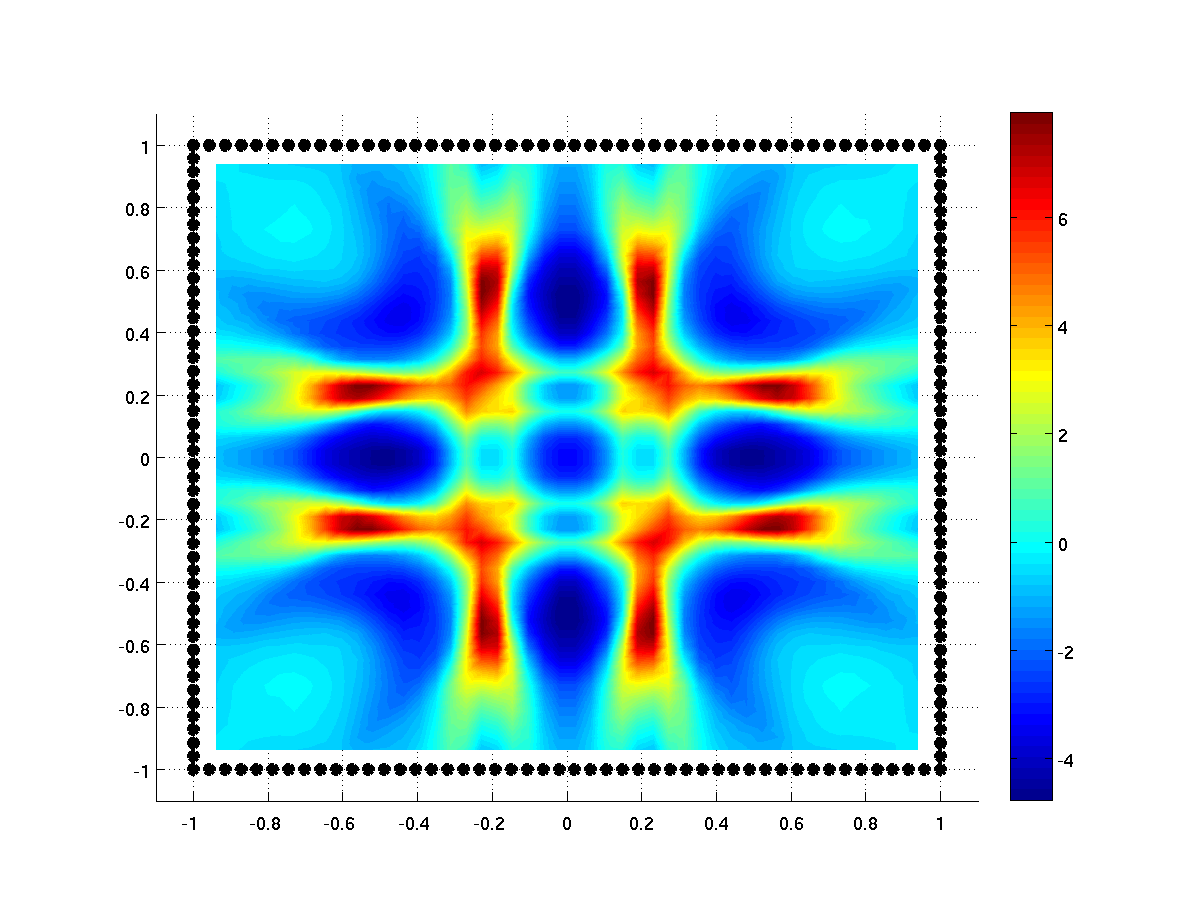}\\
\caption{Grid with 4 detectors at various SNR$=0,0.0001,0.001$}\label{4DGridSNR}
\end{center}
\end{figure}

The Figure \ref{4DGridSNR} shows the results of reconstruction procedure. The left column shows the backprojected image, while the right column shows the deviation from the (local) mean, measured in standard deviations. The top row shows the results without the source being present, i.e. just with the background and thus SNR$=0$. The middle row shows the presence of a very weak source (SNR=$0.0001$). Finally, the lowest row corresponds to our target SNR $0.001$.  Here the SNR is the ratio of the number all particles (not just ballistic) emitted by the source to the number of the background ones. The exposure time is long.
One notices that in presence of the source one gets much clearer picture of the passages between the cargo boxes. It is not only clearer geometrically, but also stronger numerically. Indeed, the color scales are different between the top and bottom rows. Otherwise the passages would not be seen well in the top row. So, the difference between the low and high intensities is about 6 times higher in the bottom picture than in the top one (with an intermediate situation in the middle row).

\begin{figure}[H]
\begin{center}
\includegraphics[width=7cm]{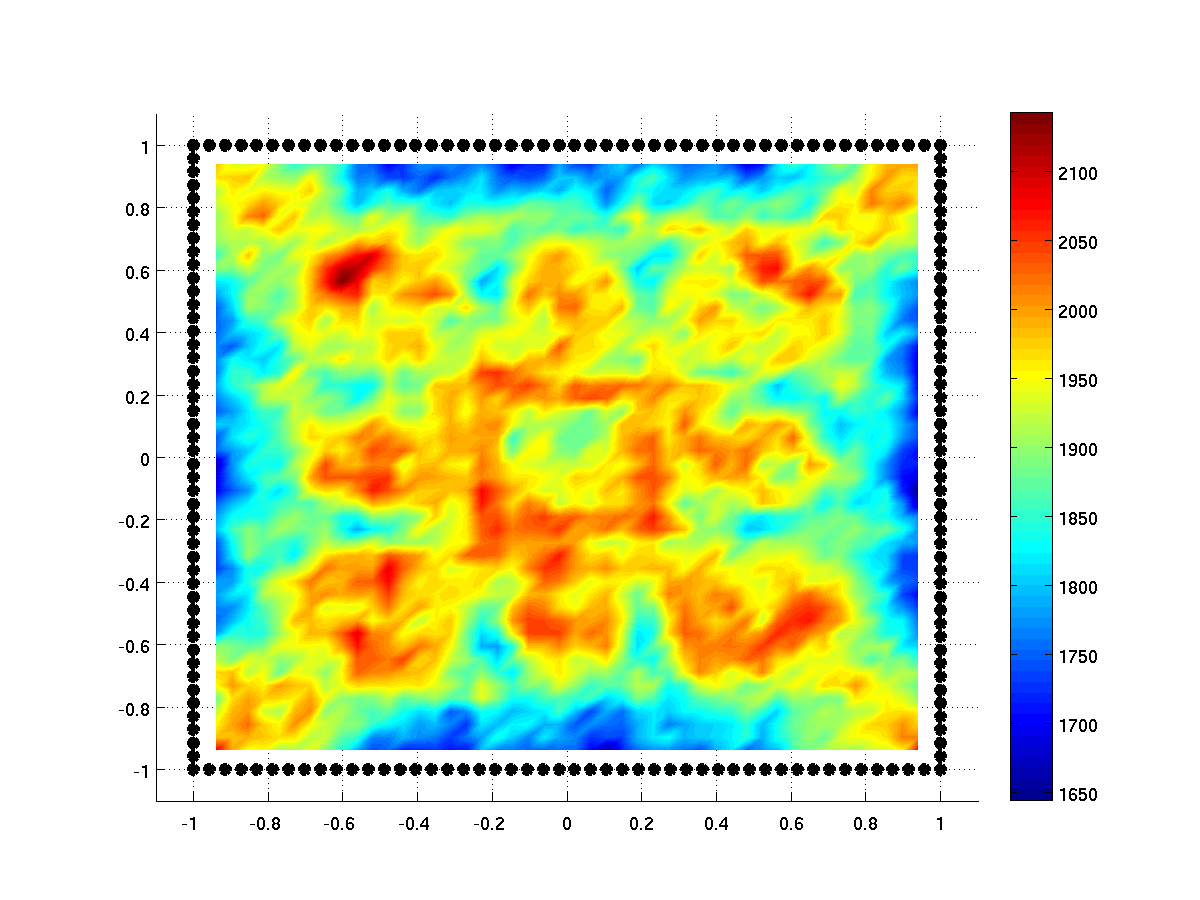}
\includegraphics[width=7cm]{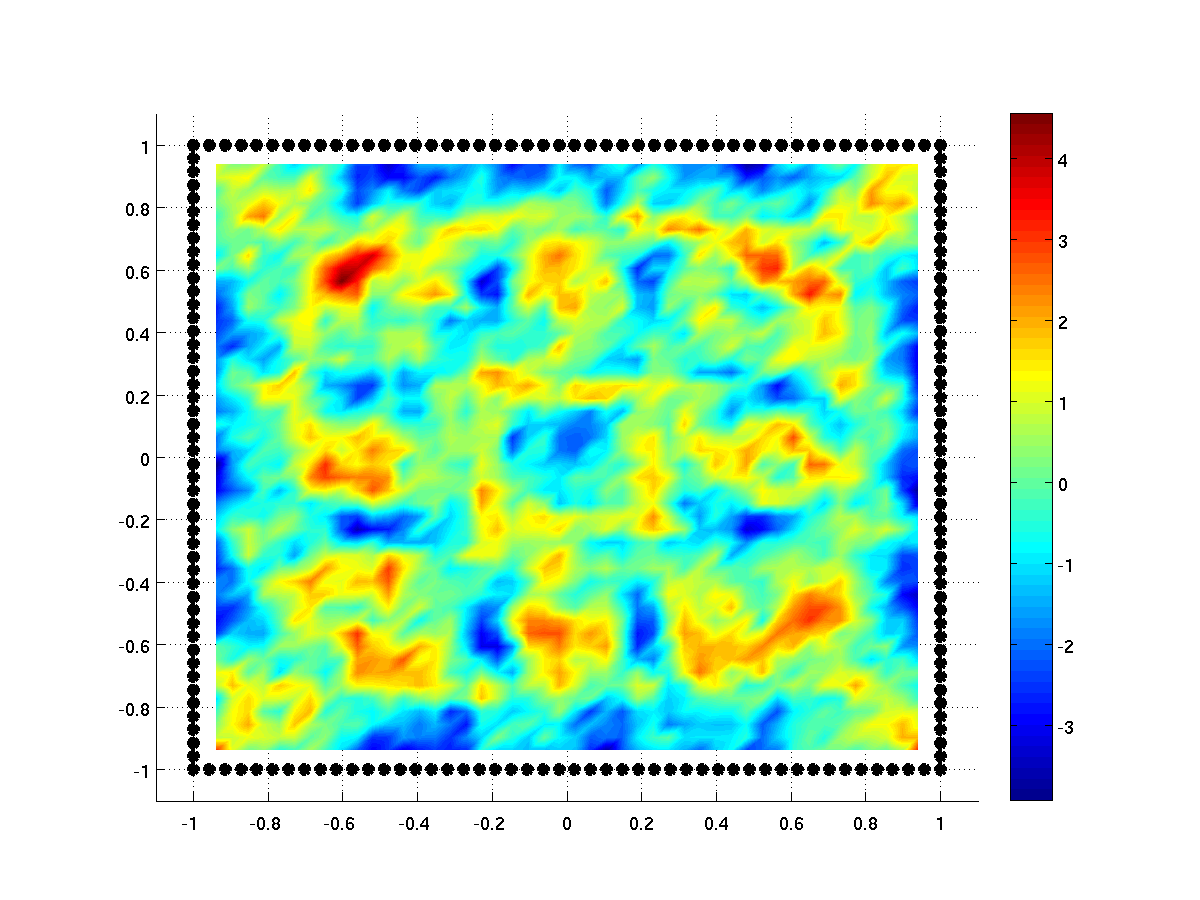}\\
\includegraphics[width=7cm]{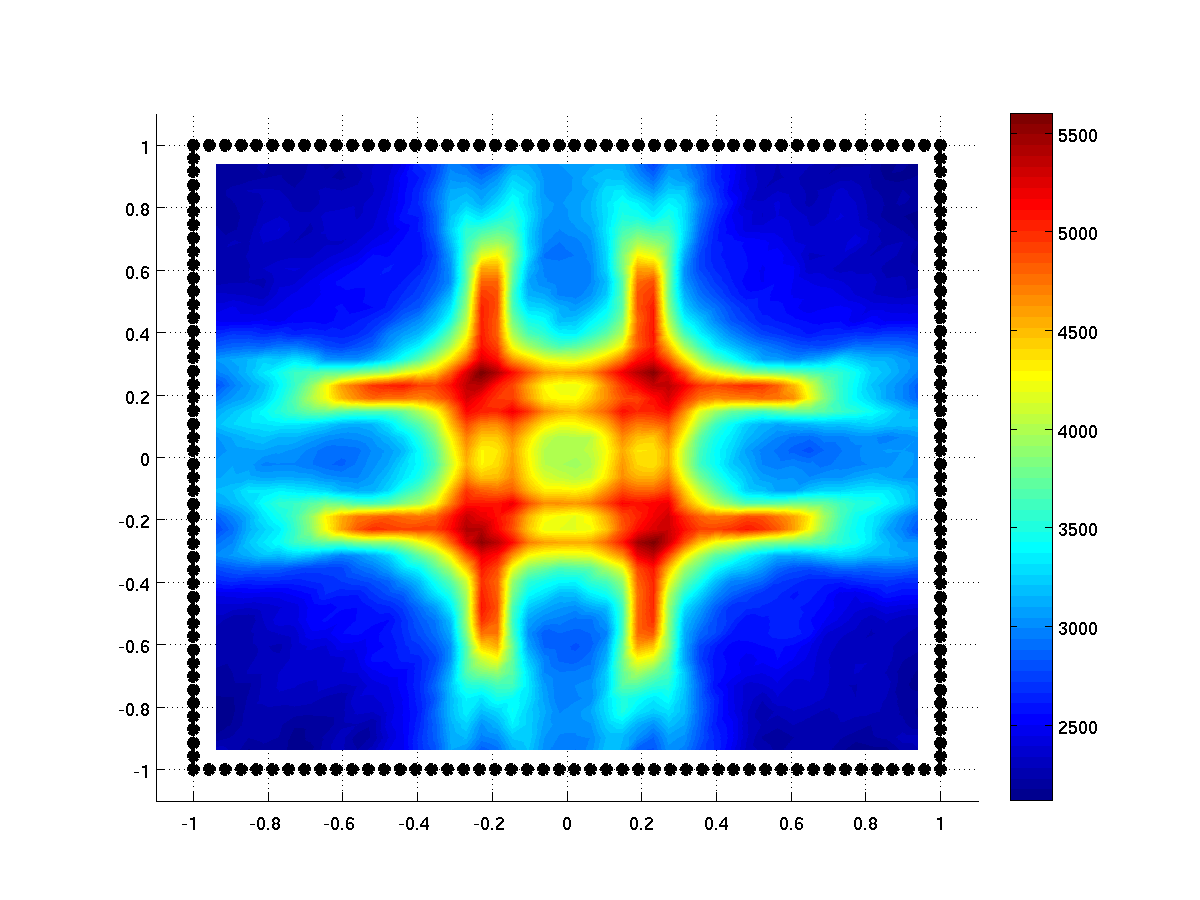}
\includegraphics[width=7cm]{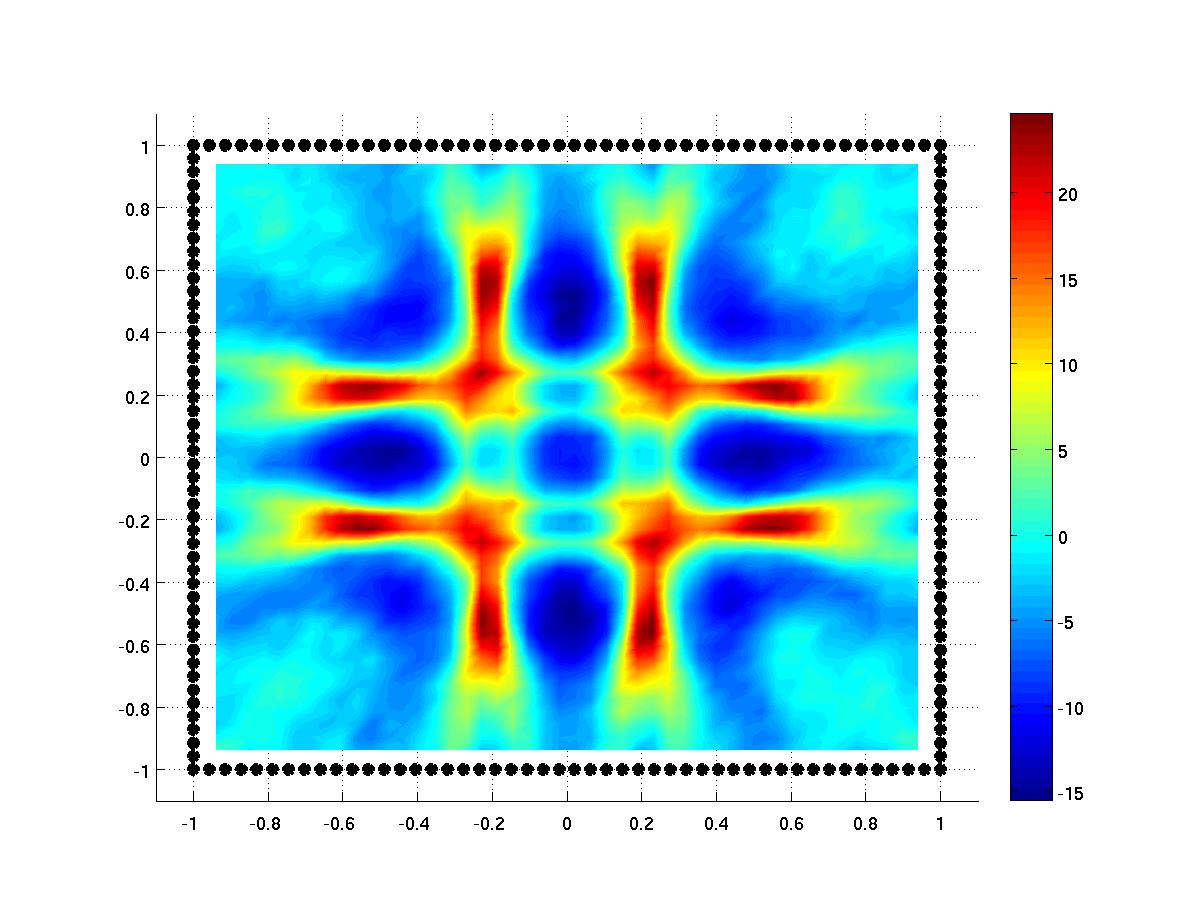}
\end{center}
\caption{Left Column: Back Projection, Right Column: local mean subtracted, SNR$=0,0.0006$. The results for a 30 second exposure time with just the background (top row) and with the HEU source (bottom row). Detected particle counts: 145,391 for background only and 210,178 for background plus HEU source (with just 134 ballistic)}\label{4DGridSNRSampled}
\end{figure}

One can ask whether the similar effect occurs for a reasonably short observation times. The Figure \ref{4DGridSNRSampled}. Shows that it does. The exposure time is $30$ seconds. The top row has only background, while the bottom one has a weak source with SNR$=0.0006$.
Here there is a much greater difference between the low and high points, which means the internal structure becomes visible at much more reasonable exposure times. (The figure below demonstrates this difference.)

\textbf{Multiple Energy Groups}

\begin{figure}[H]
\begin{center}
\includegraphics[width=7cm]{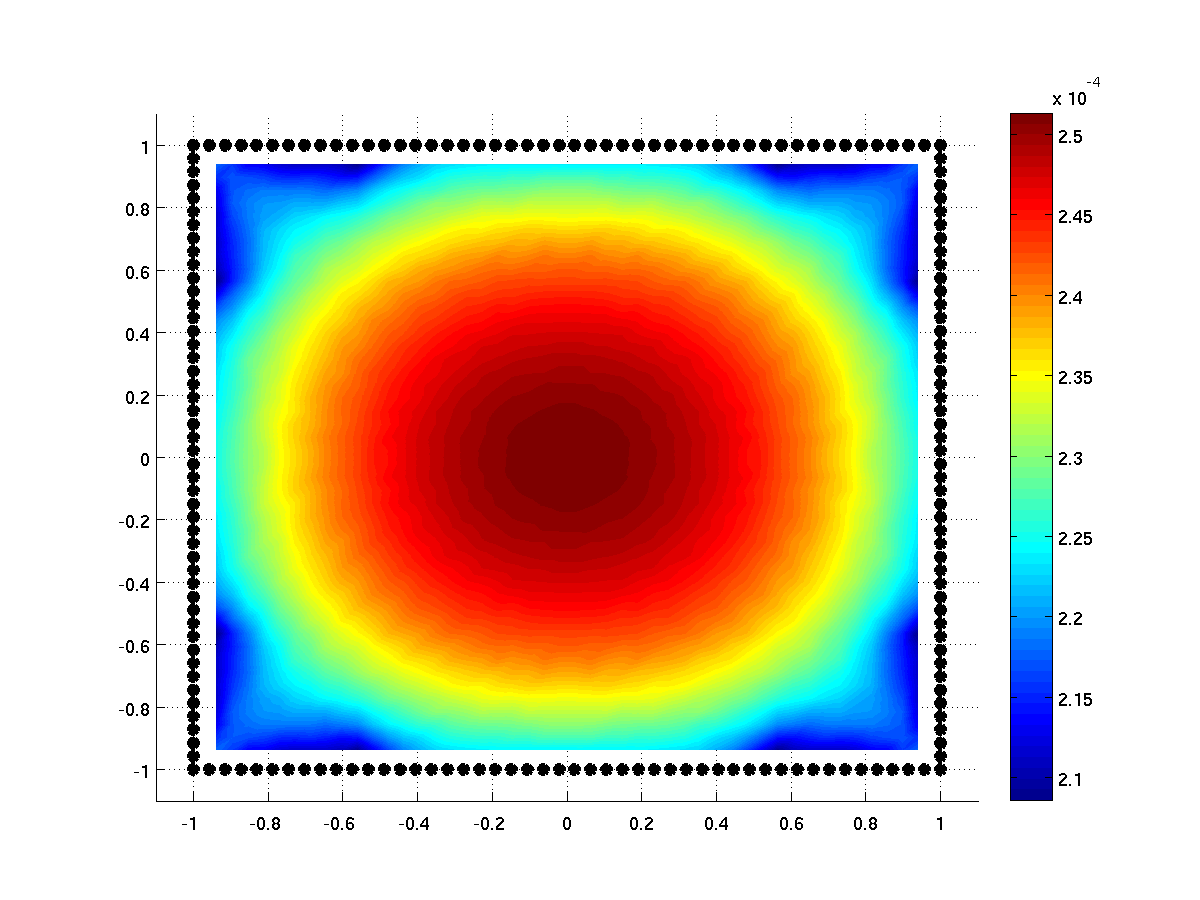}
\includegraphics[width=7cm]{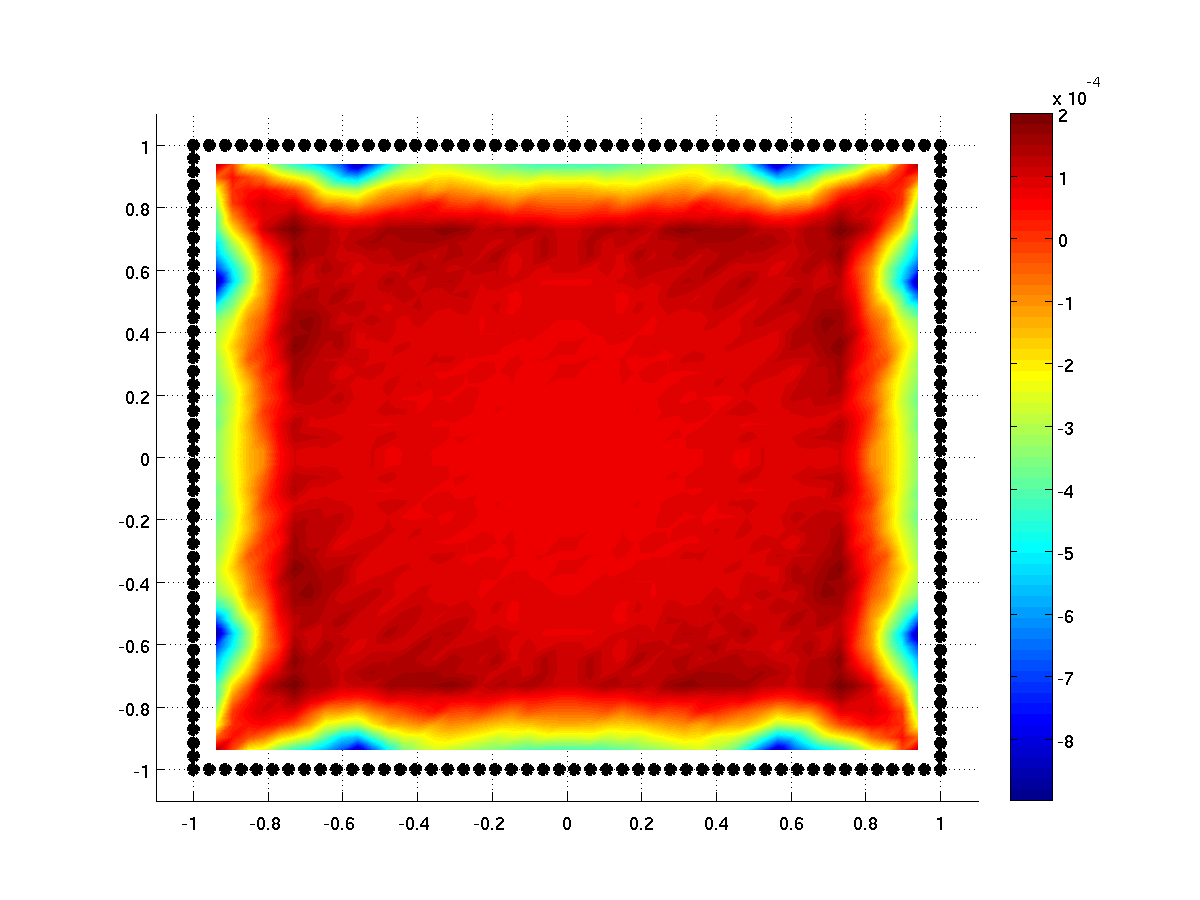}\\
\includegraphics[width=7cm]{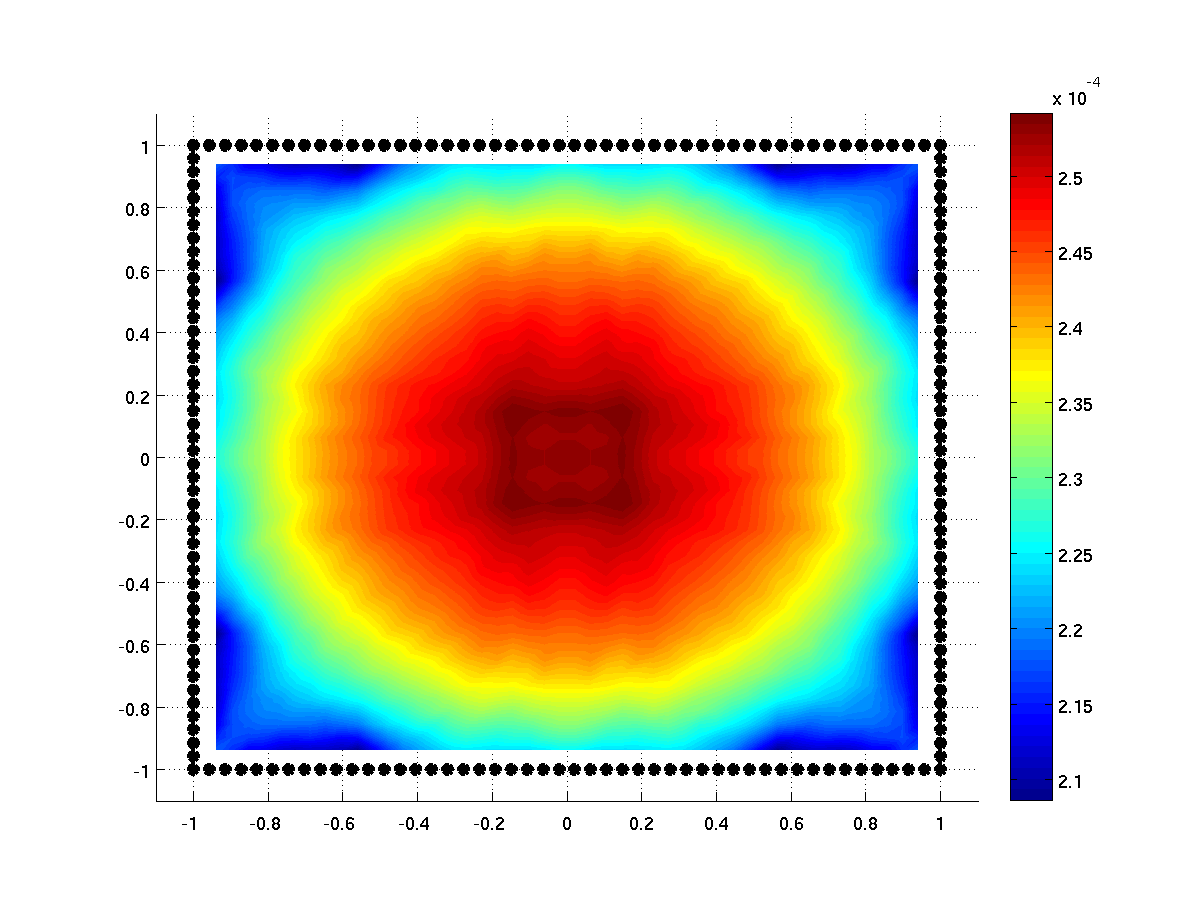}
\includegraphics[width=7cm]{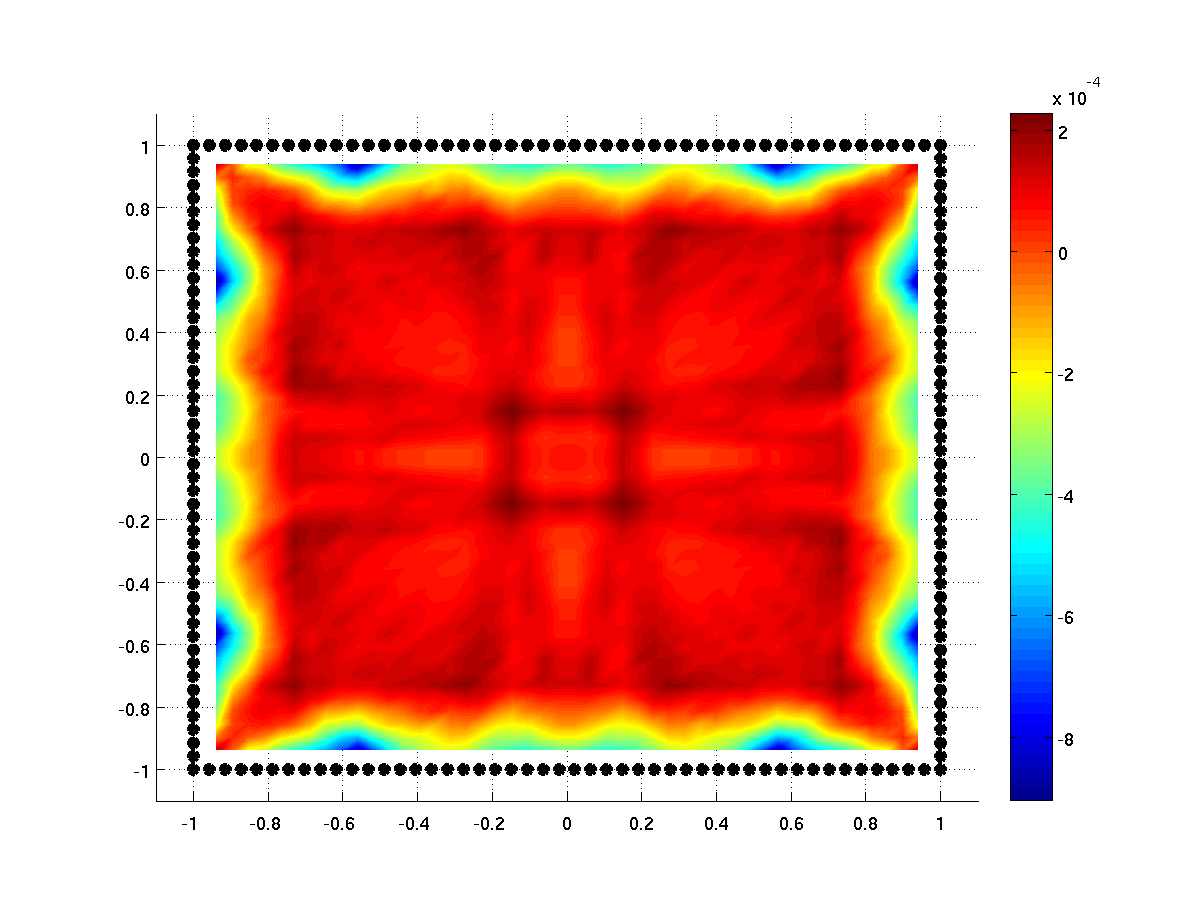}\\
\includegraphics[width=7cm]{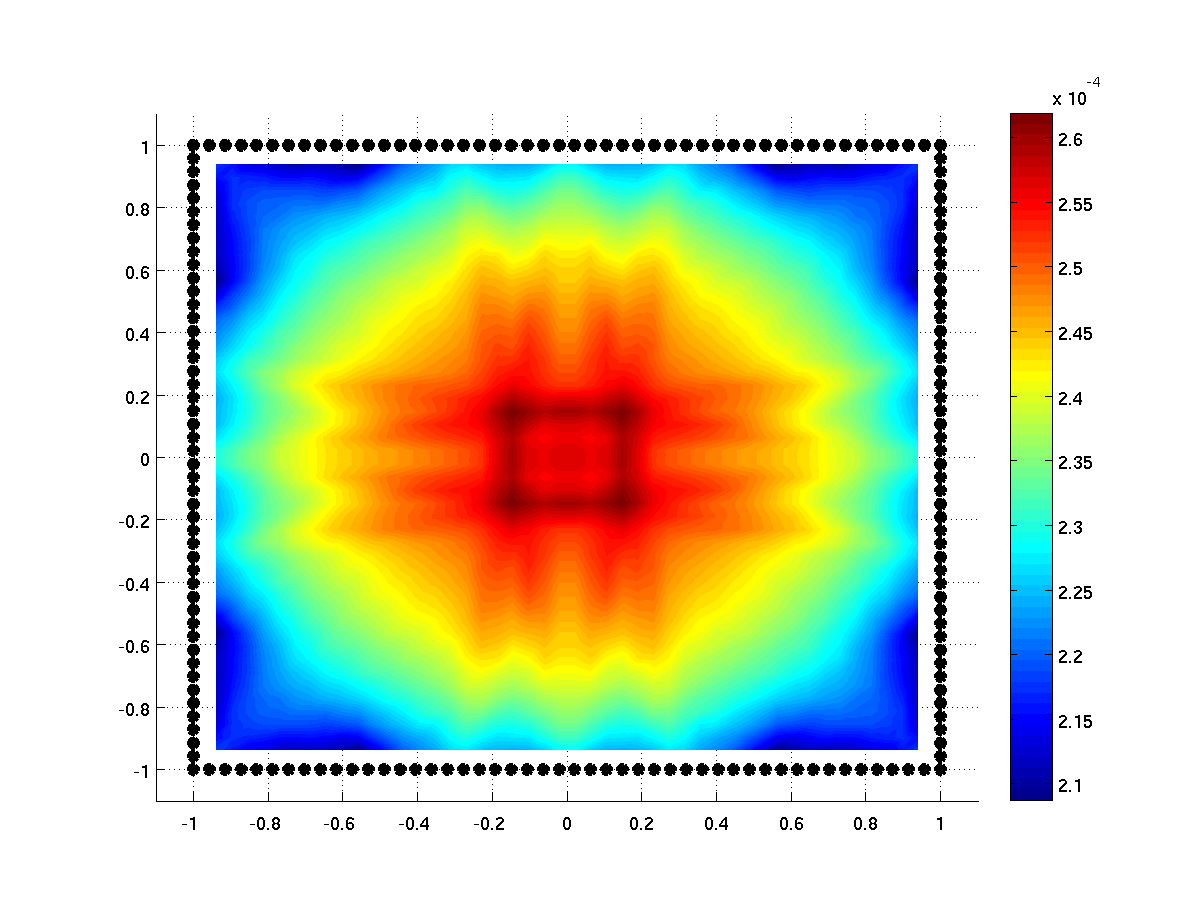}
\includegraphics[width=7cm]{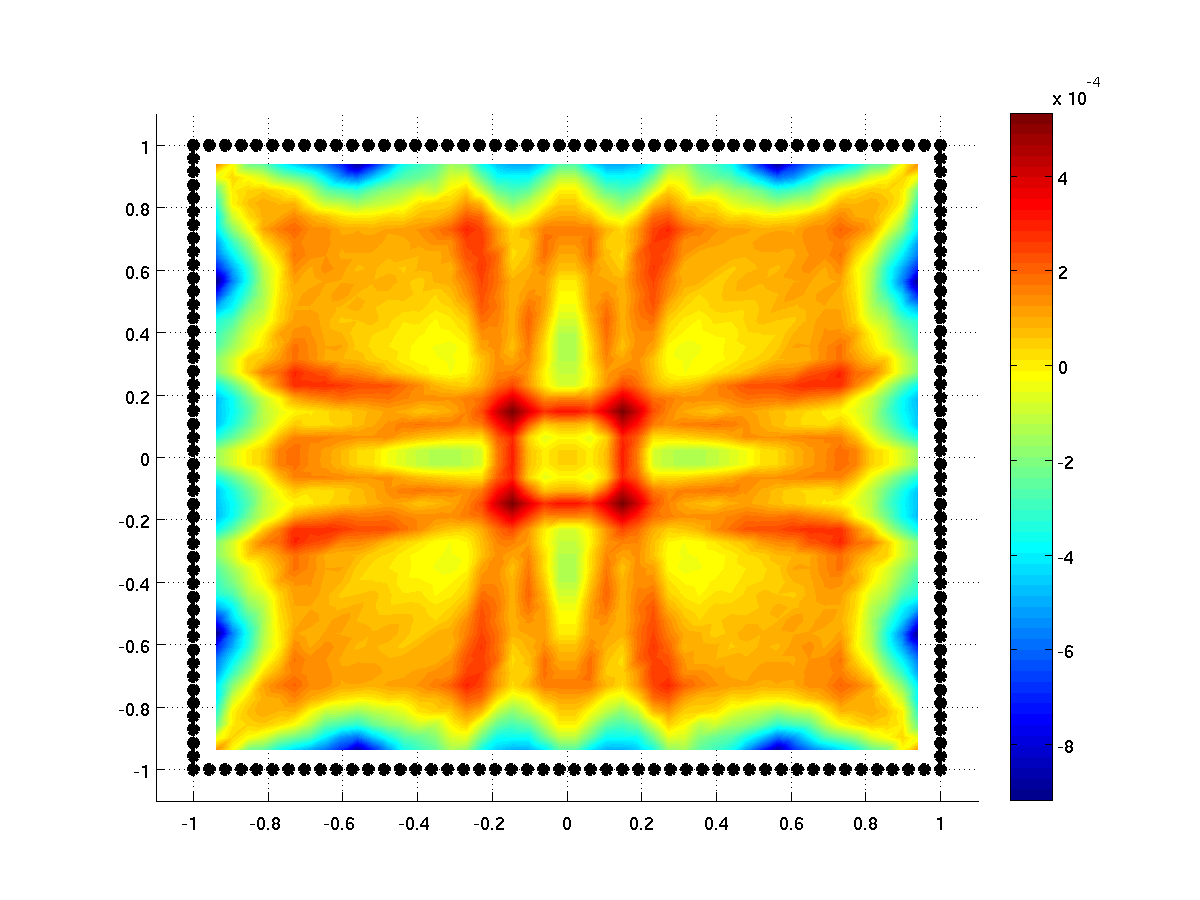}\\
\includegraphics[width=7cm]{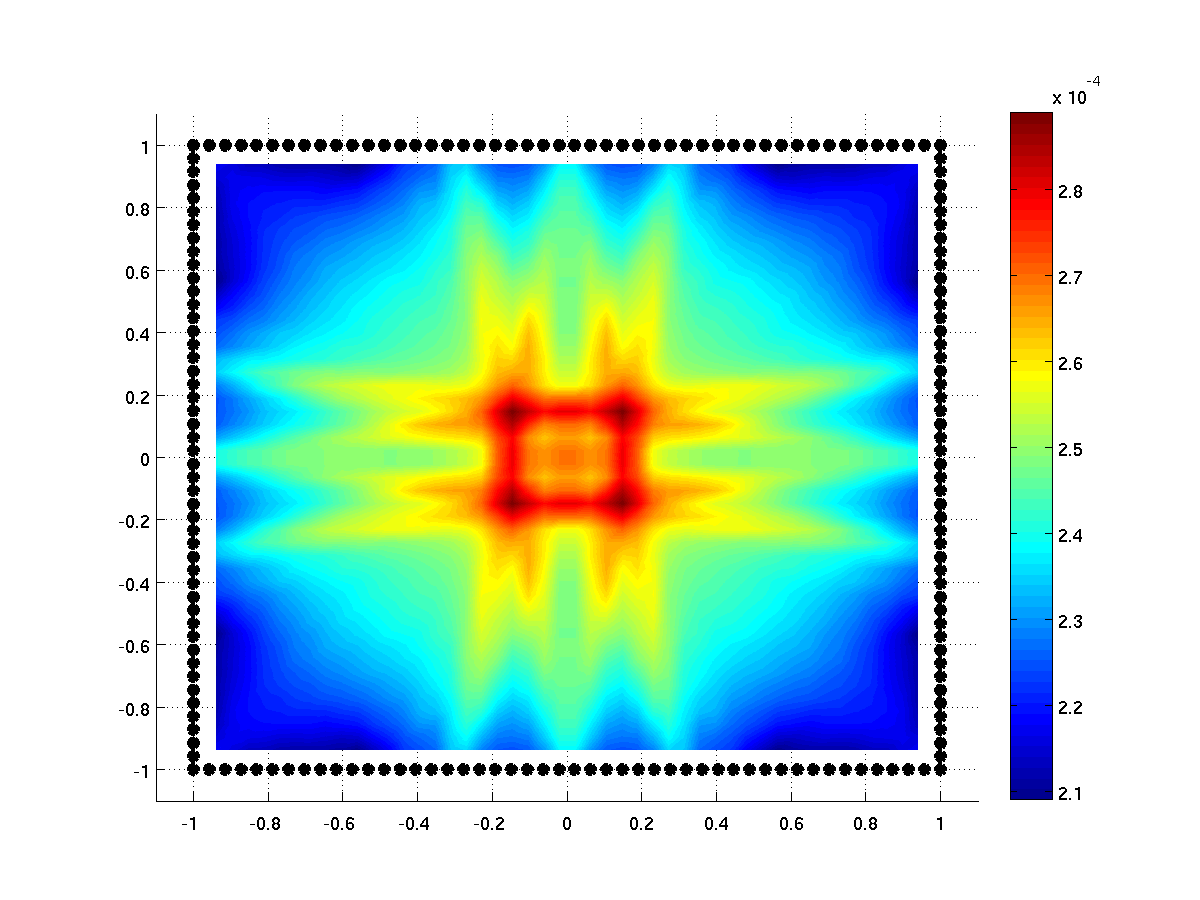}
\includegraphics[width=7cm]{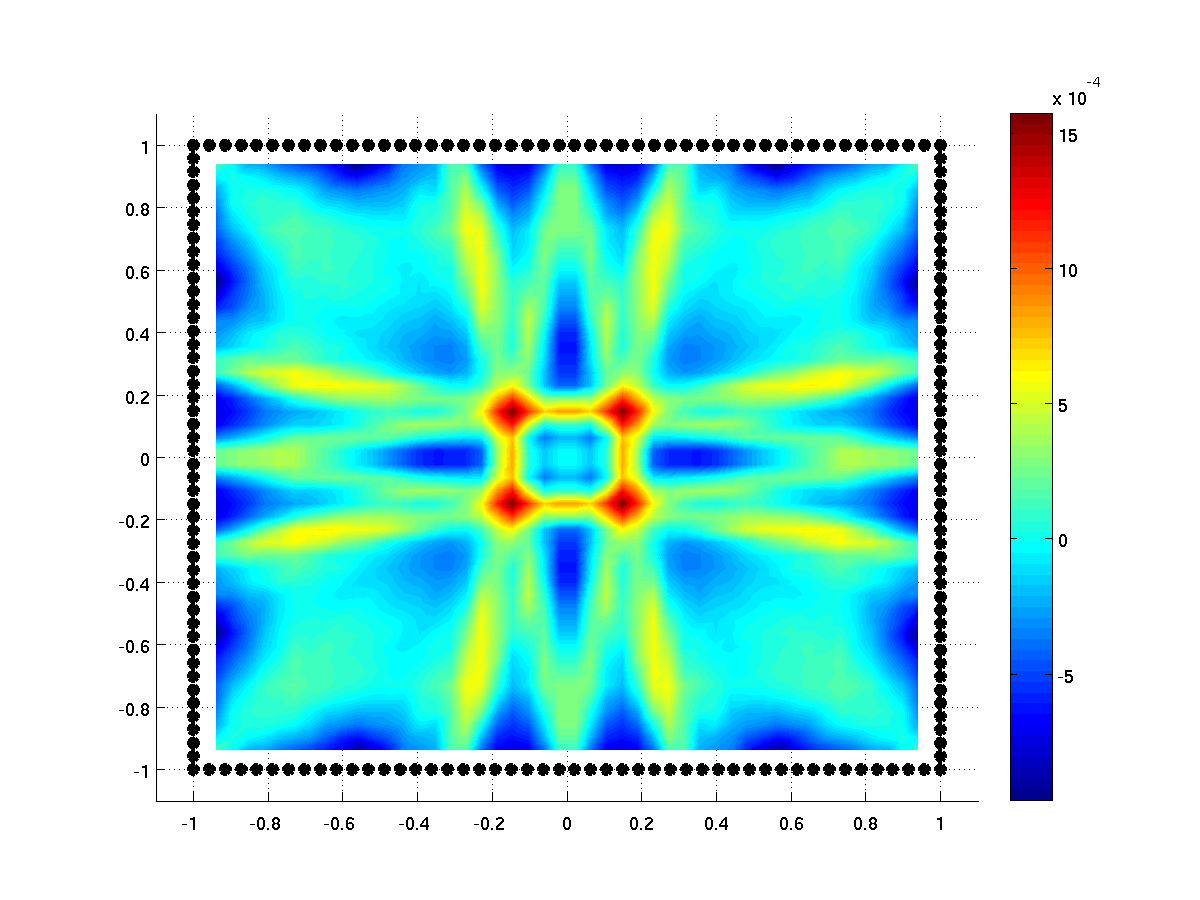}\\
\end{center}
\caption{Results at SNR$=0,0.001,0.003,0.01$. Left Column: Back Projection. Right Column: Local Mean subtracted}\label{Multiple}
\end{figure}

For this scenario (as with all the cargo container scenarios),the forward modeling was done using multiple energy groups as described in the forward modeling section. The background radiation is due to concrete and, in this case, is coming from all sides isotropically. All energy groups were modeled in the forward calculation, with energy changes allowed. For the back projection, only energy group 37 is used, that is, we only deal with the subset of data from the 1 MeV range for the inverse problem. Due to the energy changes, this group is polluted by additional noise. The Figure \ref{Multiple} shows that this time we cannot see the internal passages without an HEU source (top row). This is due to the fact that concrete does not radiate in the energy group we are back projecting. Thus any radiation due to background detected in that energy group must have been scattered.
As we increase the strength of the source (second row: SNR = 0.001, third row: SNR = 0.003, bottom row: SNR = 0.01), the empty corridors in the container become more and more apparent.

\begin{figure}[H]
\begin{center}
\includegraphics[width=7cm]{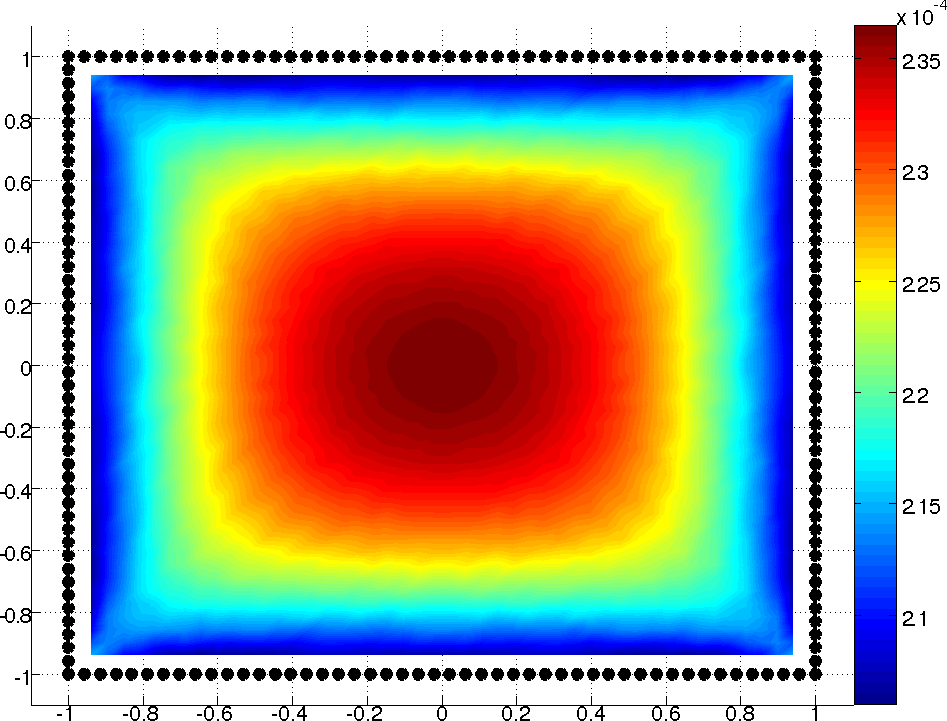}
\includegraphics[width=7cm]{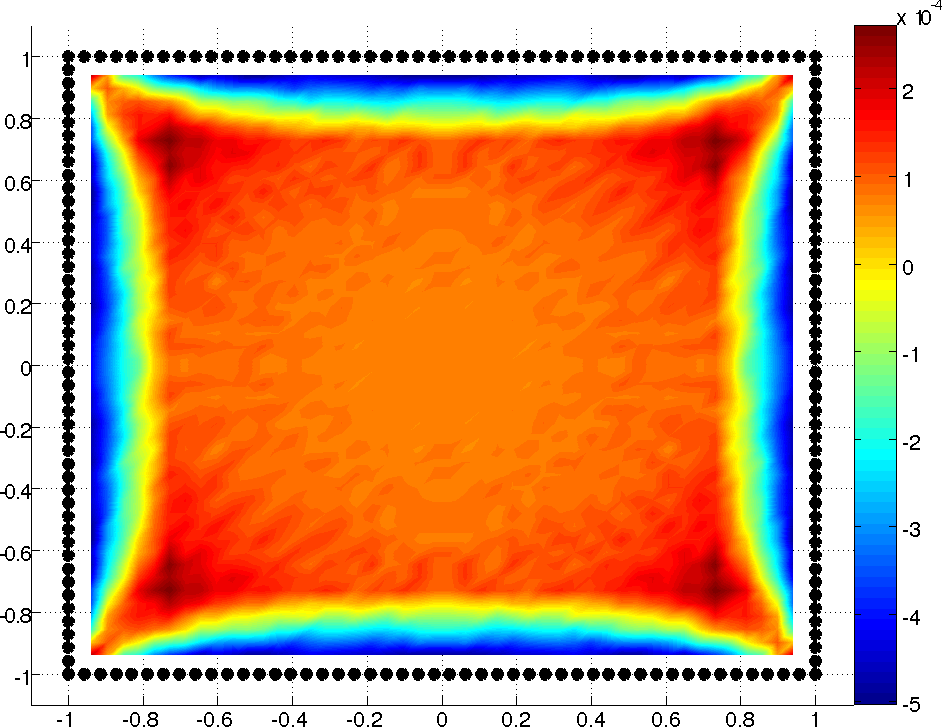}\\
\includegraphics[width=7cm]{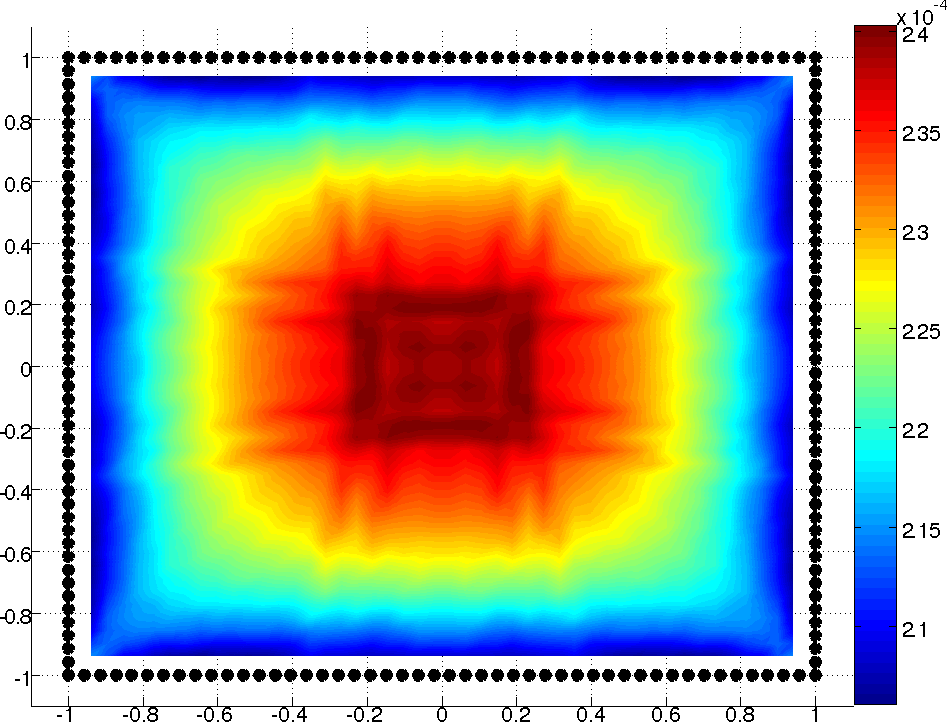}
\includegraphics[width=7cm]{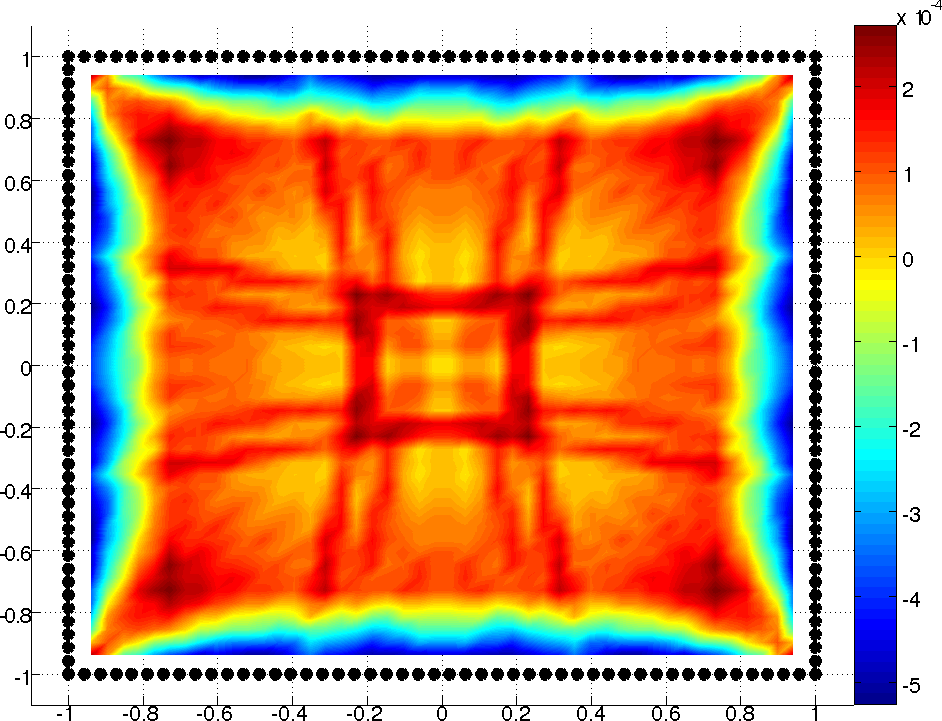}\\
\includegraphics[width=7cm]{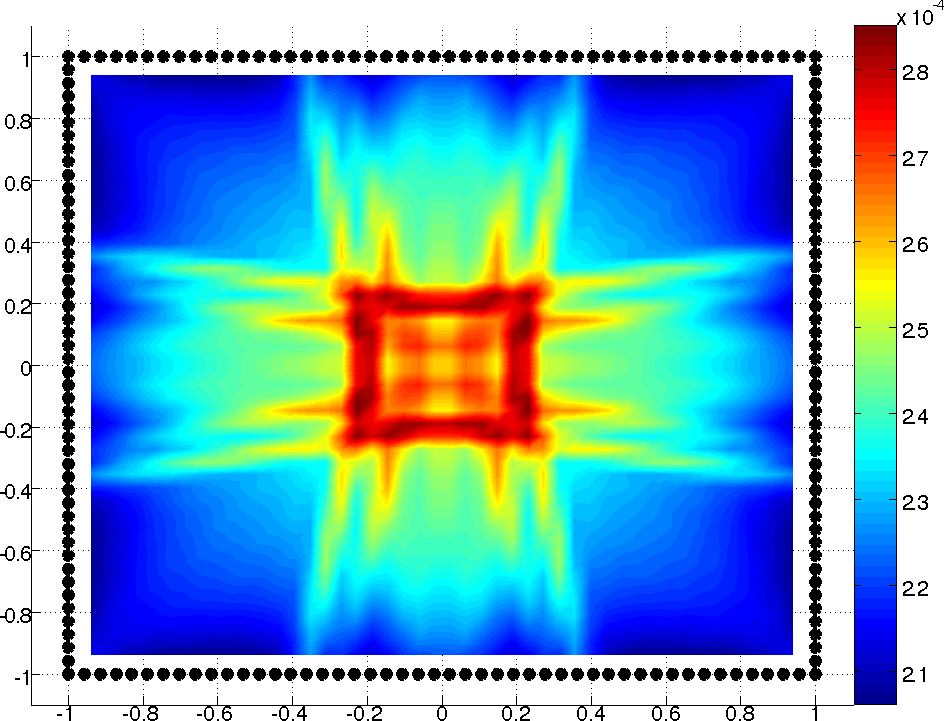}
\includegraphics[width=7cm]{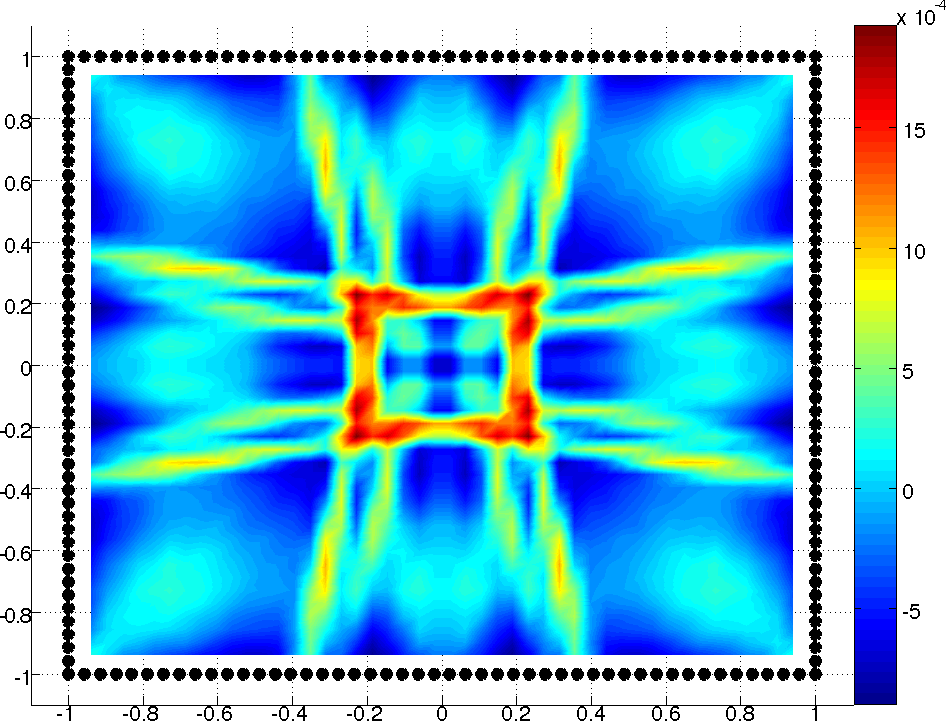}\\
\includegraphics[width=7cm]{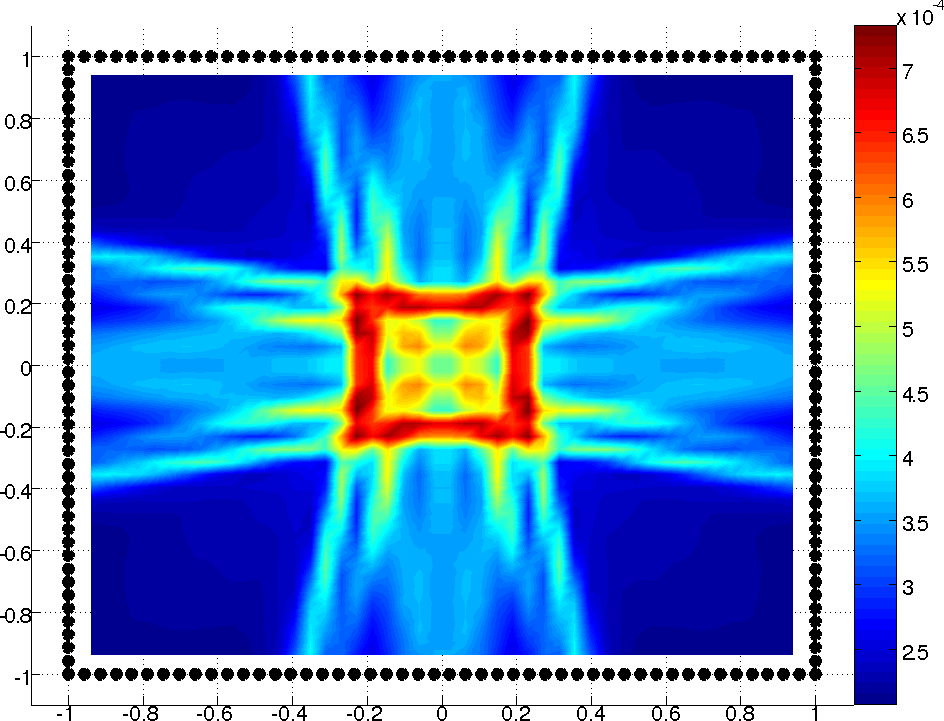}
\includegraphics[width=7cm]{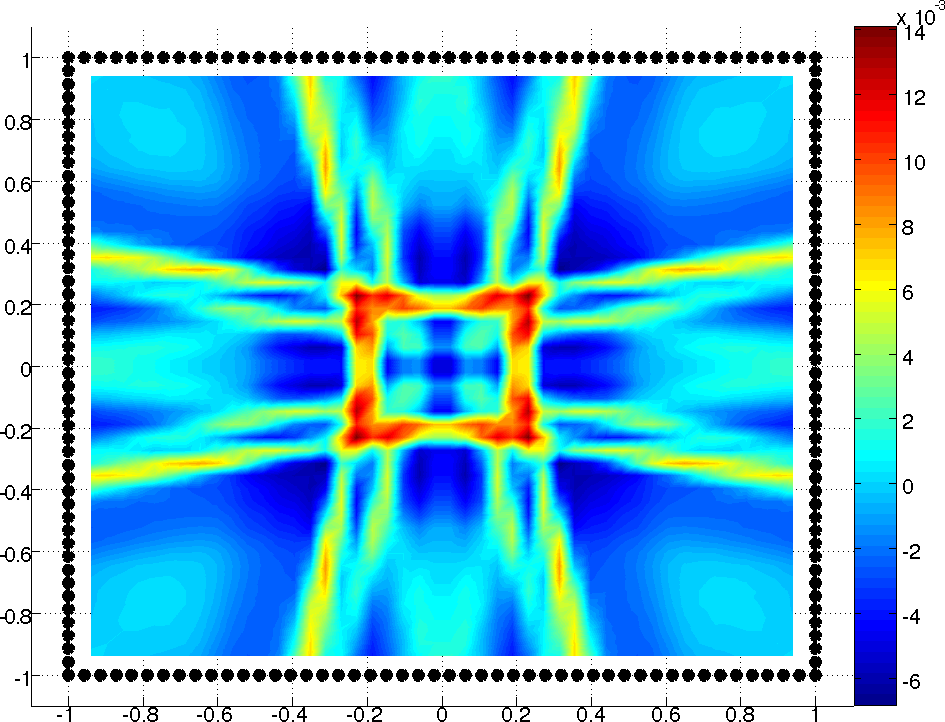}\\
\caption{Results at SNR$=0,0.0001,0.001,0.01$. Left Column: Back Projection. Right Column: Local Mean subtracted}\label{Grid_rec}
\end{center}
\end{figure}

One more example below shows a stronger iron shielding of the source (Fig. \ref{Grid}) and long reconstruction time (Fig. \ref{Grid_rec}).

\begin{figure}[H]
\begin{center}
 \includegraphics[width=5cm]{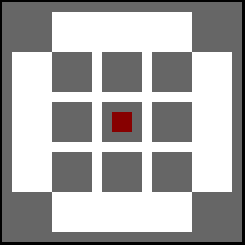}
\end{center}
\caption{Material Arrangement}\label{Grid}
\end{figure}

Again, we see that even though the actual location of the internal source is not detected, we can reasonably deduce that there is one. Clearly, additional analytic and probabilistic considerations are needed to make these results more quantitative and reliable. This is planned for another publication.

%%%%%%%%%%%%%%%%%%%%%%%%
\section{Final remarks and conclusions}\label{S:remarks}
%%%%%%%%%%%%%%%%%%%%%
The results of this study show that the method developed in \cite{AHKK} works well for detecting the presence of small amounts of special nuclear materials placed in large containers/trucks filled with various realistic types of cargo. Not only sensitivity, but also specificity of the method, as it was shown in \cite{AHKK} and in this text, can be very high even in cases of an extremely low SNR, in the ballpark of $0.001$.

The suggested technique is highly parallelizable and can be run simultaneously with the measurements being collected.

There are, however, various issues still to be addressed in the future work. Among those, one can mention the following:

\begin{enumerate}

\item The technique has been implemented to the full extent in the 2D situation only. Although some trial runs have been done in $3D$, the full implementation of numerics (for both forward modeling and detection) still needs to be done.

    One can ask whether the $2D$ implementation has any relation to the real world situation at all then, besides running the technique in numerically simpler situation. For instance, one wonders whether one could try to image a $2D$ slice of the container. It is possible in principle, but it would clearly reduce significantly the already small SNR, since the emission from the source is highly unlikely to honor the plane of the slice we choose for observation. This should render detectability to be much worse. Some modeling of this has been done, which confirms this conclusion.

    On the other hand, the $2D$ model we studied corresponds to the cylindrical $3D$ geometry of the cargo and detectors. Although this is also not a too realistic situation, it still can be considered as a useful trial run before the full $3D$ technique is implemented.

   % One can also notice that, in spite of the $3D$ being computationally somewhat more delicate, the detection
    %of a source of the same strength and same linear dimensions as in $2D$ might be easier in $3D$, as the formulas (\ref{E:rule}) and (\ref{E:rule_error}) show. Indeed, while in $2D$ the value of $p$ behaves as $r/R$, it is $(r/R)^2$, and thus smaller in $3D$.

\item The very intriguing and potentially important numerical observation made in the final section of the paper is that the presence of a source whose shielding does not allow for any significant number of ballistic particles getting to the detectors could still be implied by highlighted cargo passages. Although background radiation also can lead to this effect, our observation is that the strength of this highlighting is much stronger in the presence of a source. This observation needs to be made more precise by its analytic study, which we also intend to produce.

\item The reader should have noticed that some corner artifacts arise in many of our reconstructions. In some cases they are strong enough to compete with the detected source. Our preliminary study shows that this is just the effect of the square detector geometry that we choose. If a ring (sphere) of detectors is used, these artifacts are not present. One can deal with them easily in the square geometry as well, just by excluding the corners from the reconstruction domain. This comment will be expanded analytically and numerically in a future publications.
\end{enumerate}

%%%%%%%%%%%%%%%%%%%
\section*{Acknowledgments}
%%%%%%%%%%%%%%%%%

Work of W.~C., P.~K., A.~O., and J.~R. was partly supported by the
DHS Grant 2008-DN-077-ARI018-04. Work of M. A., Y. H., and P. K. was
partly supported by the NSF DMS grants 0604778 and 0908208. P. K, was also
partially supported by the NSF grant DMS-1211463. Work of M. A. and
P. K. was supported in part by the IAMCS (Institute for Applied Mathematics and
Computational Science of Texas A\&M University). Y. H. thanks the Institute for
Mathematics and Its Applications (IMA), where a part of the work was conducted,
for the support and hospitality. Research
at the IMA was supported by the National Science Foundation and the University
of Minnesota.

\end{document}